\documentclass[preprint,12pt]{elsarticle}

\usepackage{graphicx}
\usepackage[margin=0.5in]{geometry} 

\usepackage{epstopdf}

\usepackage{amssymb,amsmath}

\journal{Physics Reports}
 
\begin{document}

\begin{frontmatter}

\title{Quantum metrology and its application in biology}
\author{Michael~A.~Taylor$^{1,2}$ and Warwick~P.~Bowen$^1$}
\address{$^1$Centre for Engineered Quantum Systems, University of Queensland, St Lucia, Queensland 4072, Australia}
\address{$^2$Research Institute of Molecular Pathology (IMP), Max F. Perutz Laboratories \& Research Platform for Quantum Phenomena and Nanoscale Biological Systems (QuNaBioS), University of Vienna, Dr. Bohr Gasse 7-9, A-1030 Vienna, Austria}



\begin{abstract}
Quantum metrology provides a route to overcome practical limits in sensing devices. It holds particular relevance to  biology, where sensitivity and resolution constraints restrict applications both in fundamental biophysics and in medicine. Here, we review quantum metrology from this biological context, focussing on optical techniques due to  their particular relevance for biological imaging, sensing, and stimulation. Our understanding of quantum mechanics 
has already enabled important applications in biology, including positron emission tomography (PET) with entangled photons, magnetic resonance imaging (MRI) using nuclear magnetic resonance, and bio-magnetic imaging with superconducting quantum interference devices (SQUIDs). 
In quantum metrology
an even greater range of applications arise from the ability to not just understand, but to engineer, coherence and correlations at the quantum level. 
In the past few years, quite dramatic progress has been seen in applying these ideas into biological systems. 
Capabilities that have been demonstrated  include enhanced sensitivity and resolution, immunity to imaging artifacts and technical noise, and characterisation of the biological response to light at the single-photon level. New quantum measurement techniques offer even greater promise, raising the prospect for improved multi-photon microscopy and magnetic imaging, among many other possible applications. Realization of this potential will require cross-disciplinary input from researchers in both biology and quantum physics. In this review we seek to communicate the developments of quantum metrology in a way that is accessible to biologists and biophysicists, while providing sufficient detail to allow the interested reader to obtain a solid understanding of the field. We further seek to introduce quantum physicists to some of the central challenges of optical measurements in biological science. We hope that this will aid in bridging the communication gap that exists between the fields, and thereby guide the future development of this multidisciplinary research area.
\end{abstract}


\end{frontmatter}

\section{Introduction}

Fundamentally, all measurement processes are governed by the laws of quantum mechanics. The most direct influence of quantum mechanics is to impose constraints on the precision with which measurements may be performed. However, it also allows for new measurement approaches with improved performance based on phenomena that are forbidden in a purely classical world. The field of {\it quantum metrology} investigates the influence of quantum mechanics on measurement systems and develops new measurement technologies that can harness non-classical effects to their advantage. 

Quantum metrology broadly began with the discovery that quantum correlated light could be used to suppress quantum shot noise in interferometric measurements, and thereby enhance precision~\cite{Caves1981}. To this day, the development of techniques to enhance precision in optical measurements remains a primary focus of the field. Such enhancement is particularly relevant in situations where precision cannot be improved simply by increasing optical power, due, for example, to power constraints introduced by optical damage or quantum measurement back-action~\cite{Giovannetti2004}. One such situation is gravitational wave measurement, where kilometre-scale interferometric observatories operate with power near the damage threshold of their mirrors, yet still have not achieved the extreme precision required to directly observe a gravitational wave~\cite{Schnabel2010,LIGO2011,LIGO2013,Grote2013}.  Biological measurements are another prominent application area which has been discussed from the earliest days of quantum metrology~\cite{Slusher1990,Schubert1987,Slusher1989,Teich1990}, since biological samples are often highly photosensitive and optical damage is a limiting factor in many biophysical experiments~\cite{Ashkin2000,Carlton2010,Pena2012}. 

Although these applications were recognised in the 1980s~\cite{Slusher1990}, at that time the technology used for quantum metrology was in its infancy and unsuited to practical measurements. Since then, both the technology and theory of quantum metrology has advanced dramatically. 
In recent years, quantum measurement techniques based on quantum correlated photons have made in-roads in significant application areas within the biological sciences~(see, for example, Refs.~\cite{Morris2015,Nasr2009,Taylor2013_sqz}),
while 
  new biologically relevant measurement technologies are under rapid development. 
  With researchers continuing to explore the possibilities of biological quantum metrology, it may soon be possible to achieve advantages over classical measurement strategies in practical settings.
%
%
%
%
%
  The promise of quantum techniques is recognised within the bioscience community, featuring prominently in recent reviews of advances in optical instrumentation (for instance, see Refs.~\cite{Cole_live_cell, mathur2015biology, Serranho2012}). However, no dedicated review yet exists that bridges the gap between the fields. While a number of review articles are available which focus on quantum metrology  (see, for example,  Refs.~\cite{demkowicz2015quantum, arXiv:1402.3713,Giovannetti2011}), they are generally  targeted towards researchers within that community, and tend to be inaccessible to readers who are uninitiated in advanced quantum mechanics. At its broadest level, therefore, the aim of this review is to act as a bridge. We aim to explain the concepts of quantum metrology and their implications in as accessible a manner as possible, while also  introducing a range of state-of-the-art approaches to biological measurement and imaging, along with their associated challenges. Such a review cannot hope to be exhaustive. However, we do seek to introduce the techniques of most relevance to near-future developments in this multi-disciplinary area, and to identify the key technological and practical challenges which must be overcome to see those those developments realised. In this way, we hope that the review will contribute in a positive way to the field.

  
%

Although quantum metrology is generally considered to have begun in the 1980s, positron emission tomography (PET) has been utilizing entangled photon pairs in imaging since the 1960s~\cite{Rankowitz1961,Kuhl1963}. In PET, a radioactive marker undergoes $\beta^+$ decay to produce a positron. The positron annihilates with a nearby electron to produce a high energy entangled photon pair. Since the photons propagate in near-opposite directions, the position of the annihilation event can be estimated to occur along a chord connecting coincident photon detections. With sufficient coincident detection events, a full three dimensional profile of the radioactive marker density can be reconstructed. This is now used routinely in clinical applications to image cancerous tumours and to observe brain function~\cite{Phelps2000}. 

The development of modern quantum technologies allows quantum correlated states of light to be engineered, in contrast to the uncontrolled generation of entangled photons which is used in PET. This enables a far broader range of applications.
Entangled photon pairs have now been applied in tissue imaging~\cite{Nasr2009}, absorption imaging~\cite{Morris2015}, and refractive index sensing of a protein solution~\cite{Crespi2012}; squeezed states of light have been used both to measure dynamic changes~\cite{Taylor2013_sqz} and image spatial properties~\cite{Taylor2014_image} of sub-cellular structure; and single photons have been used to stimulate retinal rod cells, thus allowing the cellular response to single photons to be deterministically characterized~\cite{Phan2014}.  These experiments have demonstrated the prospects of
quantum correlations 
for new capabilities and unrivalled precision in practical biological measurements. Furthermore, a broad range of technologies have already been demonstrated in non-biological measurements which could soon have important applications in biology. These near-future applications include cellular imaging with both multi-photon microscopy~\cite{Teich1997,Upton2013} and super-resolution of fluorescent markers~\cite{Cui2013,Schwartz2013}, enhanced 
 phase contrast~\cite{Ono2013,Israel2014} microscopy, and measurement of biomagnetic fields~\cite{Wolfgramm2010}.

 The review begins with a semi-classical explanation of quantum noise in optical measurements, which allows a qualitative, but not rigorous, derivation of the limits imposed on optical measurements by the quantisation of light, including the standard quantum limit, the Heisenberg limit, and the limit imposed by optical inefficiencies (Section~\ref{semiclass}). It then proceeds to a quantum mechanical description of photodetection, quantum coherence and quantum correlations (Section~\ref{CorrelationSection}). Section~\ref{PhaseTheory} introduces the theoretical tools of quantum metrology for the example case of optical phase measurement, and introduces the commonly used squeezed and NOON states.  
 Section~\ref{Biology} then describes the unique challenges associated with practical biological experiments, including resolution requirements, optical damage, and the non-static nature of living cells. The experiments which have applied quantum metrology in biology are described in Section~\ref{Progress}. Section~\ref{Future} overviews a range of promising technologies which in future may have important biological applications. Section 
 Section~\ref{spinsection}, departs briefly from optical approaches to quantum metrology,   providing a brief overview of spin-based quantum metrology experiments that hold promise for future biological applications. Finally, Section~\ref{conclusion} concludes the review with a broad summary of the potential of quantum metrology for biological measurements in the near future.


%
%

\section{Semi-classical treatment of optical phase measurement}

\label{semiclass}

This review introduces the theory of quantum measurement in the context of interferometric phase measurements. In doing so, we seek to provide the simplest possible example quantum limits to measurements, and how they may be overcome using quantum correlations. This example is particularly relevant, both since it has been comprehensively studied in the quantum metrology literature, and due to its many applications ranging from optical range-finding to phase-contrast imaging. We would emphasise that  the concepts introduced are quite general and can be naturally applied in other contexts -- as is particularly well seen for optical nanoparticle tracking in Section~\ref{SQZ_light_Qmet_sec}. 

The measurement process quite generally involves the preparation of a probe, its interaction with a system of interest, and finally measurement of the probe to extract relevant information about the system.  In optical phase measurements, this  typically involves using a laser to produce a coherent optical field, propagating the field though an interferometer, and measuring the power in the two output ports to estimate the phase shift $\phi$ applied within one arm of the interferometer (see Fig.~\ref{MachZehnder}). More generally, the probe need not be laser light; optical measurements can be carried out with states of light ranging from thermal light~\cite{Huang1991} to non-classical states of light~\cite{Giovannetti2004}, while non-optical measurements can be carried out with probes such as coherent matter waves~\cite{Gustavson1997}, spin states of atoms~\cite{Budker2007}, or mechanical states of a cantilever~\cite{Rugar1991}. The field of quantum metrology explores the influence of the input state on the achievable precision, as well as the advantage which can be gained from use of non-classical states. 

Before introducing a full quantum treatment of the problem of phase measurement, we consider a semi-classical scenario where the optical electric field is treated classically and can, in principle, be deterministic and carry no noise, with photon quantization (or `shot noise') introduced phenomenologically in the detection process. This approach to quantization results in a stochastic output photocurrent with mean proportional to the intensity of the measured field. This semi-classical approach has the advantage of illustrating the deleterious effect of shot noise on optical measurements, and allows straightforward (though not rigorous) derivations of quantum limits to the precision of phase measurements which should be readily understood by scientists from outside the quantum metrology community.
 
\subsection{Standard quantum limit of optical phase measurement} \label{interferometry}
\begin{figure}
 \begin{center}
   \includegraphics[width=9cm]{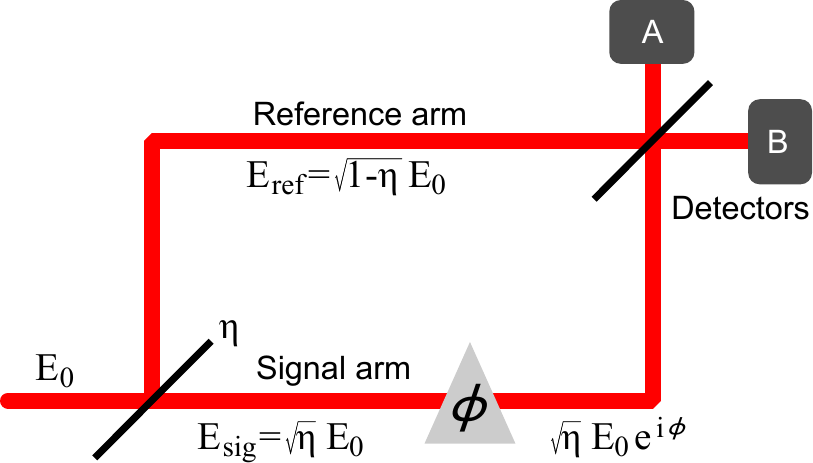}
   \caption{A classical analysis of a phase sensing experiment based on a Mach-Zehnder interferometer. An incident optical field $E_0$ is split into a reference $E_{\rm ref}$ and probe $E_{\rm sig}$ field  at the input port of the interferometer by a beam splitter with transmission $\eta$. A differential phase shift $\phi$ is then applied between the fields in the two arms. The fields are then recombined on a second beam splitter to produce two output fields (termed $E_A$ and $E_B$). Detection of these fields allows  the phase shift to be estimated. 
}
 \label{MachZehnder}  
  \end{center}
\end{figure}


%
Optical phase is generally measured via interference. Here we consider the case where an optical field propagates through the arms of a Mach-Zehnder interferometer, as shown in Fig.~\ref{MachZehnder}. We term the arm in which the system of interest is placed the ``signal arm", and the other arm of the interferometer the ``reference arm", since its role in the measurement is solely to provide a phase reference. The field in the signal arm experiences a phase shift $\phi$ from its interaction with the system, which we often refer to alternatively as the ``sample" or ``specimen". This phase shift is estimated from intensity measurements at the two interferometer outputs, labelled ``$A$" and ``$B$" here. In a classical treatment of this problem, taking the case where the two beam splitters in the interferometer each have 50\% reflectivity and no absorption, the electric fields of the light that reaches the detectors are given by
\begin{subequations}
\begin{eqnarray}
E_A(t) &=& \frac{1}{2} \left (1+ e^{i\phi} \right ) E_0(t)\\ 
E_B(t) &=& \frac{1}{2} \left (1- e^{i\phi} \right ) E_0(t),
\end{eqnarray}
\end{subequations}
where $E_0(t)$ is the incident optical field. Each field is then detected with a photodiode. Quantized photocurrents $n_A(t)$ and $n_B(t)$ are produced as valence band electrons in each photodiode are independently excited into the conduction band, with probability proportional to the optical intensity. As such, each photocurrent fluctuates stochastically about a mean that is
proportional to the intensity of light incident on the photodiode ($\langle n_A(t) \rangle \propto |E_A(t)|^2$ and $\langle n_B(t) \rangle \propto |E_B(t)|^2$).
In a quantum treatment of photodetection, each photoelectron is excited by a single photon (see Section~\ref{CorrelationSection}). Consequently, the photocurrents $n$ can equivalently be thought of as the photon flux of the detected fields. Evaluating the mean  detected intensities, one finds that
\begin{subequations}
\begin{eqnarray}
\langle n_A(t) \rangle &=& \frac{\langle n_0(t) \rangle}{2} \left( 1 + \cos \phi \right) \\ 
\langle n_B(t) \rangle &=& \frac{\langle n_0(t) \rangle}{2}\left( 1 - \cos \phi \right),
\end{eqnarray}
\end{subequations}
where $n_0$ is the photon flux of the input field to the interferometer. Information about the phase $\phi$ can be extracted from the difference photocurrent, which has a mean value of
\begin{equation}
\langle n_{A} (t) - n_{B} (t) \rangle=  \langle n_0(t) \rangle \cos \phi. \label{n_diff}
\end{equation}
The phase sensitivity is optimized when $\phi=(m+1/2) \pi$ where $m\in\mathbb{Z}$, since this maximises the derivative of the difference photocurrent with respect to a small change in $\phi$. Interferometers are generally actively stabilised to ensure operation near one of these optimal points. For small displacements about such a point, the phase shift $\phi$ is given to first order as 
%
\begin{equation}
\phi = \frac{\langle n_{A} (t) - n_{B} (t) \rangle}{ \langle n_0(t) \rangle} - \frac{\pi}{2}.
\end{equation}
Consequently, the relative phase may be estimated  as 
 \begin{equation}
\phi_{\rm estimate}(t) = \frac{n_A(t) - n_B(t)}{\langle n_0(t) \rangle} - \frac{\pi}{2}. \label{phi_estimate}
\end{equation}
 The statistical variance of the phase estimate is then given by
 \begin{equation}
V( \phi) = \frac{V ( n_A)+ V( n_B) - 2 \, {\rm cov}( n_A ,  n_B)}{\langle n_0 \rangle^2},\label{phi_noise}
\end{equation}
where $V(x) \equiv \langle x^2 \rangle - \langle x \rangle^2$ is the variance of the variable $x$, and the covariance ${\rm cov} (x,y) \equiv \langle (x - \langle x \rangle) (y - \langle y \rangle)  \rangle$ quantifies the correlations between the variables $x$ and $y$. In our semi-classical treatment, photoelectrons are generated through stochastic random processes at each photodetector. In the limit that the optical fields are stationary in time, the photon detection events are uncorrelated both on one photodetector, and between the photodetectors. The latter property means that the covariance ${\rm cov}( n_A ,  n_B)$ is zero, while the former results in Poissonian photon counting statistics on each detector.  As discussed in Section~\ref{CoherentState}, this prediction of Poissonian statistics is consistent with a fully quantum treatment of a coherent state, which is the state generated by a perfectly noise-free laser. 
%
Due to the  Poissonian counting statistics, the variance of each photocurrent is equal to its mean, i.e., $V(n_i) = \langle n_i \rangle$ with $i \in \{A,B\}$. 
Assuming that the interferometer is operating very close to its optimal point, with the phase deviation away from this point being much smaller than one, the input photon flux  $n_0$ is split approximately equally between the two interferometer outputs, so that
$V( n_A) \approx V( n_B) \approx \langle n_0 \rangle/2$. Substituting for the photocurrent variances and covariance in Eq.~(\ref{phi_noise}), the achievable phase precision is then given by
 \begin{equation}
\Delta \phi_{\rm SQL} = \sqrt{V(\phi)} = \frac{1}{\sqrt{\langle n_0 \rangle}}. \label{phi_SQL_totalpower}
\end{equation}
We see that the phase precision improves as the square-root of the photon flux input into the interferometer (see Fig.~\ref{Phase_sensing_limits}).
Even though the approach used here is semi-classical and treats the optical electric field as a perfectly deterministic quantity, Eq.~(\ref{phi_SQL_totalpower}) reproduces the {\it standard quantum limit} for phase measurements, which quantifies the best precision that can be reached without the use of quantum correlations for any optical phase measurement using a mean photon  flux of $\langle n_0 \rangle$. As discussed  in Section~\ref{CoherentState}, the standard quantum limit can be achieved using coherent states. In fact, coherent states achieve the best precision possible without quantum correlations for many forms of measurement, not just phase estimation~\cite{Giovannetti2004}.  Consequently, the sensitivity achievable using coherent light of a given power generally provides an important benchmark for quantum metrology experiments.

 Examination of Eq.~(\ref{phi_noise}) suggests that precision could be improved if the detection events are correlated, such that ${\rm cov} (  n_A , n_B )>0$. As we will see later (first in Section~\ref{HL_section}), this is the case for quantum correlated light. However, a detailed calculation shows that when classical correlations are introduced -- such as those introduced by a temporal modulation of the input optical intensity --  they  also increase the photon number variances $V( n_A)$ and $V( n_B)$, and ultimately the  precision is not enhanced. 

It is worthwhile to note that the above derivation only assumes quantization in the photocurrent which provides the electronic record of the light intensity. As such, the limit of Eq.~(\ref{phi_SQL_totalpower}) can be arrived at either 
either by considering a perfectly noiseless optical field which probabilistically excites photoelectrons, or a quantized field with each photon exciting a single electron. Violation of this limit, however, requires electron correlations in the detected photocurrents which cannot follow from probabilistic detection, and therefore necessitates  a quantum treatment of the optical fields~\cite{Fox}.

\subsection{Variations on the standard quantum limit}

 \subsubsection{The quantum noise limit}
\label{QNL_sec}

So far in the review we have considered only the restricted class of measurements that can be performed with perfect efficiency. That is, we have assumed that no photons are lost in transmission of the optical fields, both through free-space and through the sample, or their detection. In practice, this is never the case, with inefficiencies being a particular concern for biological applications of quantum measurement. 

For the moment, constraining our analysis to the case of uncorrelated photons, the analysis in the previous section can be quite straightforwardly extended to include optical inefficiencies. This results in the so-termed {\it quantum noise limit} of interferometric phase measurement. 
 Here, we consider the  particular case where the two arms of the interferometer exhibit balanced losses, each having transmissivity $\eta$. Within our semi-classical treatment, such balanced losses have the same effect as loss of the same magnitude prior to the interferometer,\footnote{Or, indeed, at the outputs of the interferometer. } and are therefore equivalent to a reduced input photon flux from $n_0$ to $\eta n_0$.  They 
  therefore
 modify the standard quantum limit of Eq.~(\ref{phi_SQL_totalpower}) to the quantum noise limit:
 \begin{equation}
\Delta \phi_{\rm QNL} =  \frac{1}{\sqrt{\eta \langle n_0 \rangle}}. \label{QNL}
\end{equation}
In the general case where unbalanced loss is present, the quantum noise limit is modified, but retains the general properties of constraining the measurement to precision inferior to the standard quantum limit, and scaling as $\langle n_0 \rangle^{-1/2}$. 

The quantum noise limit should be interpreted as quantifying the precision that is achievable in a given (imperfect) apparatus without access to quantum correlations between photons.
%
 As such, 
it is particularly relevant to biological applications where experimental non-idealities are commonly unavoidable and prevent the standard quantum limit from being reached with coherent light.
 Quantum metrology experiments often compare their achieved precision to the quantum noise limit, since violation of this limit proves that quantum resources have enabled an improvement in precision. The standard quantum limit provides a more stringent bound, defining the precision that is achievable  without quantum correlations in an ideal apparatus that has no loss and perfect detectors. Violation of this limit therefore proves that the experiment operates in a regime that is classically inaccessible not only for a given apparatus, but for any apparatus in general.
 The standard quantum limit and quantum noise limit are generally used in different contexts, with continuous measurements on bright fields often compared to the quantum noise limit~\cite{Treps2002,Taylor2013_sqz,Wolfgramm2010,Stefszky2012,LIGO2013}, and photon counting measurements typically compared to the standard quantum limit~\cite{Crespi2012,Nagata2007,Afek2010}.

Although the quantum noise limit is a widely used benchmark, it is worth noting that there is no clear consensus as to its name. It is most often referred to as either the quantum noise limit~\cite{Treps2002,Taylor2013_sqz,Stefszky2012} or the shot noise limit~\cite{Grote2013,Brida2010,Wolfgramm2010}, often interchangeably~\cite{LIGO2013}, though other names are also used~\cite{Treps2003,Horrom2012,PhysRevLett.102.040403}. It is also important to note that the phrase ``standard quantum limit'' carries two distinct meanings in different communities. While much of the quantum metrology community uses the definition introduced here, the optomechanics community defines the standard quantum limit as the best sensitivity possible with arbitrary optical power, which occurs when quantum back-action from the measurement is equal to the measurement imprecision~\cite{Giovannetti2004}.

\subsubsection{Power constraints}
\label{power_constraints_section}
 
 As can be seen from Eqs.~(\ref{phi_SQL_totalpower}) and (\ref{QNL}), the precision of an optical phase measurement can, at least in principle, be enhanced arbitrarily by increasing the optical power input to the interferometer. Consequently, quantum limits on precision are only relevant in circumstances where the optical power is constrained.

\begin{figure}[t!]
 \begin{center}
   \includegraphics[width=11cm]{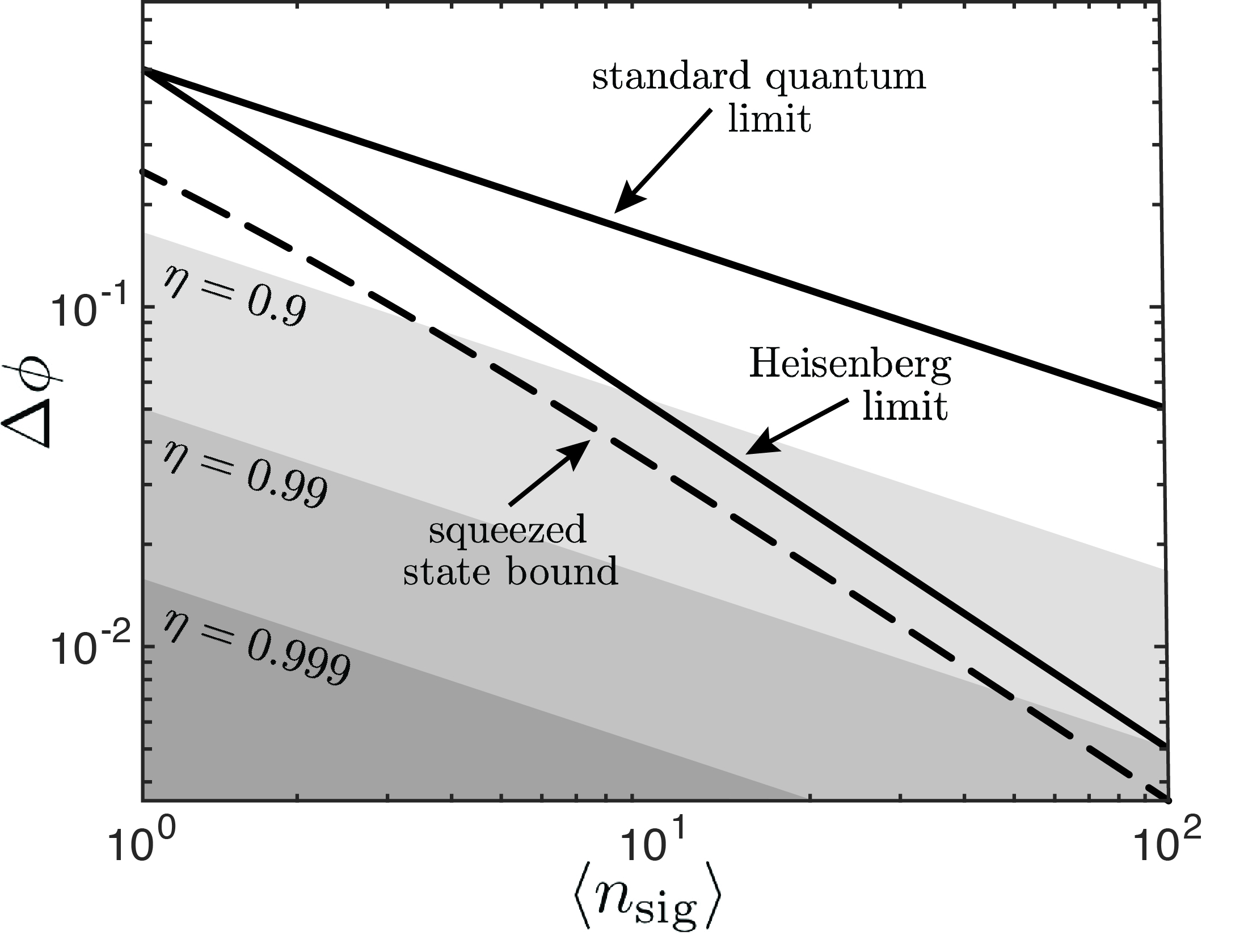}
   \caption{Sensitivity limits of interferometric phase measurement as a function of the number of photons used in the probe arm $\langle n_{\rm sig} \rangle$. Here, we consider the scenario where the  power constraint is imposed by the possibility of damage to the sample, rather than a constraint on total power within the interferometer.
   The solid black lines show the standard quantum limit (Eq.~(\ref{phi_SQL_samplepower})) and the Heisenberg limit (Eq.~(\ref{Heisenberg1})), where in the later case $\langle n_{\rm sig} \rangle = \langle n_0 \rangle/2$. The dashed line shows the Cramer-Rao Bound for squeezed states (Eq.~(\ref{CRB_sqz})), defining the best precision achievable with such states and ideal measurement. The grey shaded regions indicate levels of precision that are unattainable given a loss at a level of $1-\eta$ within the probe arm of the interferometer (see Section~\ref{inf_bound_limit_sec}). 
}
 \label{Phase_sensing_limits}  
  \end{center}
\end{figure}

 In circumstances where an experimental constraint existed on the total photon flux $\langle n_0 \rangle$, due, for example, to limitations in available laser output power or detector damage thresholds, Eq.~(\ref{phi_SQL_totalpower}) defines the standard quantum limit to precision. In many other experiments -- and particularly in biological measurements where the sample is often susceptible to photo-induced damage and photochemical intrusion (see Section~\ref{BioDamage}) -- however, the constraint is instead placed on the power incident on the sample.
  In this case, the precision can be improved by unbalancing the interferometer such that the reference arm carries more power than the signal arm. This suppresses the noise contribution from photon fluctuations in the reference arm, and thereby improves the precision achievable for a fixed power at the sample.  A similar treatment to that given in Section~\ref{interferometry} shows that, when the power incident on the sample is constrained, the standard quantum limit becomes
 \begin{equation}
\Delta \phi_{\rm SQL} = \frac{1}{2 \sqrt{\langle n_{\rm sig} \rangle} }, \label{phi_SQL_samplepower}
\end{equation}
where $n_{\rm sig}$ is the photon flux in the signal arm. As discussed briefly above and more thoroughly in Section~\ref{BioDamage}, biological specimens can be significantly influenced by the optical fields used to probe them. In this review we will therefore generally use this sample-power-constrained standard quantum limit, rather than the perhaps more conventional standard quantum limit of Eq.~(\ref{phi_SQL_totalpower}). It is shown, as a function of $\langle n_{\rm sig} \rangle$ in Fig.~\ref{Phase_sensing_limits}. 
Since, for the balanced interferometer considered in Section~\ref{interferometry}, $\langle n_{\rm sig} \rangle = \langle n_0 \rangle/2$, we see by comparison of Eqs.~(\ref{phi_SQL_totalpower})~and~(\ref{phi_SQL_samplepower}) that, with a constraint on power within the sample, 
 the standard quantum limit is improved by a factor of $\sqrt{2}$. 
 

Thus far, our discussion has centred entirely on interferometric phase measurements. A similar semi-classical analysis can also be performed to find standard quantum limits for most other optical measurements. For instance, Poissonian detection statistics also introduce noise to amplitude or intensity measurements, thus setting a lower limit to precision (we will see in Section~\ref{SQZ_light_Qmet_sec} for the case of optical particle tracking). Although the exact form of the quantum limits can depend on the type of measurement, the precision achievable with coherent light is almost always used as a benchmark in quantum metrology experiments.

\subsection{The Heisenberg limit}
\label{HL_section}

The standard quantum limit derived above cannot be violated with purely probabilistic photon detection. However, as we briefly discussed at the end of Section~\ref{interferometry}, the presence of correlations (${\rm cov}( n_A, n_B)>0$) between the output photocurrents from the interferometer provides the prospect to suppress statistical noise and therefore improve the measurement precision (see Eq.~(\ref{phi_noise})).
In particular, the two detected fields could, in principle, be entangled such that quantum correlations exist between photodetection events,\footnote{In quantum mechanics ``entangled states" are states of a system that require a nonlocal description (that is, sub-sets of the state, such as each of the two fields within an interferometer, cannot be fully described in isolation). In quantum mechanics terminology, such states are {\it inseparable}. Two optical fields whose combined state is inseparable will necessarily exhibit some form of correlation -- for instance, in the time at which photons arrive at photoreceivers placed in each field (see Section~\ref{CorrelationSection}). Purely classical correlations may exist between two fields. These could be generated, for example,  by applying an equal amplitude modulation to both fields. The term ``quantum correlation'' is reserved for fields that exhibit entanglement. }  suppressing the statistical variance in the difference signal. Taken to its extreme, it can be conceived that perfect entanglement may allow the elimination of all statistical noise in Eq.~(\ref{phi_noise}). In this case, the phase precision would only be limited by the requirement that $n_A$ and $n_B$  be integers. A phase shift can only be resolved if it changes the difference signal $n_A - n_B$ by at least one photon. Under a constraint on total power, this can be seen to limit the possible precision to
 \begin{equation}
\Delta \phi_{\rm Heisenberg} \geq \frac{1}{\langle n_0 \rangle}. \label{Heisenberg1}
\end{equation}
%
Once again, this semi-classical derivation reproduces an important and fundamental result. Eq.~(\ref{Heisenberg1}) is generally referred to as the {\it Heisenberg limit}, and is the absolute limit to precision possible in a phase measurement using exactly $n_0$ photons, as discussed in more detail in Section~\ref{NOONState}. Notice that the Heisenberg limit scales faster in $\langle n_0 \rangle$ than the standard quantum limit, as shown graphically in Fig.~\ref{Phase_sensing_limits}. This $\langle n_0 \rangle^{-1}$ scaling is generally referred to as ``Heisenberg scaling''. In principle, it promises a dramatic improvement in precision. For instance, a typical 1 mW laser has a photon flux of around $10^{16}$~s$^{-1}$. If the Heisenberg limit could be reached with this photon flux, it would be possible, within a one second measurement time, to achieve a phase sensitivity $10^8$ times superior to the best sensitivity possible without quantum correlations. Unfortunately, as it turns out, entanglement tends to become increasingly fragile as the number of photons involved increases. This places a prohibitive limitation on the absolute enhancements that are possible, as discussed in Section~\ref{Loss}.

The Heisenberg limit of Eq.~(\ref{Heisenberg1}) is often described as a fundamental  lower limit to the precision of phase measurement achievable using any quantum state~\cite{Giovannetti2004}. In fact, it is possible to outperform this limit using states with indeterminate total photon number -- we consider the specific case of squeezed states in Section~\ref{SqueezedState}. However, no linear phase estimation scheme has been found that provides {\it scaling} that is superior to $1/\langle n_0 \rangle$. For completeness, we note that nonlinear parameters can in principle be estimated with scaling that exceeds the Heisenberg limit~\cite{Beltran2005,Roy2008}. However, it is not currently clear what benefits such approaches might offer in biological applications, and they will not be discussed further in this review.

\subsection{Fundamental limit introduced by inefficiencies}
\label{inf_bound_limit_sec}

As we saw in Section~\ref{QNL_sec}, the precision of optical phase measurements is degraded in the presence of optical inefficiencies. For measurements that utilise quantum correlated photons, this degradation enters in two ways. First, inefficiencies reduce the magnitude of observed signals and, second, they degrade the correlations used to reduce the noise-floor of the measurement. 
%
We consider the effect of inefficiencies on two specific approaches to quantum measurement  in Sections~\ref{Loss}~and~\ref{Sqz_eff_eff_sec}. 
Here, without restricting ourselves to a specific class of quantum measurement, we introduce -- again via a semi-classical treatment -- a fundamental limit introduced by the presence of inefficiencies that is applicable for all phase measurements. This limit is rigorous, although our derivation is not. For a rigorous derivation of this limit we refer the reader to Refs.~\cite{Demkowicz2012,demkowicz2015quantum}.

Let us return to the balanced optical interferometer from Section~\ref{interferometry}, for which we derived the standard quantum limit with a constraint on total power. We found, there, an expression for the statistical variance of phase estimation (Eq.~(\ref{phi_noise})), which depends on the mean injected photon number $\langle n_0 \rangle$ as well as the variances $V(n_A)$ and $V(n_B)$ of the photon number arriving at the detectors placed at each interferometer output. As in the previous section, let us imagine that it was possible in some way to achieve photon number variances $V(n_A) = V(n_B)=0$ (and therefore, also, ${\rm cov}(n_A,n_B)=0$). With the caveat that $n_A$ and $n_B$ must be integers which leads to the Heisenberg limit discussed in the previous section,
this would allow an exact measurement of phase. 

Now, consider the effect of inefficiencies on this (unrealistic) perfect phase measurement. By probabilistically removing photons from the fields incident on each of the two detectors, such inefficiencies will introduce statistical uncertainty in the detected photon number, and therefore degrade the precision of the phase measurement.
Let us define the detection efficiency at each detector to be $\eta$, so that each photon arriving at the detector had a probability $1-\eta$ of not being registered. Within our semi-classical treatment, if half of the $\langle n_0 \rangle$ photons input to the interferometer are incident on each of the two detectors, the probability distribution of observed photons can be calculated straightforwardly from basic probability theory. This results in the binomial distribution for detector $A$:
\begin{equation}
p(n_A) = \eta^{n_A} (1-\eta)^{\langle n_0 \rangle/2 - n_A} \frac{(\langle n_0 \rangle/2 )!}{n_A! (\langle n_0 \rangle/2 - n_A)!},
\end{equation}
where here, $\langle n_0 \rangle$ must be even to allow a deterministic and equal number of photons to arrive at each of the two  detectors. $p(n_B)$ is identical to $p(n_A)$, except for the replacement $n_A \rightarrow n_B$, throughout. Since this is a binomial distribution, its variance is well known, and given by
\begin{equation}
V(n_A) = V(n_B) = \frac{\eta (1-\eta)}{2} \langle n_0 \rangle.
\end{equation}
Substituting this expression into Eq.~(\ref{phi_noise}), and also making the substitution $\langle n_0 \rangle \rightarrow \eta \langle n_0 \rangle$ on the bottom line to account for the reduction in the signal due to the presence of inefficiency, we arrive at the bound on phase measurement precision
\begin{equation}
\Delta \phi_{\rm loss} = \sqrt{V(\phi)} = \sqrt{\frac{1-\eta}{\eta}} \frac{1}{\sqrt{\langle n_0 \rangle}}, \label{fund_loss_limit}
\end{equation}
which applies under a constraint on total power injected into the interferometer. This exactly corresponds to the fundamental bound due to inefficiency more rigorously derived in Ref.~\cite{Demkowicz2012}. Comparing this expression to the relevant standard quantum limit in Eq.~(\ref{phi_SQL_totalpower}), we see that the scaling with input photon number is identical, but with an efficiency-dependent  coefficient introduced which, for $\eta<1/2$ fundamentally constrains the measurement precision above the standard quantum limit. 

This shows that the Heisenberg scaling discussed in the previous section cannot be achieved with arbitrarily high photon numbers, but rather the scaling returns to the usual $\langle n_0 \rangle^{-1/2}$ dependence for sufficiently high photon numbers. Equating Eq.~(\ref{fund_loss_limit}) with Eq.~(\ref{Heisenberg1}), we find that this transition back to $\langle n_0 \rangle^{-1/2}$ dependence occurs at a mean input photon number of $\langle n_0 \rangle = \eta/(1-\eta)$. 


So far in this section we have considered the case where the power constraint on the measurement is placed on the total power in the interferometer. In scenarios where the constraint is instead on the power within the sample, the measurement precision is improved by a factor of two compared to Eq.~(\ref{fund_loss_limit})~\cite{demkowicz2015quantum}, similarly to the case of the standard quantum limit discussed in Section~\ref{power_constraints_section}.

\section{Quantum coherence and quantum correlations}\label{CorrelationSection}

In the previous section, we introduced several important quantum limits on optical measurements that arise due to the quantised nature of the photon. In order to do this in the simplest possible way, we have treated optical fields from a semiclassical perspective, imagining that they consist of a train of discrete photons, or even as a noise-free electromagnetic field with quantisation only occurring via the production of electrons within the detection apparatus. While this semiclassical analysis is helpful to develop intuition, the quantum correlations at the heart of quantum-enhanced measurements require a full quantum mechanical treatment.

\begin{figure}
 \begin{center}
   \includegraphics[width=10cm]{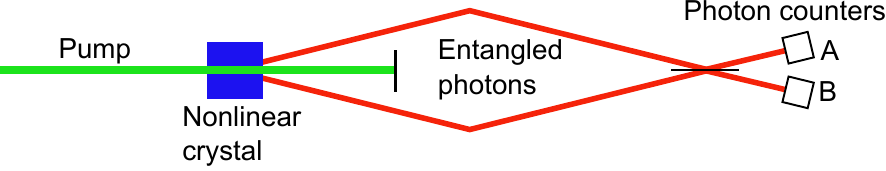}
   \caption{Layout of a Hong-Ou-Mandel interferometer~\cite{Hong1987}. A nonlinear crystal is pumped to produce entangled photon pairs by parametric down-conversion. These photons are then recombined with a beam splitter and detected. A classical treatment would predict that the relative phase would dictate the detection events, and would determine whether the photons are more likely to arrive at detector A or detector B. In reality, non-classical interference ensures that both photons arrive at one detector, with equal probabilities that the photon pair arrives at detector A and B. This violates the classical wave treatment of light, and can only be predicted with quantum theory.
}
 \label{HOM}  
  \end{center}
\end{figure}

A quantum treatment of photodetection and optical coherence was first performed by Glauber and Sudarshan in Refs.~\cite{Glauber1963_PR,Sudarshan1963}. Following such a treatment, one finds that there are some optical phenomena which exhibit classically forbidden behaviour, such as two-photon interference in a Hong-Ou-Mandel (HOM) interferometer~\cite{Hong1987} (see Fig.~\ref{HOM}). These non-classical phenomena rely on quantum correlations, which can also be used to surpass quantum limits to measurement such as the quantum noise limit and standard quantum limit introduced above. This section seeks to provide a basic introduction to the quantum theory of light, and quantum correlations, in the context of quantum metrology. We will utilise standard techniques in introductory quantum mechanics. We refer the uninitiated reader to one of many undergraduate quantum mechanics textbooks for details of these techniques (for example, Ref.~\cite{griffiths2005introduction}). A reader more interested in the applications of quantum measurements than the underlying physics could reasonably skip both this and the following section of the review.


\subsection{Quantum treatment of the electric field}\label{FieldModes}

In a full quantum treatment of light, the deterministic classical description of the optical electric field introduced earlier must be replaced with an operator description. The electric field is decomposed into a sum of contributions from a complete set of orthonormal spatial modes 
\begin{equation}
\hat E(t) = i \sqrt{\frac{\hbar \Omega_m}{2 \epsilon_0 V_m}} \sum_m \left [ \hat a_m  u_m({\bf r})e^{-i\Omega_m t} -   \hat a_m^\dagger(t)  u_m^*({\bf r})e^{i\Omega_m t}  \right ], \label{FieldModesEqn}
\end{equation}
where $u_m({\bf r})$, $\Omega_m$, and $V_m$ are, respectively, the spatial mode function, frequency, and volume of mode $m$; and $\hat a_m$ and $\hat a_m^\dagger$ are, respectively, the annihilation and creation operators. As their names suggest, when acted upon an optical field, the annihilation operator removes one photon from mode $m$, while the creation operator adds one photon to mode $m$. The spatial mode function is normalised such that $\langle |u_m({\bf r})|^2 \rangle \equiv \int_{-\infty}^\infty d{\bf r} |u_m({\bf r})|^2 = 1$. It is straightforward to show that $\langle |\hat E(t)|^2 \rangle \propto \sum_m \langle \hat n_m \rangle$, where $\hat n_m \equiv \hat a_m^\dagger \hat a_m$ is the photon number operator for mode $m$. Therefore, similarly to our previous classical description, $\langle |\hat E(t)|^2 \rangle$ is proportional to the total mean photon number in the field.






\subsection{Quantum treatment of photodetection}
\label{quantum_treat_sec1}

Following the approach of Glauber~\cite{Glauber1963_PR}, the process of photodetection can be viewed as the annihilation of photons from the optical field, with corresponding generation of photoelectrons that can be amplified to generate a photocurrent. Annihilation of one photon collapses the state of the optical field from its initial state $|i\rangle$ into a new state, defined by the initial state acted upon by the annihilation operator $\hat a$
\begin{equation}
| i \rangle \rightarrow \hat a | i \rangle,
\end{equation}
where for simplicity, here, and for the majority of the review, we limit our analysis to only one spatial mode of the field and drop the subscript $m$. We return to multimode fields when treating approaches to quantum imaging in Sections~
\ref{single_param_sec}~and~\ref{QImaging}.
 Fermi's Golden Rule tells us that the transition rate $R_{i \rightarrow f}$ to an arbitrary final state $| f \rangle$ due to photon annihilation is proportional to
\begin{equation}
R_{i \rightarrow f} \propto \left |\langle f | \hat a | i \rangle \right |^2.
\end{equation}
Since the optical field is destroyed in the photodetection process, we are ultimately uninterested in the transition rate to a specific final state, but rather the overall decay of the field. This is given by the sum over all possible final states
\begin{eqnarray}
R^{(1)}(t) &=& \sum_f R_{i \rightarrow f} \\
&\propto& \sum_f \left |\langle f | \hat a | i \rangle \right |^2
= \sum_f \langle i | \hat a^\dagger | f \rangle \langle f | \hat a | i \rangle
= \langle i |\hat  a^\dagger \hat a | i \rangle
= \langle \hat  a^\dagger \hat a  \rangle = \langle \hat n \rangle
\end{eqnarray}
where $\hat n$ is the photon number operator. We see, as expected, a direct correspondence to the classical case where the photon number in the field $n(t)$ is viewed as a classical stochastic process, with
the rate at which photons are detected being proportional to the photon number in the field.

\subsection{Higher order quantum coherence}
\label{quantum_treat_sec2}

Let us now consider a second order process where two photons are annihilated. In general, the annihilation events can occur at distinct locations in space and time. However, here we restrict our analysis to co-located events to display the essential physics in the simplest way. 
In this case
\begin{equation}
| i \rangle \rightarrow \hat a \hat a | i \rangle.
\end{equation}
Repeating a similar calculation as that performed above, we find using Fermi's Golden Rule that the rate of two-photon detection is proportional to
\begin{eqnarray}
R^{(2)}(t) &\propto& \langle \hat a^{\dagger}  \hat a^{\dagger}  \hat a  \hat a  \rangle \\
&=&  \langle \hat n^2  \rangle -   \langle \hat n \rangle,
\end{eqnarray}
where we have used the commutation relation 
\begin{equation}
[\hat a, \hat a^\dagger] \equiv \hat a \hat a^\dagger - \hat a^\dagger \hat a =1. \label{a_comm}
\end{equation}
%
%
We see that, due to the non-commutation of the annihilation operators,
 the rate of two-photon detection is fundamentally different than would be predicted for a classical stochastic variable $n(t)$, since it includes the additional term $-\langle \hat n \rangle$.
 
  To quantify the second order coherence it is conventional to define the normalised second order coherence function
\begin{equation} 
g^{(2)}_{11} = \frac{R^{(2)}(t) }{R^{(1)}(t)^2} = \frac{ \langle \hat n^2  \rangle }{\langle  \hat n \rangle^2} -  \frac{1}{ \langle \hat n \rangle} = 1 + \frac{V(\hat n)}{\langle  \hat n \rangle^2} -  \frac{1}{ \langle \hat n \rangle}, \label{g2}
\end{equation}
where $V(\hat n) = \langle \hat n^2  \rangle - \langle  \hat n \rangle^2$ is the variance of $\hat n$ and the sub-script `11' is used to indicate that the annihilation events are co-located in space and time. In the more general case of non-coincident annihilation, both in time and space, the second order coherence function can be easily shown to be
 \begin{equation}
g^{(2)}_{12} = \frac{\langle \hat a^\dagger_1 \hat a^\dagger_2 \hat a_2 \hat a_1 \rangle}{\langle \hat  n_1 \rangle \langle \hat n_2 \rangle}, \label{g2_general}
\end{equation}
where the subscript $j \in \{1,2 \}$ is used to label an annihilation event at some spatial location ${\bf r}_j$ and time $t_j$.
When the annihilation events are spatially co-located (${\bf r}_1 = {\bf r}_2$), this expression reduces to the well known $g^{(2)}(\tau)$ with the substitution $t_1 \rightarrow t$ and $t_2 \rightarrow t+ \tau$
\begin{equation}
g^{(2)} (\tau) = \frac{\langle \hat a^\dagger(t) \hat a^\dagger(t+\tau) \hat a(t+\tau) \hat a(t) \rangle}{\langle \hat  n(t) \rangle \langle \hat n(t+\tau) \rangle}. \label{g2_tau}
\end{equation}
Higher order coherence functions may be defined analogously to Eq.~(\ref{g2_general}) for annihilation events involving more than two  photons (see Ref.~\cite{Glauber1963} for further details).

\subsection{Classically forbidden statistics}

\label{class_forb}

It is illuminating to compare the second order correlation functions in Eq.~(\ref{g2})~and~(\ref{g2_general}) to those obtained by modelling the photon number $\hat n$ as a classical stochastic process $n(t)$ described by a well defined probability distribution $|E(t)|^2$. For a field which can fluctuate in time, a classical treatment finds that $R^{(1)} \propto \langle   |E|^2 \rangle$ and $R^{(2)} \propto \langle   |E|^4 \rangle$. The classical second order coherence function is then
\begin{equation}
g^{(2)}_{11,\, \rm classical} = \frac{\langle   |E|^4 \rangle}{\langle   |E|^2 \rangle^2} = 1+\frac{V(|E|^2 )}{\langle   |E|^2  \rangle^2}=1 + \frac{V(n)}{\langle   n \rangle^2}. \label{g2class}
\end{equation}
Similarly, the two-point second order correlation function would classically be given by
 \begin{eqnarray}
g^{(2)}_{12, \, \rm classical}  = \frac{\langle n_1 n_2   \rangle}{\langle   n_1 \rangle \langle  n_2 \rangle}. \label{g2class_general}
\end{eqnarray}
As pointed out by Glauber~\cite{Glauber1963_PR}, there exist rigorous bounds on the values that these classical coherence functions can take. Firstly, since $V(n) \ge 0$ it is immediately clear from Eq.~(\ref{g2class}) that
\begin{equation} 
g^{(2)}_{11, \, \rm classical} \ge 1.
\end{equation}
A classical field can only saturate this limit if it is perfectly noise-free. By inspection of Eq.~(\ref{g2}) it is clear that a quantum mechanical field can violate this limit. Similarly, the Cauchy-Schwarz inequality $ \langle n_1 n_2 \rangle^2 \le \langle n^2_1 \rangle \langle n^2_2 \rangle $ can be applied to Eq.~(\ref{g2class_general}) to show that
\begin{equation}
g^{(2)}_{12, \,\rm classical} \le \left [ g^{(2)}_{11, \, \rm classical} \, g^{(2)}_{22, \, \rm classical} \right ]^{1/2}. \label{g2class_2mode}
\end{equation}
No process with a well defined classical probability distribution can exceed either of these bounds. However, both may be violated with a non-classical field due to the additional term in Eq.~(\ref{g2}) which acts to reduce the second order coherence function in  co-incident detection. Higher order coherence functions exhibit similar classical bounds, which also may be violated with non-classical fields. Violation both indicates that the field cannot be described fully by a classical probability distribution (and consequently is ill-behaved in phase space in the Glauber--Sudarshan $P$--representation~\cite{mandel1995optical}), and is an unambiguous signature that quantum correlations are present. Importantly, the inability to describe such non-classical states via a classical probability distribution provides the prospect for measurements whose performance exceeds the usual bounds introduced by classical statistics.

\subsection{Photon bunching and anti-bunching}

\begin{figure}
 \begin{center}
   \includegraphics[width=10cm]{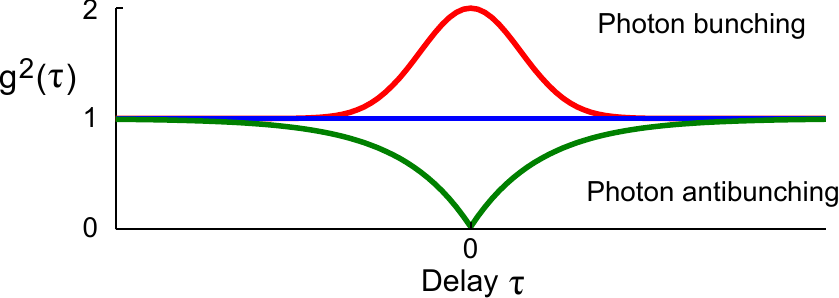}
   \caption{Photon statistics can be formally quantified with higher order correlation functions, such as $g^{(2)}(\tau)$ (see Eq.~(\ref{g2_tau})), as shown here. The example profiles here correspond to (upper red) thermal light with photon bunching ($g^{(2)}(\tau) < g^{(2)}(0)$), (blue) coherent light with $g^{(2)}(\tau)=1$, and (bottom green) anti-bunched light  ($g^{(2)}(\tau) > g^{(2)}(0)$). 
}
 \label{PhotonBunching}  
  \end{center}
\end{figure}

The second order coherence function quantifies pair-wise correlations between photons in an optical field. If no pair-wise correlations are present, $g^{(2)}(\tau)=1$ for all $\tau$. If the average intensity fluctuates, due, for example, to temperature fluctuations in a thermal source, temporally co-incident photons are more likely at times of high intensity. This is known as photon bunching, and is quantified by a second order coherence function for which $g^{(2)}(\tau) < g^{(2)}(0)$. Photon co-incidences can also be suppressed, which is known as photon anti-bunching and characterised by $g^{(2)}(\tau) > g^{(2)}(0)$. Photon anti-bunching occurs naturally in atomic emission, since an atom is a single-photon emitter. Such a system enters the ground state upon emission of a photon, and must first be excited before it can emit another photon. A non-classical field with $g^{(2)}(0) < 1$ (see Eq.~(\ref{g2class})) must also exhibit anti-bunching, since at sufficiently long delays $\tau$ any correlations between photons must decay away with $g^{(2)}(\tau)$ approaching unity. Photon bunching and anti-bunching, and their corresponding second order coherence functions are illustrated in Fig.~\ref{PhotonBunching}.

\subsection{Phase space representations of optical states}

Similarly to classical fields, it is often convenient to represent quantum fields as a vector in phase space. However, there exist fundamental differences in the phase space representations of classical and quantum fields. Classical fields may, in principle, be noiseless and represented as a deterministic vector in phase space. Furthermore, in the presence of noise they can be represented as a positive-definite well behaved probability distribution. As we will see in this section, quantum fields, by contrast, exhibit a fundamental minimum level of phase space uncertainty due to the well known Heisenberg uncertainty principle, and have quasi-probability distributions (the quantum analog of a classical probability distribution) that can be badly behaved and even negative over small regions of phase space. Several different but closely related quasi-probability distributions are commonly used to describe quantum fields, including the $P$-representation~\cite{Drummond1980}\footnote{It can be shown that there is a one-to-one correspondence between a badly-behaved $P$-representation and the existence of non-classical correlations discussed in the previous section~\cite{mandel1995optical}.}, the $Q$-representation~\cite{Husimi1940}, and the Wigner representation~\cite{Wigner1932}. In this review we focus on the Wigner representation, since it is most analogous to a usual classical probability distribution. For simplicity, we avoid a mathematical definition of the Wigner representation, but rather discuss qualitative aspects of the distribution.

\subsubsection{Optical amplitude and phase quadratures}

\begin{figure}[t!]
 \begin{center}
   \includegraphics[width=14cm]{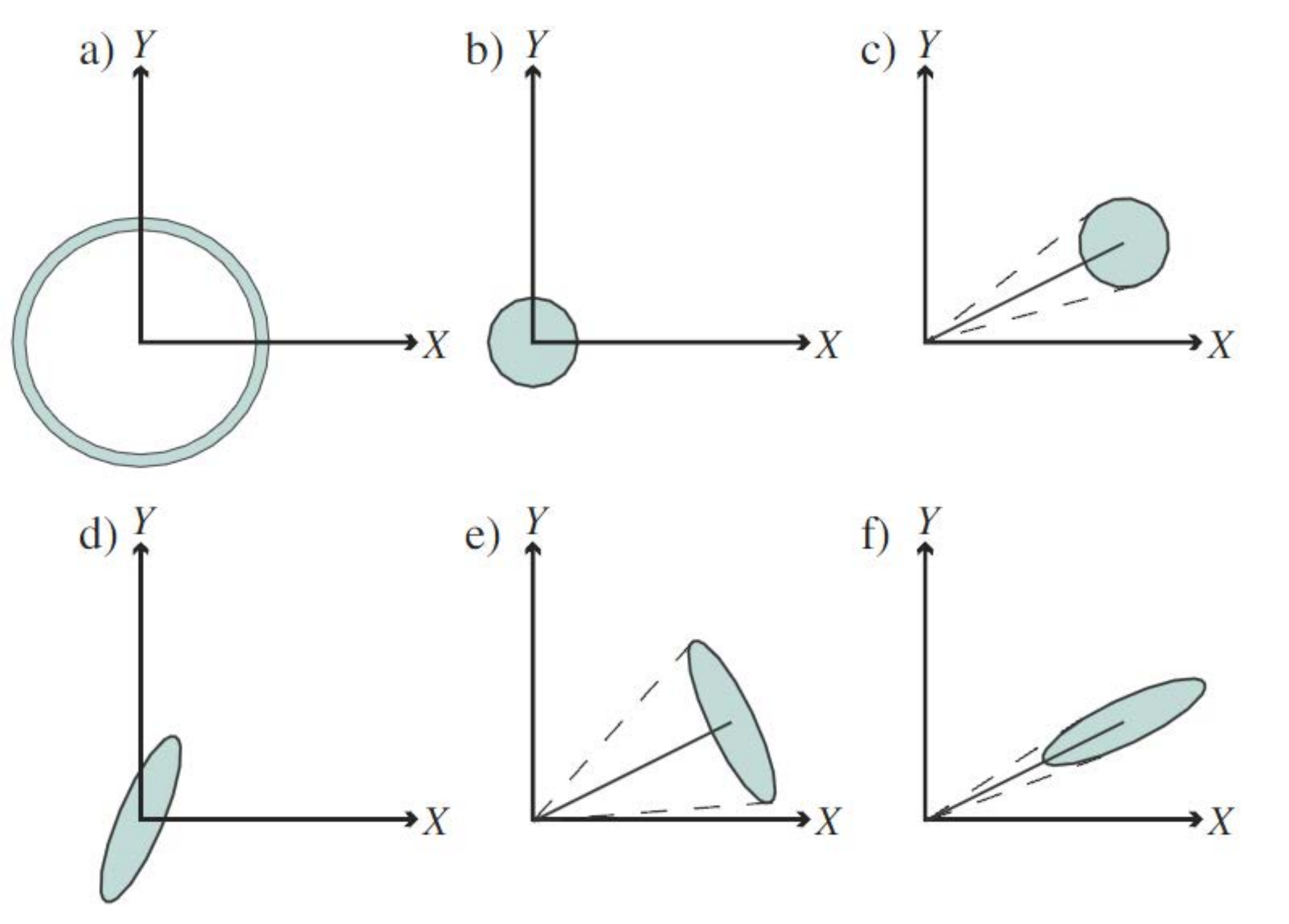}
   \caption{Phase space representations of some important quantum states of light. Dashed-lines illustrate the characteristic phase uncertainty of state. a) Fock state (Note: Wigner-negativity associated with Fock states is not depicted), b) vacuum state, c) coherent state, d) squeezed vacuum state, e) amplitude squeezed state, f) phase squeezed state.
}
 \label{PhaseSpace}  
  \end{center}
\end{figure}
The amplitude and phase quadratures which form the axes of the Wigner representation are the quantum mechanical analog of the real and imaginary parts of a classical electric field. They may be defined in terms of the creation and annihilation operators as
\begin{eqnarray}
\hat X \equiv \hat{a}^\dagger + \hat{a}^\dagger \label{Xquad}\\
\hat Y \equiv i \left( \hat{a}^{\dagger} - \hat{a} \right) \label{Yquad}.
\end{eqnarray}
Using the boson commutation relation given in Eq.~(\ref{a_comm}), it is possible to show that the optical amplitude and phase quadratures do not commute, and satisfy
\begin{equation}
[\hat X, \hat Y] \equiv \hat X \hat Y - \hat Y \hat X = 2i. \label{XY_comm}
\end{equation}
As a result, it is not possible to simultaneously measure both quadratures with arbitrary precision. For any optical field, the  quadratures are subject to an uncertainty principle given by 
\begin{equation}
V( \hat X) V (\hat Y) \geq \frac{1}{4} \left< [ \hat X, \hat Y ] \right>^2\\ 
=1.\label{HeisenbergX}
\end{equation}
 Within the Wigner phase space representation, this enforces a minimum area for the fluctuations of any optical field~\cite{Glauber1963}. This places a fundamental constraint on measurements of both the amplitude and phase of the field.

\subsubsection{Ball and stick diagrams}

Wigner distributions of an optical state may be represented qualitatively on a ball and stick diagram, such as those shown in Fig.~\ref{PhaseSpace}.  Diagrams of this kind illustrate in a clear way the consequences of quantum noise on optical measurements. This is particularly true for Gaussian states that are displaced far from the origin (see Fig.~\ref{PhaseSpace}c,~e,~and~f), such as the coherent and squeezed states to be discussed in Sections~\ref{CoherentState}~and~\ref{SqueezedState}, where the extent of the noise balls graphically illustrates the precision constraints quantum noise introduces to both amplitude and phase measurements.

\section{Quantum treatment of optical phase measurement}\label{PhaseTheory}


In this section we introduce the quantum Fisher information, which allows a full quantum treatment of the precision achievable in general measurements. Its use is illustrated for the phase estimation experiment considered in Section~\ref{semiclass}. Sections~\ref{SqueezedState}~and~\ref{NOONState} then extend the treatment to squeezed states and NOON states, which are quantum correlated states that are regularly applied in the context of quantum metrology.


\subsection{Quantum Fisher Information}
\label{QuantumFI}

\begin{figure}
 \begin{center}
   \includegraphics[width=9cm]{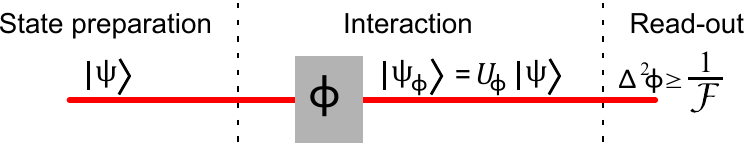}
   \caption{ A quantum treatment of a general measurement process.
}
 \label{QFIplot}  
  \end{center}
\end{figure}

One of the key goals of quantum metrology is to study the fundamental limits to precision. The optimal precision achievable with any measurement is quantified by the Cramer-Rao Bound, which states that the Fisher information $\mathcal{F}$ limits the precision with which a general parameter $\phi$ may be determined,
\begin{equation}
\Delta \phi \ge \frac{1}{\sqrt{\mathcal{F}}} \label{cram-rao}.
\end{equation}
For a classical measurement system, the Fisher information is defined by the probability distribution of measurement outcomes. As discussed in the previous section, however, both quantum states and the quantum measurement process are fundamentally different than their classical counterparts. For quantum measurement processes, the quantum Fisher information is determined by the quantum state of the probe.  For any quantum measurement procedure, an input probe state $|\Psi\rangle$ interacts with the system of interest.  The quantum Fisher information $\mathcal{F}$ quantifies the information contained within the  final state $|\Psi_\phi\rangle$ about the parameter $\phi$  (see Fig.~\ref{QFIplot}) and is defined for pure states as~\cite{Braunstein1994} 
\begin{equation}
\mathcal{F} \equiv 4\left( \langle \Psi'_\phi | \Psi'_\phi \rangle - |\langle \Psi'_\phi | \Psi_\phi \rangle|^2 \right),\label{QFI}
\end{equation}
where primes denote derivatives with respect to $\phi$.  The achievable precision improves when a small change in $\phi$ induces a large change in the final state, as this maximizes the first term $\langle \Psi'_\phi | \Psi'_\phi \rangle$. The second term can be understood by considering a first order Taylor expansion of the final state,
\begin{equation}
|\Psi_{\phi+\delta\phi}\rangle \approx | \Psi_\phi \rangle +\delta\phi| \Psi'_\phi \rangle.
\end{equation}
with the small perturbation $\delta\phi$ estimated via measurement of the occupation of the state $| \Psi'_\phi \rangle$. The precision of this estimation is limited by $\langle \Psi'_\phi | \Psi_\phi \rangle$, which defines the overlap of the state $| \Psi'_\phi \rangle$ with the unperturbed state and therefore establishes a noisy background to the measurement.

In the specific  case of phase measurement, a phase shift $\phi$ within the signal arm of the interferometer is generated by the unitary operator
\begin{equation}
U_\phi \equiv e^{i \phi  \hat{n}_{\rm sig}}, \label{phase_shift_unitary}
\end{equation}
where $\hat n_{\rm sig}$ is the photon number operator for the field in the signal arm. This transforms an arbitrary probe state to
\begin{equation}
|\Psi_\phi\rangle=U_\phi |\Psi\rangle.
\end{equation}
To evaluate the quantum Fisher information, we first note that
\begin{equation}
| \Psi'_\phi \rangle = \frac{d}{d\phi} \left(U_\phi |\Psi\rangle \right) =i \hat{n}_{\rm sig} U_\phi |\Psi\rangle = i \hat{n}_{\rm sig}  |\Psi_\phi\rangle.
\end{equation}
Using this relation, it is evident that 
\begin{equation}
\langle \Psi'_\phi | \Psi'_\phi \rangle =  \langle \Psi_\phi | \hat{n}_{\rm sig}^2| \Psi_\phi \rangle = \langle \hat{n}_{\rm sig}^2 \rangle,
\end{equation}
and
\begin{equation}
\langle \Psi'_\phi | \Psi_\phi \rangle =-i \langle \Psi_\phi |\hat{n}_{\rm sig} | \Psi_\phi \rangle =-i \langle \hat{n}_{\rm sig} \rangle.
\end{equation}
The quantum Fisher information is therefore given by
\begin{equation}
\mathcal{F} = 4 \left (  \langle \hat{n}_{\rm sig}^2 \rangle - | \langle\hat{n}_{\rm sig} \rangle|^2 \right )= 4 V (\hat{n}_{\rm sig}). \label{phase_QFI}
\end{equation}
Since the photon number is conserved by the phase shift operator $U_\phi$, the Fisher information  can be evaluated either for the input state $|\Psi\rangle$ or the final state $|\Psi_\phi\rangle$. This very elegant result shows that, fundamentally, the sensitivity of phase measurement on a pure quantum state is dictated solely by the photon number variance of the state in the signal arm of the interferometer. Of course, technical limitations can also limit the sensitivity. Most significantly, the Cramer-Rao Bound may only be reached through an optimal measurement on the optical field, and this measurement may prove intractable, for example requiring perfect photon number resolving detectors.

%
%
%
%

\subsection{Phase sensing with coherent states}\label{CoherentState}

We begin our quantum treatment of phase sensing by considering interferometry with coherent states. 

\subsubsection{Coherent states}

Coherent states were introduced to the field of quantum optics in 1963 simultaneously by Sudarshan~\cite{Sudarshan1963} and Glauber~\cite{Glauber1963_PR,Glauber1963,PhysRevLett.10.84}, who were motivated by the goal to provide an accurate quantum mechanical description of the field emitted by a laser. Coherent states are eigenstates of the annihilation operator
\begin{equation}
\hat a | \alpha \rangle = \alpha | \alpha \rangle, \label{def_coh_state}
\end{equation}
where $\alpha$ is the coherent amplitude of the state, given by a complex number. As can be easily shown from Eq.~(\ref{def_coh_state}), the coherent amplitude is related to the mean photon number in the field by
\begin{equation}
\langle \hat n \rangle = |\alpha|^2. \label{mean_coh}
\end{equation}
Of all quantum states, coherent states most closely resemble semi-classical fields such as those considered in Section~\ref{semiclass}. For instance, photon detection on coherent states exhibits the Poisson distribution expected of random arrivals of uncorrelated photons or random production of photoelectrons. This can be seen by expanding the state in the Fock basis~\cite{orszag2000quantum}
\begin{equation}
| \alpha \rangle = e^{-|\alpha|^2/2} \sum_{N=0}^{\infty} \frac{\alpha^N}{\sqrt{N!}} | N \rangle, \label{fock_basis}
\end{equation}
where $| N \rangle$ is a Fock state consisting of exactly $N$ photons. The photon number distribution $p(N)$ is then given by
\begin{eqnarray}
p(N) &=& | \langle N | \alpha \rangle|^2\\
&=&  e^{-|\alpha|^2} \sum_{j=0}^{\infty} \frac{|\alpha|^{2j}}{j!}  | \langle N | j \rangle|^2\\
&=& e^{-\langle \hat n \rangle}   \frac{\langle \hat n \rangle ^{N}}{N!}, \label{photon_num_dist_coh_sdgs}
\end{eqnarray}
where in the last step, we have used Eq.~(\ref{mean_coh}) and the orthogonality of the Fock states $\langle N | j \rangle = \delta_{Nj}$. This is the usual form of a Poisson distribution, with variance
\begin{equation}
V(\hat n) = \langle \alpha | \hat n^2 | \alpha \rangle - \langle \alpha | \hat n | \alpha \rangle^2= \langle \hat n \rangle. \label{var_coh}
\end{equation}
Eqs.~(\ref{mean_coh})~and~(\ref{var_coh}) allow the second order coherence function of a coherent state to be determined from Eq.~(\ref{g2}), with the result that $g_{11}^{(2)} = 1$. This indicates that coherent states exhibit no pair-wise correlations between photons, and indeed their pair-wise statistics may be fully understood within a classical framework. 

The variances of the amplitude and phase quadratures of a coherent state are
\begin{eqnarray}
V(\hat X) &=& \langle \alpha | \hat X^2 | \alpha \rangle  -  \langle \alpha | \hat X | \alpha \rangle^2  = \langle \alpha | ( \hat a^\dagger + \hat a )^2 | \alpha \rangle  -  \langle \alpha | \hat a^\dagger + \hat a | \alpha \rangle^2 = 1\\
V(\hat Y) &=& \langle \alpha | \hat Y^2 | \alpha \rangle  -  \langle \alpha | \hat Y | \alpha \rangle^2  = - \langle \alpha | ( \hat a^\dagger - \hat a )^2 | \alpha \rangle  +  \langle \alpha | \hat a^\dagger - \hat a | \alpha \rangle^2 = 1,
\end{eqnarray}
where we have used the Boson commutation relation (Eq.~(\ref{a_comm})), and the definitions of the quadrature operators (Eqs.~(\ref{Xquad})~and~(\ref{Yquad})). We see that the coherent state is a minimum uncertainty state, whose quadrature variances are equal and saturate  the uncertainty principle (Eq.~(\ref{HeisenbergX})). As discussed earlier in Section~\ref{CorrelationSection}, coherent states provide the minimum uncertainty possible without quantum correlations, and therefore establish an important lower bound on measurement precision.  Indeed, a coherent state  can be thought of as a perfectly noiseless classical field with additional Gaussian noise introduced with the statistics of vacuum noise. For this reason they are often referred to as ``classical light"~\cite{Giovannetti2004,Treps2002,Taylor2013_sqz}, with the standard quantum limit described as the best precision which can be achieved classically~\cite{Dowling2008,Taylor2013tutorial}, even though truly classical light could, in principle, be entirely noise free.

\subsubsection{Cramer-Rao Bound on phase sensing with coherent states}

From Eqs.~(\ref{cram-rao}),~(\ref{phase_QFI}),~and~(\ref{var_coh}) it is possible to immediately determine the Cramer-Rao Bound for phase sensing with coherent states. This, unsurprisingly, reproduces the semi-classical result given in Eq.~(\ref{phi_SQL_samplepower}), and can be reached using the same linear estimation strategy given in Eq.~(\ref{phi_estimate}). Since this represents the best precision that is classically achievable, it follows that any state with higher quantum Fisher information than a coherent state must necessarily exhibit quantum correlations; though the value of the quantum Fisher information does not necessarily quantify the degree of entanglement~\cite{erol2014analysis}.

It is important to note that the Cramer-Rao Bound, as derived here, establishes the fundamental limit to the precision achievable for a given photon number at the phase-shifting sample. However, achieving this bound generally requires that additional photons are introduced within the measurement device. 
 If the total power is constrained, these additional photons must be included in the derivation of the quantum Fisher information.

\subsection{Phase sensing with squeezed states}\label{SqueezedState}

 The field of quantum metrology broadly began when Caves showed theoretically that squeezed states of light could be used to suppress quantum noise in an inteferometric phase measurement~\cite{Caves1981}.  This principle is currently used in gravitational wave observatories, with squeezed vacuum used to enhance precision beyond the limits of classical technology~\cite{LIGO2011,LIGO2013}.  In this sub-section we introduce the concept of phase sensing with squeezed states.

\subsubsection{Squeezed states}

As we have already shown, the uncertainty principle places a fundamental constraint on the product of amplitude and phase quadrature variances of an optical field (see Eq.~(\ref{HeisenbergX})). Coherent states spread this uncertainty equally across both quadratures. By contrast, as their name suggests, squeezed states trade-off reduced uncertainty in one quadrature with increased uncertainty in the other. This is an important capability in quantum metrology, and can be applied to enhance the precision of a broad range of optical measurements. If the amplitude is squeezed as illustrated in Fig~\ref{PhaseSpace}e, for instance, the photons tend to arrive more evenly spaced than in a coherent field, which is a classically forbidden phenomena known as photon anti-bunching (see Section~\ref{class_forb}). This can be used to reduce the variance in amplitude or intensity measurements, and thus enable precision better than that possible with coherent states. Alternatively, as illustrated in Fig.~\ref{PhaseSpace}f, the precision of phase sensing can be improved by orientating the squeezed quadrature to be orthogonal  to the coherent amplitude of the field.


\subsubsection{Mean photon number and photon number variance}

%
%
%
%

The mean photon number in a squeezed state is given by
\begin{eqnarray}
\langle \hat n \rangle &\equiv& \langle \hat a^\dagger \hat a \rangle \nonumber\\
&=& \frac{1}{4} \left \langle \left ( \hat X - i \hat Y  \right ) \left ( \hat X + i \hat Y \right ) \right \rangle \nonumber\\
&=& \frac{1}{4} \left ( \left \langle  \hat X^2 \right \rangle + \left \langle \hat Y^2 \right \rangle + i \left \langle  \hat X \hat Y - \hat Y \hat X  \right \rangle \right )\nonumber \\
&=& |\alpha|^2 + \frac{1}{4} \left (V( \hat Y ) + V( \hat X )  -2  \right ) \label{mean_n},
\end{eqnarray}
where we have used the definitions of $\hat X$ and $\hat Y$ in Eqs.~(\ref{Xquad})~and~(\ref{Yquad}), and the commutation relation between them (Eq.~(\ref{XY_comm})). We see that, unlike the coherent state, squeezed states have non-zero photon number even when their coherent amplitude $\alpha$  is zero (see Fig.~\ref{PhaseSpace}d). While we do not derive it here, in general, the photon number variance of a squeezed state is given by~\cite{orszag2000quantum}
\begin{equation}
V(\hat n) =|\alpha|^2 \left [V(\hat X) \cos^2 \theta + V(\hat Y) \sin^2 \theta  \right ]  + \frac{1}{8} \left [ V(\hat Y)^2 + V(\hat X)^2 -2 \right], \label{general_Vn}
\end{equation}
where $\theta$ is the angle between the squeezed quadrature of the state and its coherent amplitude.

\subsubsection{Cramer-Rao Bound for squeezed vacuum}

As shown in Section~\ref{QuantumFI}, to maximise the precision of a phase measurement the photon number variance should be made as large as possible. Without loss of generality, we take the $\hat Y$ quadrature to be squeezed, with the $\hat X$ quadrature then maximally anti-squeezed. For fixed mean photon number $\langle \hat n \rangle$, it can be shown from Eqs.~(\ref{mean_n})~and~(\ref{general_Vn}) that the maximum photon number variance is achieved when $\alpha=0$.  Rearranging Eq.~(\ref{mean_n}), we then find that
\begin{subequations}
\begin{eqnarray}
V(\hat X) &=& 2 \langle \hat n \rangle + 1 + 2 \sqrt{\langle \hat n \rangle^2 + \langle \hat n \rangle  } \\
V(\hat Y) &=& 2 \langle \hat n \rangle + 1 - 2 \sqrt{\langle \hat n \rangle^2 + \langle \hat n \rangle  },
\end{eqnarray}
\end{subequations}
where we have used the relation $V(\hat X) V(\hat Y) = 1$ which is valid for pure squeezed states. Substituting these expressions into Eq.~(\ref{general_Vn}) and simplifying results in the photon number variance:
\begin{equation}
V(\hat n) = 2 \left ( \langle \hat n \rangle^2 + \langle \hat n \rangle \right ).
\end{equation}
Substitution into Eq.~(\ref{phase_QFI}) for the quantum Fisher information then yields
\begin{equation}
\mathcal{F}  =  8 \left ( \langle \hat n \rangle^2 + \langle \hat n \rangle  \right ),
\end{equation}
so that the Cramer-Rao Bound (Eq.~(\ref{cram-rao})) on phase sensing with squeezed light is given by
\begin{equation}
\Delta \phi_{\rm squeezed} \ge \frac{1}{2 \sqrt{2}} \left [ \frac{1}{ \langle  \hat n \rangle^2+ \langle \hat n  \rangle} \right ]^{1/2}. \label{CRB_sqz}
\end{equation}
%
This squeezed state bound is compared with the Heisenberg and standard quantum limits in Fig.~\ref{Phase_sensing_limits}.  It exceeds the Heisenberg limit at all photon numbers, though only by a relatively small margin.
For large photon number $\langle \hat n \rangle$, it has Heisenberg scaling ($\Delta \phi \propto \langle \hat n  \rangle^{-1} $) as has been shown previously in Ref.~\cite{PhysRevLett.100.073601}. Furthermore,  Ref.~\cite{Lang2013} shows that of all possible choices of input state, squeezed vacuum achieves the optimal sensitivity. It should be emphasized, however, that proposals to reach the bound given in Eq.~(\ref{CRB_sqz}) require perfectly efficient photon-number resolving detectors and nonlinear estimation processes, and are not achievable with existing technology~\cite{PhysRevLett.100.073601}. The influence of inefficiencies on the precision of squeezed-state based phase measurement is considered for the specific case of linear detection and estimation in Section~\ref{Sqz_eff_eff_sec}.

\subsubsection{Cramer-Rao Bound in the large coherent amplitude limit}
\label{large_alp_lim_sec}

In general, increasing levels of quantum correlation are required to achieve a large photon number variance in a pure  quantum state that has no coherent amplitude ($\alpha=0$). As a result such states quickly become fragile to the presence of loss, which removes photons and therefore degrades the correlations between them (see Sections~\ref{Loss}~and~\ref{single_param_sec}). As a result, most state-of-the-art measurements rely on coherent states with large coherent amplitudes rather than non-classical states. However, with squeezed states it is possible to benefit from both large coherent amplitude and quantum correlations. 
Considering the limit $|\alpha|^2 \gg \{V(\hat X), V(\hat Y)\}$, Eqs.~(\ref{mean_n})~and~(\ref{general_Vn}) become:
\begin{eqnarray}
\langle \hat n \rangle & \approx & |\alpha|^2 \label{Sqz_n}\\
V(\hat n)  & \approx & |\alpha|^2 \left [V(\hat X) \cos^2 \theta + V(\hat Y) \sin^2 \theta  \right ].\label{Sqz_V}
\end{eqnarray}
These expressions 
allow us to calculate the second order coherence function for bright squeezed states, as described in Eq.~(\ref{g2}), with the result that 
\begin{equation}
g^{(2)}_{11} = 1+\frac{1}{\langle \hat n \rangle} \left( V(\hat X) \cos^2 \theta + V(\hat Y) \sin^2 \theta  -1\right).
\end{equation}
Remembering that we have chosen to orientate our squeezed state  such that $V(\hat Y) = 1/V(\hat X) <1$, we can thus see that phase squeezed light with $\theta=0$ is bunched ($g^{(2)}_{11}>1$); by contrast, amplitude squeezed light with $\theta=\pi/2$ is anti-bunched ($g^{(2)}_{11}<1$), which is one clear indicator  of quantum correlations (see Section~\ref{class_forb}).

Using Eqs.~(\ref{Sqz_n}) and (\ref{Sqz_V}) the quantum Fisher information for a bright squeezed state can be calculated from Eq.~(\ref{QFI}). It is given by
\begin{equation}
\mathcal{F}_{|\alpha|^2 \gg  \{V(\hat X), V(\hat Y)\}} = 4 \langle \hat n \rangle \left[V(\hat X) \cos^2 \theta + V(\hat Y) \sin^2 \theta  \right ].
\end{equation}
The optimal phase precision is clearly achieved when the antisqueezed quadrature $\hat X$ is aligned in phase space with the coherent amplitude $\alpha$ of the state (i.e., $\theta=0$), such that the phase variance is minimized while the amplitude variance is maximized. In this limit, the achievable phase precision is given by
\begin{equation}
\Delta \phi_{\rm squeezed} \ge \frac{1}{2 \sqrt{\langle \hat n \rangle V(\hat X)}} =\sqrt{V(\hat Y) } \Delta \phi_{\rm SQL},\label{SqueezedSQL}
\end{equation}
where the relevant standard quantum limit for phase measurement  $\Delta \phi_{\rm SQL}$ is given in Eq.~(\ref{phi_SQL_samplepower}). Since here $V(\hat Y) < 1$, we see that squeezed light allows precision beyond the standard quantum limit. 

\begin{figure}[t!]
 \begin{center}
   \includegraphics[width=14cm]{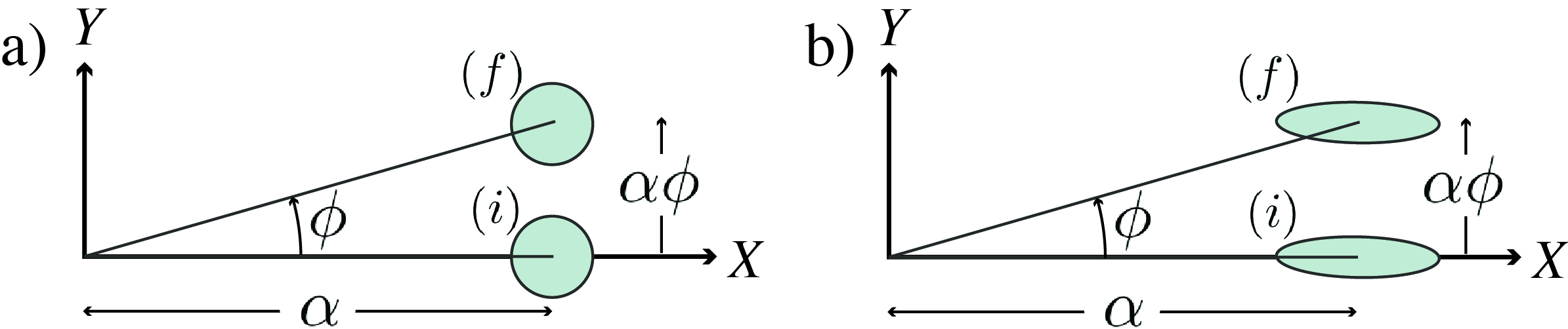}
   \caption{Phase space representations of the first-order effect of a phase shift $\phi$ on  (a)  bright coherent and (b) phase squeezed  fields. ($i$) and ($f$) label the states of the fields prior to, and after, the phase shift, respectively.
}
 \label{PhaseSpaceBrightSQZ}  
  \end{center}
\end{figure}

Similarly to the case of phase sensing with coherent states, the optimum phase precision in Eq.~(\ref{SqueezedSQL}) may be reached using a straight-forward linear estimation strategy. Let us describe the squeezed probe field in the Heisenberg picture via the annihilation operator $\hat a_{\rm sig} = \alpha + \delta \hat a_{\rm sig}$, where as usual the coherent amplitude $\alpha = \langle \hat a_{\rm sig} \rangle$ and the quantum noise operator $\delta \hat a_{\rm sig}$ has zero expectation value.  The action of a phase shift $\phi$ is to transform the field $\hat a_{\rm sig}$ to the output field $\hat a$  given by
\begin{eqnarray}
\hat a &=& (\alpha + \delta \hat a_{\rm sig}) e^{i \phi}\\
&\approx&  (\alpha + \delta \hat a_{\rm sig}) (1 + i \phi)\\
&=& \alpha  (1 + i \phi) +  \delta \hat a_{\rm sig}, \label{phase_sqz_a}
\end{eqnarray}
where in the first approximation, we have taken a first order Taylor expansion of the exponential, constraining our analysis to a small phase shift, $\phi \ll 1$; and in the second approximation, since we are considering the case of a bright field with $\alpha$ large, we neglect the product of two small terms $i \phi \, \delta \hat a_{\rm sig}$. This expression shows that, to first order, the action of the phase shift is to introduce an imaginary component to the coherent amplitude of the field -- i.e., it displaces the field in phase space along the axis of the phase quadrature $\hat Y$.  This is illustrated in Fig.~\ref{PhaseSpaceBrightSQZ}, which compares coherent and phase squeezed fields. 

Using Eq.~(\ref{Yquad}), the phase quadrature of the output field is given by
\begin{equation}
\hat Y = \hat Y_{\rm sig} + 2 \alpha \phi,
\end{equation}
so that, rearranging,
 \begin{equation}
\phi = \underbrace{\hat Y/2 \alpha}_{\text{detected signal}} - \underbrace{\hat Y_{\rm sig}/2 \alpha}_{\text{measurement noise}}. \label{phi_eq_bri_sqz}
\end{equation}
We see that measurement of $\hat Y$, which can be achieved quite straightforwardly, for instance, using homodyne detection, allows an estimate of the phase $\phi$. The uncertainty of the estimate is  determined, in combination, by the uncertainty of the phase quadrature of the probe field $\hat Y_{\rm sig}$, and the magnitude of the coherent amplitude $\alpha$ of the probe field, which, in the bright field limit taken here, is simply related to the probe photon number via Eq.~(\ref{Sqz_n}). 
Taking the standard deviation of the measurement noise term in Eq.~(\ref{phi_eq_bri_sqz}),
we find that this measurement exactly saturates the phase precision inequality of Eq.~(\ref{SqueezedSQL}), demonstrating that a standard noise-free homodyne measurement can -- in principle -- reach the absolute Cramer-Rao Bound to phase precision for a bright squeezed field.

\subsubsection{Interferometry combining squeezed vacuum with a coherent field}

\begin{figure}
 \begin{center}
   \includegraphics[width=8.25cm]{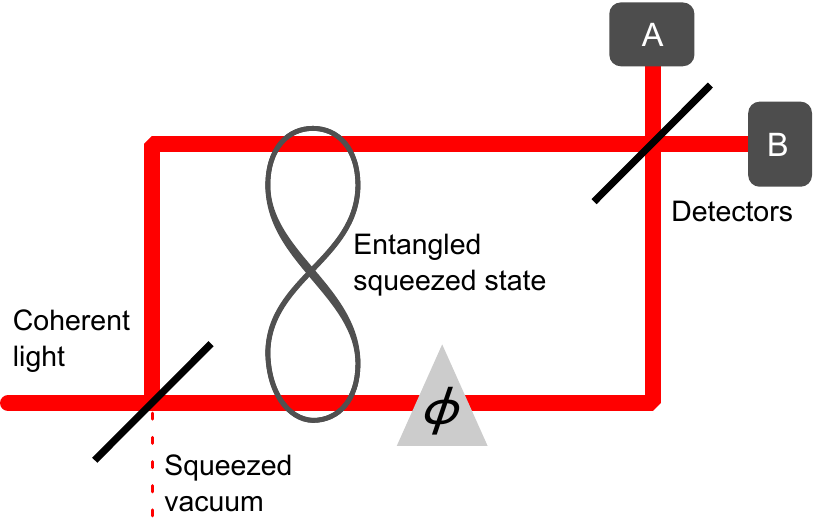}
   \caption{A Mach--Zehnder phase sensing experiment enhanced via the injection of a squeezed vacuum. The squeezed vacuum entangles the fields within the interferometer, and can be used to suppress quantum noise.
}
 \label{MachZehnder_sqz}  
  \end{center}
\end{figure}

 The scenario described in the previous sub-section corresponds to a single bright squeezed state propagating through a phase shifting element. Most experimental demonstrations apply an alternative approach, in which  a squeezed vacuum and coherent state are combined on a beam splitter and then propagate through an interferometer (see Fig.~\ref{MachZehnder_sqz}).  The resulting two fields in the interferometer are an entangled pair of bright squeezed states, with anticorrelated quantum fluctuations. The phase of the input fields is chosen to align the coherent input to the antisqueezed quadrature, which maximizes the quantum Fisher information for a differential phase shift measurement. A full analysis of this approach shows that the phase precision can be enhanced by $\sqrt{V(\hat X) }$, similar to Eq.~(\ref{SqueezedSQL})~\cite{Caves1981}.

 This is the basic approach considered by Caves in the first quantum metrology proposal~\cite{Caves1981}, and which is now used in gravitational wave observatories~\cite{LIGO2011,LIGO2013}. It is particularly useful since the quantum state preparation can be separated from the generation of a large coherent amplitude, allowing both states to be independently optimized.  Within this framework, the non-classical state provides a fixed enhancement in precision for the arbitrarily bright coherent state. It is of note that, for constrained photon occupation in the non-classical state, of all possible choices of input state squeezed vacuum provides the largest possible precision enhancement to the bright interferometric phase measurement~\cite{Lang2013}.

\subsubsection{Generation of squeezed states}
\label{Gen_sqz_st_sec}

Squeezed states of light are produced via nonlinear optical interactions. The essential feature of such nonlinear interactions is a reversible phase sensitive amplification of the optical field, such that the fluctuations of one quadrature are noiselessly amplified while the fluctuations of the orthogonal quadrature are noiselessly de-amplified. Phase sensitive amplification can be achieved using a wide range of nonlinear processes, such as optical parametric oscillation and amplification~\cite{Mehmet2010,Stefszky2012}, optical Kerr nonlinearities~\cite{Silberhorn2001}, second harmonic generation~\cite{Pereira1988}, and four-wave mixing in an atomic vapour~\cite{Boyer2008,Corzo2012}; and each of these approaches has been used to generate squeezed light. To take one example, amplitude squeezed light can be generated via second harmonic generation, where two photons combine to produce a single frequency-doubled photon with a probability proportional to the square of the intensity of the incident field. The quantum fluctuations in the intensity of the field then translate to fluctuations in the probability of second harmonic generation. Thus, when the field intensity fluctuates toward a larger value, the second harmonic generation becomes more efficient and more of the light is lost. A fluctuation of field intensity toward a smaller value will result in less efficient second harmonic generation, and consequently smaller loss. The net effect is to suppress the intensity fluctuations, which can result in fluctuations below the vacuum level, with a consequent increase in the phase fluctuations.

Much of the effort in the development of squeezed light sources over the past decade has focussed on achieving sources useful to enhance measurements in gravitational wave observatories~\cite{Schnabel2010}. One important requirement for this application is a high degree of squeezing which can allow an appreciable enhancement in precision. In absolute terms, the strongest single-mode squeezing is currently achieved in optical parametric oscillators that are  pumped with light at 532~nm or 430~nm and which produce vacuum squeezed fields at 1064~nm~\cite{Eberle2010,Mehmet2010,Stefszky2012} and 860~nm~\cite{Takeno2007}, respectively.  Such sources are capable of squeezed quadrature variances as small as $V(\hat X) = 0.054$~\cite{Eberle2010}. A second key requirement is for the squeezed light source to allow quantum enhanced precision at low frequencies~\cite{Bowen2002,McKenzie2004,Vahlbruch2006}. Squeezed light has a finite frequency band of enhancement, and many sources only provide squeezing in the MHz regime~\cite{Taylor2013_sqz,Zhai2012}. This is inadequate for gravitational wave detection, where most of the signals are expected  at hertz and kilohertz frequencies~\cite{Schnabel2010}. This frequency band is also important to biological experiments, as it encompasses most biophysical dynamics studied to date (see Section~\ref{BioFreq}). Consequently, biological quantum metrology could directly benefit from the advances toward gravitational wave detection. Squeezed light has now been demonstrated at frequencies as low as 1~Hz~\cite{Vahlbruch2007}, though light which is squeezed to 10~Hz is still considered state-of-the-art~\cite{LIGO2013,Stefszky2012}.


In a typical squeezed light source, the nonlinear medium is placed within an optical resonator to increase the strength of the nonlinear interaction and thus the magnitude of the squeezing. This generally allows only a single mode of the optical field to be squeezed, as is required to enhance single-parameter measurements such as phase or intensity measurements. However, efforts have also been made to generate strong multimode squeezing for applications such as quantum enhanced imaging. In particular, multimode degenerate optical resonators have been developed to allow resonant enhancement of as many as eight modes~\cite{Armstrong2012}; while the strong optical nonlinearity near the optical resonances of an atomic vapour allows strong squeezing without use of an optical cavity, allowing generation of a multimode squeezed vacuum state  over hundreds of orthogonal spatial modes~\cite{Boyer2008,Corzo2012}. Another source of highly multimode quantum correlated light is parametric down-conversion, which produces a photon pair from a single high-energy photon. To conserve momentum, the  photons always propagate with opposing directions, such that they occupy coupled spatial modes.

\subsection{Phase sensing with NOON states}\label{NOONState}



As discussed in Section~\ref{QuantumFI}, in general, the achievable precision of optical phase measurements improves as the  photon number variance of the field in the signal arm of the interferometer increases (see Eq.~(\ref{QuantumFI})). In the limit that exactly $N$ photons are used,\footnote{That is, the variance of the {\it total} photon number in the interferometer (the sum of the photon number in the signal and reference arms) is zero. This contrasts circumstances where the total photon number is indeterminate, such as  the squeezed state based phase measurements we considered earlier. In that case, while the mean photon number is well defined, the photon number variance is non-zero.} the precision is maximized for a NOON state, in which the constituent photons are in a superposition of all occupying either the signal or reference arm of the interferometer, with the other unoccupied~\cite{Dowling2008}.  In the number state basis, the NOON state can be represented as 
\begin{equation}
| \Psi_{\rm NOON} \rangle  = \frac{1}{\sqrt{2}}\left(| N \rangle_{\rm sig} | 0 \rangle_{\rm ref} +| 0 \rangle_{\rm sig} | N \rangle_{\rm ref} \right),
\end{equation}
where $| j \rangle$ signifies a state containing $j$ photons, and the subscripts ``sig" and ``ref" label the signal and reference arms of the interferometer, respectively. When using a NOON state, the quantum Fisher information for phase estimation (Eq.~(\ref{phase_QFI})) is determined by
\begin{eqnarray}
\langle \Psi_{\rm NOON} |\hat{n}_{\rm sig}  | \Psi_{\rm NOON} \rangle &=& \frac{1}{\sqrt{2}} \langle \Psi_{\rm NOON}| \hat n_{\rm sig} |N \rangle_{\rm sig} | 0 \rangle_{\rm ref} -\frac{1}{\sqrt{2}}\langle \Psi_{\rm NOON} | \hat n_{\rm sig} | 0 \rangle_{\rm sig} | N \rangle_{\rm ref}\\
&=& \frac{N}{\sqrt{2}} \langle \Psi_{\rm NOON} |N \rangle_{\rm sig} | 0 \rangle_{\rm ref}\\
&=& \frac{N}{2}\\
\langle \Psi_{\rm NOON} |\hat{n}_{\rm sig} ^2| \Psi_{\rm NOON} \rangle & =&\frac{1}{\sqrt{2}} \langle \Psi_{\rm NOON} | \hat n_{\rm sig}^2 | N \rangle_{\rm sig} | 0 \rangle_{\rm ref} +\frac{1}{\sqrt{2}}\langle \Psi_{\rm NOON}  | \hat n_{\rm sig}^2 | 0 \rangle_{\rm sig} | N \rangle_{\rm ref}\\
& =&\frac{N^2}{\sqrt{2}} \langle \Psi_{\rm NOON} |N \rangle_{\rm sig} | 0 \rangle_{\rm ref} \\
&=&\frac{N^2}{2},
\end{eqnarray}
such that 
\begin{equation}
\mathcal{F} = 4 \left ( \langle \Psi_{\rm NOON} |\hat{n}_{\rm sig} ^2| \Psi_{\rm NOON} \rangle - \left | \langle \Psi_{\rm NOON} |\hat{n}_{\rm sig}  | \Psi_{\rm NOON} \rangle \right |^2 \right ) = N^2
\end{equation}
and
\begin{equation}
\Delta \phi = \frac{1}{N}. \label{asdfadsgag}
\end{equation}
This is exactly the Heisenberg limit on phase estimation which we derived via a semi-classical approach earlier (see  Eq.~(\ref{Heisenberg1})). It has been shown that the NOON state achieves the optimal phase precision for states which contain exactly $N$ photons; although states with indeterminate photon number such as the squeezed vacuum have been shown to also achieve Heisenberg scaling and allow  superior precision~\cite{PhysRevA.61.043811,PhysRevLett.85.2733, PhysRevA.66.013804, Lee2002} (compare Eq.~(\ref{asdfadsgag}) with Eq.~(\ref{CRB_sqz}) in  Section~\ref{SqueezedState}). 


To understand why NOON states allow enhanced precision, it is instructive to consider the evolution of the state through the interferometer. When the phase shifting unitary of Eq.~(\ref{phase_shift_unitary})
is applied to a number state, it evolves as
\begin{equation}
e^{i \phi \hat{n}} |N\rangle = \sum _{j=0}^\infty \frac{(i \phi \hat{n})^j}{j!} |N\rangle =e^{i \phi N} |N\rangle. \label{PhaseShiftOperator}
\end{equation}
Consequently, the phase shifting operator applies a phase shift of $N \phi$ to a number state. NOON states  acquire this amplified phase shift. By contrast, using the Fock state expansion of coherent states given in Eq.~(\ref{fock_basis}), we see that a coherent state $|\alpha\rangle$ evolves as
\begin{equation}
e^{i \phi \hat{n}} |\alpha\rangle = e^{i \phi \hat{n}} e^{-|\alpha|^2/2} \sum _{j=0}^\infty \frac{\alpha^j}{\sqrt{j!}}|j\rangle=e^{-|\alpha|^2/2} \sum _{j=0}^\infty \frac{(e^{i \phi}\alpha)^j}{\sqrt{j!}}|j\rangle = |e^{i \phi}\alpha\rangle,
\end{equation}
which corresponds to the same phase shift $\phi$ for any mean photon number. The enhanced phase precision achievable with a NOON state originates from this amplification of the relative phase shift~\cite{Dowling2008}. Based on this behaviour, one might expect that a NOON state would achieve sub-wavelength interference features. This intuition is correct, but not in the manner that may be expected. NOON states do not allow sub-wavelength interference of {\it intensity}, but rather allow sub-wavelength interference features in the higher order photon statistics, such as two-photon coincidence detection. 
 
\subsubsection{Generation of  NOON states}
\label{gen_noon_state_subsubsec}

A single photon NOON state can be generated simply by splitting a single photon on a 50/50 beam splitter, which places the photon into an equal superposition of reflecting and transmitting, i.e.,  $| \Psi \rangle = (| 1 \rangle_{\rm sig} | 0\rangle_{\rm ref}+ | 0\rangle_{\rm sig} | 1\rangle_{\rm ref} )/\sqrt{2}$. However, the Heisenberg limit coincides with the standard quantum limit for $N=1$ (compare Eq.~(\ref{phi_SQL_totalpower})~with Eq.~(\ref{Heisenberg1})), so that such a state cannot be used to enhance precision. Most sensing applications, instead, apply two photon NOON states, which are generally  produced from entangled photon pairs via two-photon interference in a Hong-Ou-Mandel (HOM) interferometer~\cite{Hong1987} (see Fig.~\ref{NOON_interferometer}, and also Fig.~\ref{HOM}). In principle this allows the standard quantum limit to be surpassed by a factor of $\sqrt{2}$. However, this is only true for constrained {\it total} power. An important and often overlooked distinction arises if we only constrain power through the sample, and allow arbitrary power in the reference arm; in this case the relevant limit is Eq.~(\ref{phi_SQL_samplepower}) and a two-photon NOON state can only equal the standard quantum limit. Put another way, NOON states must contain at least three photons before they can outcompete coherent states of similar mean intensity at the sample.

The entangled photon pairs can be generated via spontaneous parametric down conversion in a nonlinear crystal that is pumped with light at twice the optical frequency. The spatial modes of the photons generated by this down conversion process are dictated by energy and momentum conservation. In general, many pairs of spatial modes are each populated with correlated photons. These photons are termed to be ``hyper-entangled"~\cite{barreiro2005generation}, since they can exhibit not only temporal correlations, but also correlations in their momentum, position, and polarization.  This enables a range of imaging applications, some of which are discussed in Sections~\ref{Q_OCT_section},~\ref{ab_im_secasf}, and~\ref{diif_inf_mic_sec}. For simplicity, here, however, we restrict our analysis to a single pair of energy and momentum conserving spatial modes (labelled with the subscripts ``1" and ``2"), in which case the ideal state generated by the process is 
~\cite{kok2007linear}
 \begin{equation}
 | \Psi \rangle = \sqrt{1-\epsilon} \sum_{N=0}^\infty \epsilon^{N/2} |N \rangle_1 | N \rangle_2, \label{state-para_down}
 \end{equation}
 where $\epsilon$ characterises the strength of the nonlinear interaction.
 It can be seen that this state includes components of vacuum ($|0 \rangle_1 | 0 \rangle_2$),  the desired photon pairs ($|1 \rangle_1 | 1 \rangle_2$), as well as higher order terms. While, in some circumstances, the higher order terms can be useful, generally they are undesirable, introducing noise to the measurements. Consequently, most quantum metrology experiments operate with very weak nonlinear interaction strength, such that $\epsilon \ll 1$. Furthermore, since photon counters will clearly register no clicks if the state is vacuum, the vacuum term in the expansion can be neglected. In this case, the output state from a parametric down converter can be well approximated as the ideal photon pair
\begin{equation}
| \Psi \rangle =  | 1 \rangle_1 | 1 \rangle_2.
\end{equation}

Interfering the photon pairs generated via parametric down conversion on a balanced beam splitter results in 
Hong-Ou-Mandel interference~\cite{Hong1987}. 
For generality, we include a relative phase shift $\phi_0$ on the input arm $1$ via the phase shifting unitary operator (Eq.~(\ref{phase_shift_unitary})), though, ultimately, we will see that this does not affect the outcome. The input quantum state  therefore becomes 
\begin{equation}
| \Psi \rangle =e^{i \phi_0 \hat{n}_1} | 1 \rangle_1 | 1 \rangle_2 = e^{i \phi_0 } | 1 \rangle_1 | 1 \rangle_2,
\end{equation}
%
When this state is incident on the beam splitter, the two output modes each contain components from both input $1$ and input $2$, which can be represented as a mapping of the annihilation operators. Specifically, the signal and reference mode operators are given, respectively, by
\begin{eqnarray}
\hat{a}_{\rm sig} = \frac{\hat{a}_{1} + \hat{a}_{2}}{\sqrt{2}} ,  \\
\hat{a}_{\rm ref} = \frac{\hat{a}_{1} - \hat{a}_{2}}{\sqrt{2}}  ,
\end{eqnarray}
where the minus sign represents the $\pi$ phase shift on reflection from a hard boundary. The mean photon flux in the signal arm is given by 
\begin{equation}
\langle \Psi | \hat a^\dagger_{\rm sig}  \hat{a}_{\rm sig} | \Psi \rangle =1.\label{NOON_mean}
\end{equation}
We can also evaluate the two-photon coincidences, with
\begin{equation}
 \langle \hat{a}^\dagger_{\rm sig} \hat{a}^\dagger_{\rm sig} \hat a_{\rm sig} \hat a_{\rm sig} \rangle = 1.\label{NOON_2p}
\end{equation}
A similar result holds for the reference arm. The beam splitter outputs have on average one photon each (Eq.~(\ref{NOON_mean})); and the two-photon coincidence rate (Eq.~(\ref{NOON_2p})) requires that the two photons always propagate in the same arm. We see, therefore, that the interference of two single photons produces a two photon NOON state of the form 
\begin{equation}
| \Psi \rangle = \frac{1}{\sqrt{2}} \left (  | 2 \rangle_{\rm sig} | 0 \rangle_{\rm ref} + | 0 \rangle_{\rm sig} | 2 \rangle_{\rm ref} \right ) .
\end{equation}
Although the phenomena used to generate this NOON state is based on interference, its characteristics are completely foreign to classical interferometry. The interference has no effect on the mean intensity; and equally strange, the output state is independent of the relative phase $\phi_0$. 

As can be seen from Eq.~(\ref{Heisenberg1}), the phase precision achievable with NOON states improves as the number of photons in the superposition increases. This motivates the generation of larger NOON states. However, NOON states that are larger than two photons cannot currently  be generated deterministically. Such states can be produced through spontaneous parametric down conversion via the higher order terms in Eq.~(\ref{state-para_down}). However, the first order term (and other lower order terms) interferes with the process, complicating matters. 
The largest NOON state generated to date contained five photons, with an average photon flux below 1~s$^{-1}$~\cite{Afek2010}. At such low flux the absolute precision achievable with such states cannot compete with routine measurements with modest power coherent states. Furthermore, the NOON states produced in Ref.~\cite{Afek2010}  co-propagated with well over 1000~s$^{-1}$ photons in other states.  In practical scenarios where the sample, detectors, or other apparatus are sensitive to photo-damage, these unwanted photons must be accounted for in the total optical power that can be used. Consequently, this degrades the phase precision that can be achieved per incident photon~\cite{Combes2014}. In Ref.~\cite{Afek2010} post-selection was required to select the 5-photon coincidences, which allowed observation of the properties of 5-photon NOON states. Although interesting for its observation of larger-scale entanglement, this could not be extended to practical metrology applications without accounting for the full photon flux, which drastically degrades the achievable precision~\cite{Combes2014}.




 \subsubsection{An example NOON state experiment}\label{ExampleNOON}

 \begin{figure}
 \begin{center}
   \includegraphics[width=10cm]{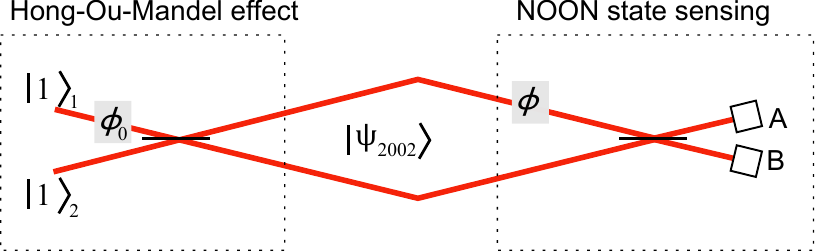}
   \caption{An example phase sensing experiment with a two photon NOON state. This schematic is divided into two sections: state preparation via the Hong--Ou--Mandel effect, followed by interferometric read-out. The input is two single photons, which are generally produced via parametric downconversion. These photons arrive simultaneously,  violating the Cauchy--Schwarz inequality for classical photon correlations given in Eq.~(\ref{g2class_2mode}). These photons interfere to form a two photon NOON state. Interference at the output beam splitter has only two physically allowed outcomes: either both photons are transmitted, or both photons are reflected. Consequently,  one photon always exits from each of the two outputs of the interferometer, so that the average intensity at the two detectors is identical; it is only the photon coincidence rate that distinguishes these two outcomes. As such, nowhere in this experiment does interference modulate the mean intensity. Rather, interference is only observable in the second order photon statistics. This is strikingly different to classical optics.
}
 \label{NOON_interferometer}  
  \end{center}
\end{figure}

To provide a more concrete basis for our discussion of NOON state-based optical phase sensing, we consider a simple NOON state experiment with two photons, as shown in Fig.~\ref{NOON_interferometer}. Once the NOON state is generated as described in the pervious section, propagation of the signal mode through the sample results in a phase shift $\phi$ applied via the unitary operator $e^{i \hat{n}_{\rm sig} \phi}$. The state is then given by
\begin{equation}
| \Psi \rangle = \frac{1}{\sqrt{2}}\left( e^{i 2 \phi}|2 \rangle_{\rm sig} | 0 \rangle_{\rm ref} +| 0 \rangle_{\rm sig} | 2 \rangle_{\rm ref} \right).
\end{equation}
The signal and reference modes are then combined on a beam splitter, with the annihilation operators at detectors $A$ and $B$ given as  
\begin{eqnarray}
\hat{a}_A = \frac{\hat{a}_{\rm sig} e^{i \hat{n}_{\rm sig} \phi} + \hat{a}_{\rm ref}}{\sqrt{2}} ,  \\
\hat{a}_B = \frac{\hat{a}_{\rm sig} e^{i \hat{n}_{\rm sig} \phi} - \hat{a}_{\rm ref}}{\sqrt{2}} .
\end{eqnarray}
As before, 
we can use these operators to evaluate the mean photon number and coincidence rate at the detectors, which yields for detector $A$:
\begin{eqnarray}
 \langle \Psi | a^\dagger_{A}\hat{a}_A | \Psi \rangle &=& 1,\\
 \langle \hat a^\dagger_{A} a^\dagger_{A} \hat a_{A} \hat a_{A} \rangle &=& \cos^2 \phi = \frac{1}{2} \left (1 + \cos 2\phi \right ).
\end{eqnarray}
Once again, the mean photon number in each arm is one, and is independent of the phase.  However, the probability of finding two photons in the same arm  varies sinusoidally with the phase $\phi$ acquired in the signal mode via propagation through the sample. This interference is only observable on the higher order photon correlations, and cannot be observed from a measurement of the mean intensity.

Since the photon coincidences vary with phase, it is possible to estimate the phase shift applied by measuring the ratio of coincident to non-coincident photon pairs. These statistics exhibit two full oscillations over a $2\pi$ phase shift, while an interferometer using coherent light will only show one full oscillation. Since the signal changes more rapidly in response to small changes in phase, NOON state sensing allows  phase precision beyond the standard quantum limit; this is often referred to as super-sensitivity.

In addition to super-sensitivity, the decreased spacing between interference fringes is sometimes referred to as super-resolution. This can cause significant confusion to those outside the field of quantum metrology. Classically, the term super-resolution denotes technologies which allow the diffraction limit on imaging resolution to be overcome (see Section~\ref{DiffractionLimit}). By contrast, super-resolution of the interference in NOON states is usually only discussed in non-imaging experiments, and is not applicable to linear imaging. The interference fringe spacing is closely related to the smallest features that can be achieved at the focus, so sub-wavelength interference can allow sub-wavelength imaging. However, the sub-wavelength interference with NOON states is achieved in  higher order photon correlations rather than the intensity, such that it is only applicable to nonlinear imaging schemes (as discussed in Section~\ref{QLithography}).

\subsection{Effects of experimental imperfection}\label{experimental_imperfection}

 The preceding sections have assumed ideal experimental conditions with no loss or noise; a situation which is not encountered in real experiments. As discussed in Section~\ref{inf_bound_limit_sec}, loss introduces an additional fundamental constraint (Eq.~(\ref{fund_loss_limit})). Experiments are thus bounded by both the quantum Cramer-Rao Bound  (Eq.~(\ref{cram-rao})) as well as this loss limit. 
 The degradation due to loss is analyzed further for the specific cases of NOON states and squeezed states in Sections~\ref{Loss} and \ref{Sqz_eff_eff_sec}, respectively. Noise is also an important limitation. If technical noise dominates the quantum noise on the light  -- due, for instance, to vibrations or laser fluctuations --  it is difficult to achieve any real advantage through quantum enhancement. Even when technical noise is below the standard quantum limit, it constrains the achievable enhancement.  Similar to loss, uncorrelated Markovian dephasing sets an upper bound to the achievable precision enhancement over the standard quantum limit~\cite{Escher2011,huelga1997improvement}; though superior scaling can in principle be  achieved in the presence of correlated noise~\cite{chin2012quantum,jeske2014quantum}.

\section{Challenges associated with biological measurement}\label{Biology}

In the previous sections we have seen that quantum mechanics introduces fundamental limits to the sensitivity of optical measurements, and that some of these limits can be overcome using quantum correlated photons. We now turn to the challenges associated with applying optical measurements within biological systems.  There are many such challenges,
both fundamental challenges associated to measurement precision and resolution, and technical challenges due, for example, to the effect of the light on the biological system, environmental fluctuations, and the growth, complexity, and dynamics of the system itself. This section seeks to introduce these challenges to the uninitiated quantum scientist, as well as describing some common approaches that are used in the biophysics community to address them. We do not aim to be exhaustive, but, rather, to give an overview of the general capabilities necessary to achieve practical biology measurements, and to provide guidance into the sort of problems that are likely to exist when applying quantum measurement techniques into biological systems.

 
\subsection{Resolution requirements}\label{ResolutionBio}

 \begin{figure}
 \begin{center}
   \includegraphics[width=14cm]{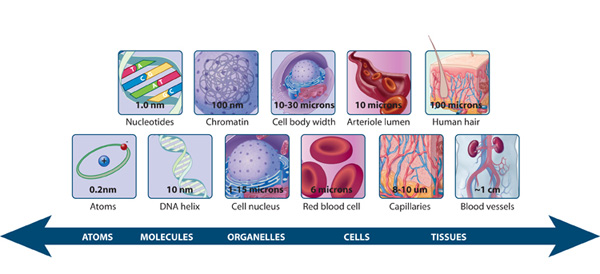}
   \caption{Length scales in biology. The diffraction limit of visible light restricts far-field imaging resolution to around 200~nm, which is the approximate scale of organelles. Single proteins or other biomolecules are not typically observable with a light microscope, and super-resolution techniques are required to study their function and structure. As discussed in Section~\ref{BioFreq}, the size scale is also strongly correlated with the characteristic frequencies of biological dynamics. Larger structures exhibit slower dynamics, with tissue dynamics typically occurring at Hz frequencies, while single-molecule dynamics can reach MHz frequencies. Reprinted by permission from Macmillan Publishers Ltd: Scitable, Ref.~\cite{ScitableFig} Copyright (2010)
}
 \label{BioScale}  
  \end{center}
\end{figure}

 
 The overarching goal of biophysical research is to understand the mechanisms by which living organisms function. This encompasses a vast range of different research areas, ranging from studies of the mechanics and utility of single biomolecules, to understanding how cells function collectively as organs (see Fig.~\ref{BioScale}). When imaging organs, the primary difficulty is often to accurately image through surrounding tissue which obscures the view, while maintaining sufficient resolution to observe features which can be as small as a few cells. While such structures can easily be observed by removing the surrounding tissue, this is a highly destructive process. Imaging techniques which can accurately and non-destructively image through tissue are extremely important for studies of changes and growth within living creatures, and also have important applications in medical diagnosis.

\begin{figure}
 \begin{center}
   \includegraphics[width=10cm]{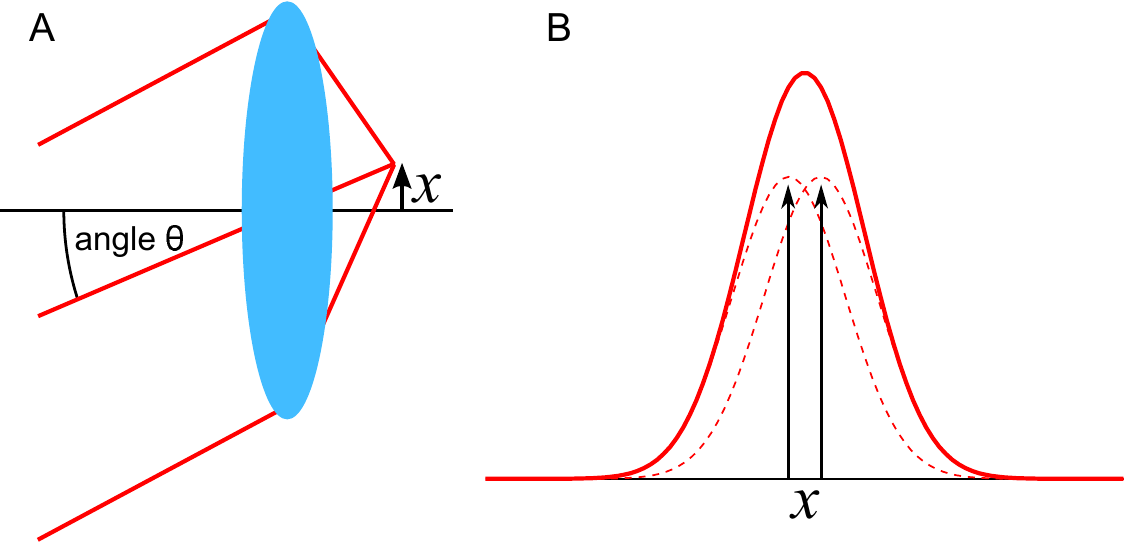}
   \caption{ (A) When light is focused through a lens, the incident angle $\theta$ determines the focal position $x$ (see Eq.~(\ref{DiffractionLimitEq})). However, diffraction establishes a minimum area in which a field can be focused. Considered in reverse, this same principle establishes the minimum angular range outside of the lens which a point source at $x$ will emit into. This has the effect of blurring an optical image, with point features broadened by the point-spread function. (B) This blurring limits the resolution with which two point sources can be resolved. In this case, the two individual sources (dashed lines) cannot be resolved from the total detected light (solid line), though its broadened profile allows inference that more than one point source is within the detected bright spot.
}
 \label{DiffractionLimit}  
  \end{center}
\end{figure}

 In contrast, optical measurements of sub-cellular components are often limited by diffraction. Diffraction constrains the  resolution that is achievable with all linear far-field optical imaging techniques, including absorption imaging, phase contrast microscopy, and differential interference microscopy. The {\it diffraction limit}\footnote{The criterion described here is also known as the Abbe limit. In some cases the term diffraction limit is used to describe the closely related Rayleigh limit or Sparrow limit, which each differ from the Abbe limit only by a small constant factor.} states that the minimum resolvable separation between features in an image is
 \begin{equation}
 x_{\rm min} = \frac{\lambda}{2\,{\rm NA}}, \label{DiffractionLimitEq}
\end{equation}
where $\lambda$ is the optical wavelength and NA is the numerical aperture of the imaging lens  (see Fig.~\ref{DiffractionLimit}). Optical microscopes operating at visible wavelengths are therefore restricted to resolution above approximately 200~nm. This length-scale is similar, for example, to the width of a typical mitochondria -- a relatively large organelle. 
%
%
 Many other organelles are much smaller, and almost no proteins within a biological system are sufficiently large to  be individually resolved (see Fig.~\ref{BioScale}). Consequently, sub-cellular features cannot be reliably separated with a conventional light microscope, and techniques which can resolve such structures are particularly useful.  Such techniques are the topic of Section~\ref{ClassicalSuperResolution}.


As well as defining the best resolution that is achievable in a conventional optical microscope, the diffraction limit of Eq.~(\ref{DiffractionLimitEq}) also quantifies the smallest spot size\footnote{This is referred to as the point-spread function.} to which light can be focused given the NA of the lens system that is used, with smaller focal spots prevented by diffraction of the optical waves. This has consequences, for instance, for optical manipulation of biological systems using tightly focussed laser fields (termed optical tweezers, see Section~\ref{SQZ_light_Qmet_sec}), and on the optical power required to achieve a fixed intensity of illumination of nanoscale biological samples.

Both for classical and quantum imaging systems, it is quite often  convenient to consider the diffraction limit within the framework of orthonormal spatial modes (introduced for quantum states of light in Section~\ref{FieldModes}).  A lens maps spherical wavefronts that are centered on the focal plane to flat wavefronts at the back focal plane (Fig.~\ref{DiffractionLimit}A). For two point emitters to be distinguishable, the propagating fields at the back focal plane should be in orthogonal spatial modes. These fields share similar intensity, but have a relative phase offset that is proportional to the emitter separation.  If the emitters are separated by ${\bf x}$, the relative phase of the light is shifted by ${\bf x} \cdot {\bf k}$, with ${\bf k}$ the wavevector. The phase shift is clearly largest for light that propagates at the widest possible angles, where $|{\bf x} \cdot {\bf k}|$ is maximized. However, in a finite-sized imaging system, not all angles of light are collected, with the numerical aperture limiting the maximum collection angle to ${\rm NA}=n {\rm sin}\theta_{\rm max}$; such that the relative phase shift spans a range of $2{\bf x} \cdot {\bf k}_{\rm max} = 4 \pi {\rm NA} x/\lambda$. The emitted fields are orthogonal provided that this range is at least $2 \pi$ across the aperture, from which we find that the emitter separation must meet the condition stated in Eq.~(\ref{DiffractionLimitEq}). If the emitters are closer together than this, their emitted fields are in non-orthogonal spatial modes. In this case, they can only be resolved if their emitted light is distinguishable in some other manner, such as via the wavelength, polarization, or time of emission. We will return to these ideas, from a quantum perspective, in Section~\ref{quantum_super_flour_mic_sec}, where we will see that quantum correlations provide one effective means to distinguish emitters.

\subsubsection{Classical super-resolution imaging techniques}\label{ClassicalSuperResolution}

As discussed above, optical diffraction limit restricts the resolution of all linear direct far-field optical imaging techniques. However, a range of techniques do exist that allow the  diffraction limit to be overcome -- both based on classical and quantum physics. Here, we consider classical techniques, returning to discuss quantum approaches to super-resolution later in the review.

Perhaps most obviously, if the point-spread function of the imaging objective is precisely known, it is possible to infer details of an optical image with resolution below the diffraction limit (see Fig.~\ref{DiffractionLimit}B). Details which cannot be directly observed in the image can be statistically reconstructed in a process known as image deconvolution~\cite{Park2003}.  Although this might appear to allow unlimited resolution, in reality the resolution is constrained because the reconstructed images become unreliable for particularly small feature sizes~\cite{Lucy1992}. This can be understood because the field leaving the microscope objective is closely approximated as the spatial Fourier transform of the field at the focal plane. From this perspective, the diffraction limit can be considered a natural consequence of the objective aperture, which prevents measurements of high spatial frequencies. Consequently, high spatial frequency details in the sample are fundamentally missing in the recorded image; and deconvolution is unable to reliably estimate the spatial frequencies which have not been sampled. An alternative method to break the diffraction limit is use of structured illumination. As noted above, a microscope only samples a finite range of spatial frequencies.  With structured illumination, the spatial frequencies at the focus can be modulated, which shifts some high frequency information to frequencies below the diffraction limit and thus allows the resolution to be improved by a factor of two~\cite{Langhorst2009}.

While the linear techniques discussed above allow super-resolution, in practice it has proved more effective to leverage optical nonlinearities within the sample. Two-photon absorption, for instance, occurs classically at a rate proportional to the square of the intensity (see Section~\ref{class_forb}). This allows the two-photon absorption to be localized with improved resolution. In general, any process which depends on the $N$th power of the optical intensity can be classically resolved with a resolution of  $x_{\rm min} N^{-1/2}$~\cite{Fabrizio2010}. This is used to achieve superior resolution in multi-photon microscopy, in which multi-photon absorption processes are driven~\cite{Helmchen2005}. Recently, it has been demonstrated that entangled photons pairs, as introduced in Section~\ref{gen_noon_state_subsubsec}, offer substantial advantages to such techniques. This is the topic of Section~\ref{ent_two_phot_mic_sec}.


 In recent years, a wide range of techniques have been developed which utilize the nonlinearities of fluorescent particles to achieve super-resolution; most notably, stimulated emission depletion (STED) microscopy~\cite{Willig2006,Schmidt2008} and photoactivated localization microscopy (PALM)~\cite{Shroff2008}. In STED microscopy, an intense pump field populates the fluorophores into a non-fluorescing state, such that any observed fluorescence must originate from a region where the field is absent. This suppression is highly nonlinear, with fluorescence completely suppressed above an intensity threshold. Consequently, fluorescence can be localized to a sub-wavelength null in the pump field, even though the field itself exhibits no sharp sub-wavelength features. PALM also provides a highly nonlinear measurement scheme, though based on photo-switchable probes rather than saturation of a transition. In this scheme the probes have a non-fluorescing and a fluorescing state, and an optical pump can switch the fluorophores between these states. PALM relies on an extremely weak pump beam, such that each individual fluorophore has a low chance of activation. Once activated, fluorescence can be excited repeatedly thus providing sufficient information to localize the fluorophore far below the diffraction limit. The different fluorophores are sequentially activated and localized, thereby allowing a full super-resolved image of the fluorescence distribution. This relies on photo-activation, for which the relation between incident intensity and output fluorescence is extremely nonlinear. Techniques in the vein of STED and PALM have enabled great progress in biological research by allowing cellular structure to be characterized at the nanometre level. 
 
A number of super-resolution techniques also rely on saturation. When imaging a saturable absorber, the absorption can scale with extreme nonlinearity near the saturation threshold,
allowing resolution far superior to the diffraction limit. This principle has enabled the recent development of saturated structured illumination microscopy, with fluorescent particles pumped into the saturation regime to allow resolution of structures with size
smaller than 50~nm~\cite{Gustafsson2005}. Saturation in non-fluorescent materials can also be used to achieve super-resolution. Recently, a photoacoustic microscopy experiment achieved resolution of less than half the diffraction limit when imaging gold nanoparticles~\cite{Yao2014}.

Nanoscale imaging techniques based on optical nonlinearities have enabled dramatic advances in biophysics, for example, allowing the direct observation of protein clustering in neural synapses~\cite{willig2006sted}. Even so, many biomolecules remain too small to be directly imaged. For instance, nanometre resolution would be required to observe the nucleotides in a DNA molecule. 
%
As we have already seen in the Sections~\ref{gen_noon_state_subsubsec}~and~\ref{Gen_sqz_st_sec} for two specific examples, quantum measurement techniques generally rely on optical nonlinearities to generate quantum correlated photons. It is therefore natural to ask whether such techniques could be integrated with 
nonlinear microscopy to simultaneously provide both
super-resolution imaging and quantum enhanced measurement. This turns out to be possible, in principle, as discussed for some specific examples in Sections~\ref{scan_probe_q_sec},~\ref{quantum_super_flour_mic_sec},~and~\ref{QLithography}.
 
\subsubsection{Challenges specific to quantum imaging systems}\label{QImaging}

 The discussion above describes some of the challenges associated with classical imaging at, and beyond, the diffraction limit.
 Additional complexities arise in quantum imaging that are specifically due to the quantum nature of light.
The images formed in a camera or eye typically involve thousands to millions of pixels, each sampling a different spatial mode of the electromagnetic field. 
%
%
 As with other optical measurements, quantum correlations can allow parameters in an image to be extracted with precision superior to the standard quantum limit.  For instance, a simple parameter might be the optical intensity on one particular pixel. Such a measurement could be enhanced by amplitude squeezing the optical field which is incident on that pixel. Of course, this is but one of many possible parameters that one ay wish to determine. You might be interested in, for example, the orientation of an organelle in a cell, or some other parameter; or you may wish to improve the contrast on every pixel in the array. The analysis in Section~\ref{PhaseTheory} can be directly extended to this sort of problem (see Section~\ref{single_param_sec}). What may be clear already, however, is that,  to track $m$ parameters with quantum enhanced precision, one must controllably populate $m$ orthogonal optical modes with 
quantum correlated photons (see Section~\ref{FieldModes}).



A typical imaging application might sample thousands of spatial modes. 
However, the most powerful squeezed light sources are based on optical cavities which confine all the generated correlations to a single mode. It is possible to utilize multiple single-mode squeezed light sources to produce multi-mode enhancement, though this approach is not scalable to a large numbers of modes. To date, this has only been demonstrated with a maximum of eight spatial modes~\cite{Armstrong2012}, and such sources have not yet been applied in multimode imaging. They have, however, recently been applied to scanning probe-based imaging systems which only required single-mode measurement, including photonic force microscopy~\cite{Taylor2014_image} (see Section~\ref{scan_probe_q_sec}), and atomic force microcopy~\cite{pooser2015ultrasensitive}. Scanned imaging only requires a single sensor, but can only form a reliable image of a sample if both the imaging equipment and sample are static over the scanning time. In many cases, this makes it challenging to use scanning measurements to image sample dynamics. However, some scanning modalities offer extremely high speed, with a recently developed scan-based technology able to capture optical images on a single pixel at a rate far faster than a high-speed multi-pixel camera~\cite{Mahjoubfar2013}. 

  

 
 By contrast to squeezed light sources, spontaneous parametric down conversion naturally populates many thousands of electromagnetic spatial modes with entangled photon pairs  (see Section~\ref{gen_noon_state_subsubsec}).  
 This has enabled
 proof-of-principle demonstrations of sub-shot noise absorption imaging~\cite{Brida2010}, ghost imaging via photon correlations~\cite{Pittman1995}, improved image reconstruction against a noisy background~\cite{Lopaeva2013},  and noiseless image amplification~\cite{Mosset2005,Lopez2008}. Similarly, multimode squeezed light produced via four-wave mixing in an atomic vapour  has been used to generate entangled images~\cite{Boyer2008}, and to estimate the shape of an absorbing mask with enhanced precision~\cite{Clark2012}. However, these approaches are so far limited to the photon counting regime, with low flux limiting their practicality for many applications (see discussion in Sections~\ref{flux_oct_sec}~and~\ref{flux_cons_noon_sec}, for example).  Furthermore, such experiments require more elaborate detection strategies to extract the information from the non-classical state, which currently presents a barrier to their widespread use~\cite{Lassen2007}. Resolution of these technical barriers could enable a wide range of quantum technologies to see practical imaging application, and would signify a crucial advance in quantum imaging.


\subsection{Sensitivity requirements}
\label{sen_rec_sec}

 While the size scale of biological structures sets natural resolution requirements that are broadly relevant throughout biological imaging, there is no corresponding sensitivity limit which can be commonly applied in biological measurements. A number of applications are currently limited by the achievable sensitivity, such as single-molecule sensing~\cite{Vollmer2008,Swaim2011} and biological mass measurements~\cite{Burg2007}. By contrast, the sensitivity requirements of some applications has already been saturated, with accuracy now limited by external factors. For instance, most methods to distinguish healthy and cancerous cells operate with accuracy limited by the cell-to-cell variations rather than instrument sensitivity~\cite{Mahjoubfar2013}.  Likewise, thermal fluctuations introduce stochastic motion to small particles which can place a precision limit on the characterization of many biomechanical forces~\cite{Bustamante2005}. A common approach used to observe the dynamics of small subcellular structures is to attach large marker particles to enhance the interaction with light. Such markers slow the natural biomolecular mechanics; this approach is somewhat analogous to studying the movement of a runner by attaching a ball-and-chain to them, and tracking its motion. Furthermore, a large marker particle is subject to a large stochastic thermal forces, which constrains the achievable force precision. Consequently, these observations can be improved with use of smaller markers or label-free detection~\cite{Neuman2008}. In many cases, however, label-free measurements require precision beyond the limits of existing technology~\cite{Neuman2008}. Consequently, although many existing biomolecular force measurements are currently limited by thermal fluctuations, an improvement in sensitivity is still highly relevant if it could allow smaller markers or label-free detection.

\subsection{Typical frequencies of interest}\label{BioFreq}

 Biological structures can exhibit dynamic behaviour over a wide range of frequencies. While the characteristic frequencies associated with particular structures and functions vary widely, as a general rule, larger structures exhibit slower dynamics (see Fig.~\ref{BioScale}), and active processes such as movement and control occur far faster than other functions such as digestion and growth. At the level of tissues and organs, many active processes occur around the hertz timescale. Examples include the heartbeat of all animals, which range from 0.1~Hz in blue whales to 21~Hz in hummingbirds; which reflects an $M^{-1/4}$ scaling with body mass $M$~\cite{West1997}.  At the cellular scale, many dynamics occur high in the hertz range; for instance, muscle cell contraction occurs at timescales of order 100~Hz, though nerve cells can transmit a neural signal at rates into the kilohertz~\cite{Grewe2010}. Subcellular dynamics are usually even faster, often occurring at kilohertz timescales; the molecular motor kinesin takes steps with a characteristic time of 5~ms (200~Hz), while the motor myosin steps with a characteristic time of 660~$\mu$s (1.5~kHz)~\cite{Bustamante2004}. As studies progress to single molecules, the frequencies can reach megahertz; for instance, light detection in a retinal cell is triggered by a conformational change of rhodopsin which occurs over a few tens of nanoseconds~\cite{Weidlich1997}. Similarly, protein folding can occur over timescales that range between 50~ns and 10~$\mu$s~\cite{Snow2002}. 
 
Compared with the active biophysical processes discussed above, growth and respiration, for instance, occur on much slower timescales. Likewise, biological structures can also have important mechanical relaxation properties which are only observable at hundreds of kilohertz or megahertz frequencies -- much higher than any related biophysical process~\cite{Koenderink2006}.  

The characteristic frequencies of biological processes have ramifications both on the required bandwidth of the measurement device, and on its noise floor. While hertz to kilohertz dynamics can often be followed straightforwardly via optical measurement techniques, many technical noise sources exist in this frequency range, including acoustic noise, vibrational noise, laser frequency noise, etc. These noise sources commonly degrade the noise floor of the measurement to such an extent that the precision of the measurement is far inferior to the quantum noise limit (Eq.~(\ref{QNL}) for the case of phase measurement). On the other hand, the technical noise sources are greatly reduced when studying dynamics in the megahertz frequency range, such that the quantum noise limit is often readily reached. As discussed above, however, megahertz dynamics generally only exist in tiny structures, such as single molecules. It is often challenging to resolve such small dynamics, and to also achieve sufficient bandwidth to follow the dynamics. For quantum enhanced measurements, it is generally crucial that the dominant sources of noise are not technical in nature -- they must arise from the quantum nature of light itself for there to be any real benefit in suppression of quantum noise (see Section~\ref{experimental_imperfection}). This introduces substantial challenges in measurements of low-frequency dynamics, as discussed in Section~\ref{Gen_sqz_st_sec}~and~\ref{SQZ_light_Qmet_sec}.



\subsection{Optical damage in biology}\label{BioDamage}

Light which is used to observe a cell can also damage it, or even cause cell death (for a good recent review, see Ref.~\cite{Cole_live_cell}). Optical destruction of a cell during experiments is sometimes referred to as ``opticution", and is a serious concern in many studies~\cite{Svoboda1994,Ashkin2000}. Even if the cell remains alive, optical damage can permanently disrupt cellular function. Some studies find that cells cannot divide after being studied, even if they are not visually damaged and continue to respire~\cite{Carlton2010}. In addition to damage, incident light can also heavily disrupt cellular function. Altogether, optical damage and biophysical disruptions can complicate efforts to observe the state of a healthy cell, and place stringent constraints on the optical power levels that can be used in biological measurements.

\subsubsection{Optical heating}

There are two major pathways by which light can disrupt cellular function. Firstly, optical heating increases the local temperature, which can cause vast changes to cellular respiration and division~\cite{Peterman2003,Liu1995}. With sufficient heating, this can destroy the cell. Laser heating also induces thermal gradients, which influence the diffusive transport of nutrients and biomolecules in the cell.

The biophysical effect of optical heating has been shown to vary with the ambient temperature. Although some cellular experiments are performed at room temperature, this significantly reduces the respiration of most cell types, and over an extended period of time can cause permanent damage to some mammalian cells~\cite{Sonna2002}. To avoid this, many cellular experiments use incubators to raise the temperature to a biologically active regime. For experiments operating at incubation temperatures, heating resulting from focused laser fields has been found to induce substantial damage within 10 seconds~\cite{Pena2012}. When operating at room temperature, however, much less laser heating damage is observed even after extended periods of study~\cite{Pena2012}. 


 \begin{figure}
 \begin{center}
   \includegraphics[width=10cm]{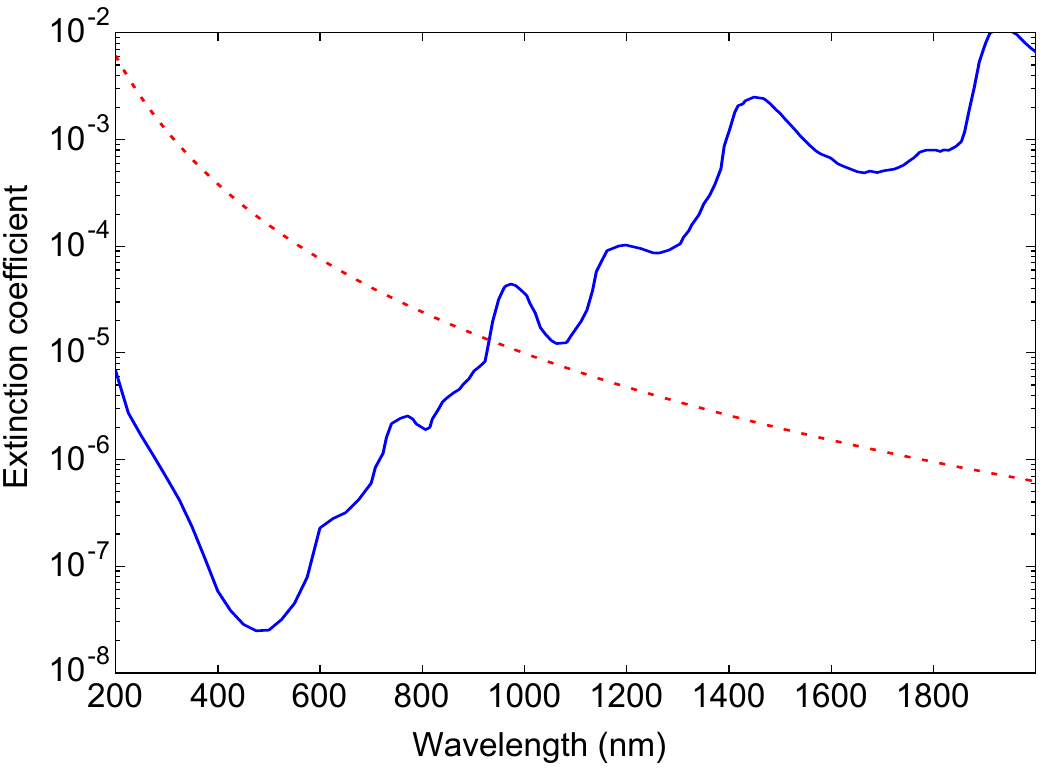}
   \caption{ 
 The sources of optical loss in a typical biophysics experiment. These are dominated by the absorption of water (blue line) and dipole scattering loss (dashed red line). Above 500~nm, the absorption spectra of water tends to increase with wavelength, such that optical heating becomes increasingly severe. However, photochemical damage decreases with wavelength. The most common wavelengths for squeezed light generation are in the near-infrared (see, e.g., Section~\ref{SqueezedState}). This is also a popular window for laser-based biological measurements.
  While the absorption of water is low in visible wavelengths, dipole scattering loss scales as  $\lambda^4$ and can be very large. The total scattering loss depends on the sample, though it is common for optical loss to be minimized in the near-infrared~\cite{Wang2012book}. Water absorption spectral data taken from Refs.~\cite{Hale1973,Curcio1951}.
}
 \label{AbsorptionH20}  
  \end{center}
\end{figure}

 Although a cell itself absorbs some of the light, much of the optical heating is introduced from heating of the water in and around a cell. Consequently, the optical heating rate is approximately proportional to the absorption coefficient of water~\cite{Peterman2003}.   This coefficient is strongly wavelength dependent, with relatively high absorption in the infra-red, and low absorption in the visible spectrum (shown in Fig.~\ref{AbsorptionH20}). As such, optical heating is a serious constraint when operating in infra-red, but can be largely neglected in cellular studies when operating in visible wavelengths. When imaging within tissue, however, absorption of visible light is generally dominated by other contributions; for instance, hemoglobin absorbs strongly at wavelengths below 600~nm, and melanin absorbs strongly for wavelengths up to 1000~nm~\cite{Jacques2013}. For a detailed overview of the optical properties of tissue, we refer readers to Ref.~\cite{Jacques2013}.

\subsubsection{Photochemical effects} 

Although operation at visible wavelengths reduces the potential for photothermal damage, it  increases photochemical intrusion upon the cell. One particularly toxic photochemical effect is introduced when light disassociates molecules and produces reactive oxygen species~\cite{Lubart2006}. It is known that the cell division rate is substantially changed by photochemical effects due to illumination~\cite{Lubart1992}, as is enzyme activity~\cite{DaSilva2010} and many other processes. When too much light is used, these photochemical effects are fatal for the cell~\cite{Neuman1999}. The damaging chemical effects are dramatically increased as the wavelength decreases, since this provides more energy to each photon and allows access to a greater range of photochemical pathways~\cite{Svoboda1994}.

Incident light can also induce damage by disrupting the cellular control mechanisms~\cite{Pena2012}. For instance, it can induce pore formation in a cell membrane, which is referred to as optoporation. These pores are an important part of a cells regulatory system, controlling the flow of water and nutrients through the membrane. When they are overstimulated with an optical probe, however, the membrane becomes permeable. Water then flows into the cell to approach equilibrium water concentrations, which pressurizes the cell and can cause it to rupture~\cite{Ibey2011}.

 To minimize such effects, biophysical experiments can be performed with infra-red optical wavelengths. The collective damage of optical heating and photochemical stimulation is minimized in the near infra-red~\cite{Liu1995,Svoboda1994,Bowman2013}. This wavelength window also corresponds to a broad minima in the absorption of light by biological tissue which is known as the ``therapeutic window"~\cite{Wang2012book}. Since minimizing the optical loss also minimizes the energy imparted to the biological sample, it might seem unsurprising that damage is reduced as the loss is minimized. However, it should be noted that optical damage and optical loss are not fundamentally related at visible wavelengths, with damage dominated by photochemical effects and optical loss dominated by dipole scattering~\cite{Wang2012book}. In biophysical experiments, 780~nm or 1064~nm wavelength lasers  are often chosen for operation within the therapeutic window, with the additional benefit that low noise lasers are readily available at these wavelengths, and 1064~nm lies close to a local minima in the optical absorption. These wavelengths are similar to those used in squeezed light generation, with the best single-mode squeezed light sources to date operating at 860~nm or 1064~nm (see Section~\ref{SqueezedState}). Such sources are therefore already well suited to biophysical studies.  However, we would emphasise that, while minimizing the damage can make the light non-fatal to the cell, it still perturbs a wide range of cell functions which can influence the biological parameter under study.


\subsection{Practical considerations}
\label{prac_cons_bio_sec}

 Biological experiments introduce unique challenges which are foreign to typical quantum metrology experiments. While most quantum metrology experiments explore static systems, cells are constantly respiring and growing. This complicates both data analysis and interpretation. Growth and division rates vary wildly with the strain of cells. 
 To take a  specific example, consider {\it Saccharomyces cerevisiae} yeast cells, which are commonly used in baking and beer-brewing. With sufficient food, these yeast cells will divide after approximately 100 minutes~\cite{Herskowitz1988}. In the absence of food, cells begin to starve, which induces substantial cell changes; for instance, yeast cells grow spores to spread their progeny to new regions~\cite{Herskowitz1988}. The level of nutrients present, therefore, influences the cellular behavior, and is an important parameter in biophysical studies. Furthermore, cellular activity varies wildly in time with respiration occurring in periodic bursts~\cite{Tu2005}. The period of cyclic processes varies substantially; {\it Saccharomyces cerevisiae} yeast exhibits both 40~minute and daily cycles in activity~\cite{Tu2005}. Consequently, the results of a wide range of cellular experiments can be expected to fluctuate with similar period. As such, analysis of a measurement conducted over a short time does not necessarily allow determination of the average cellular properties. Furthermore, substantial cell-to-cell variations must be accounted for if attempting to estimate average cellular parameters. Cell-to-cell variations can be severe for a wide range of measurements. For instance, measurements of cell adhesion to a substrate find the variance between cells to be comparable to the mean adhesion force~\cite{Bowen2001}.

 Quantum metrology experiments are also typically performed in environments which are optimized for the optical apparatus. The fields usually propagate through free-space or optical fiber, with optical losses and spatial distortion minimized throughout. However, any truly practical application of quantum metrology to biology will need to operate in biologically relevant conditions. In many instances, cellular or sub-cellular samples cannot survive being dried out in air, and should be studied while in an aqueous environment. Most focusing optics suffer aberrations when focusing into water, as they are not designed to operate with an elevated refractive index. This degrades precision and resolution, and is an important limitation in many classical studies~\cite{Neuman2008}. Furthermore, cells have a highly inhomogeneous refractive index~\cite{Choi2007,Uchida2011}, which leads to both substantial distortion of transmitted fields~\cite{Taylor2013_sqz} and relatively large scattering loss~\cite{Taylor2013darkfield}. This loss degrades non-classical states, and limits the achievable enhancement in precision (see Sections~\ref{Loss} and \ref{Sqz_eff_eff_sec}). Quantum measurements also require that the spatial mode of the probe field is efficiently sampled~\cite{Treps2002}
%
  (see Sections~\ref{QImaging}~and~\ref{single_param_sec}). The presence of a large unknown distortion can lead to non-optimal sampling of the field, with a consequent reduction in the effective measurement efficiency. This is equivalent to the addition of loss, which further degrades the achievable quantum enhancement.

\section{Progress in biological quantum metrology}\label{Progress}

The previous three sections introduced the context for quantum metrology of biological systems, including the concepts behind, and challenges associated with, both quantum and biophysical measurements. We now present a summary of experimental progress, to date, in this area.

\subsection{Quantum optical coherence tomography}\label{Q_OCT_section}

\begin{figure}
 \begin{center}
   \includegraphics[width=8.25cm]{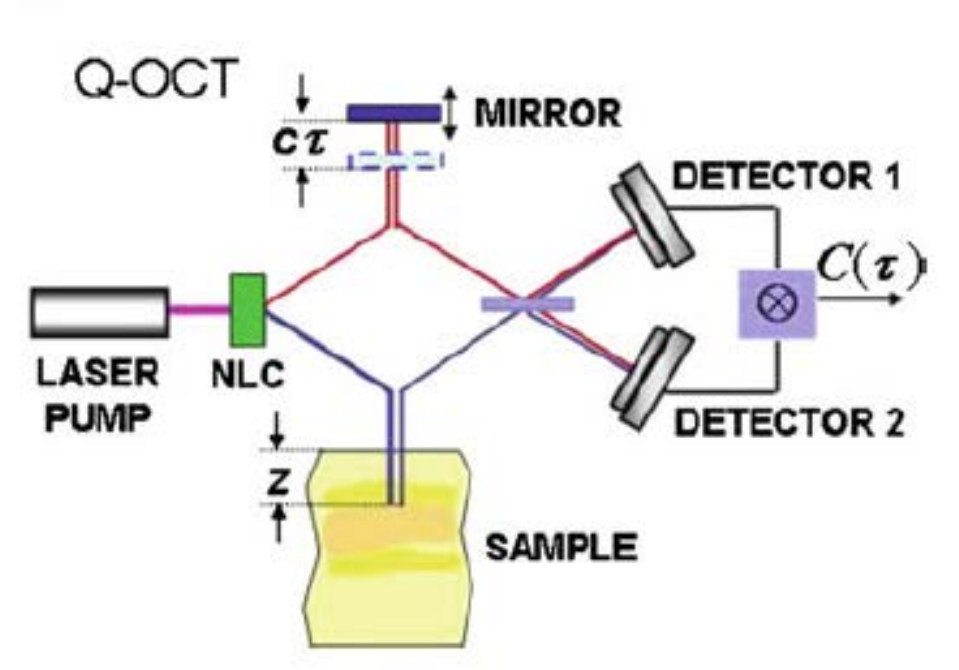}
   \caption{Layout of quantum OCT. The nonlinear crystal (NLC) generates entangled photon pairs via spontaneous parametric down conversion, with one photon passing through a reference arm, while the other illuminates the scattering sample. When the photons are recombined, two-photon interference only occurs for photons which experience a similar optical path length. Measurement of this interference thus allows an image to be constructed using only photons which have scattered from a specific depth in the sample. Conventional OCT applies light with a short  coherence length to achieve a similar depth resolution. Reproduced from Ref.~\cite{Teich2012} with kind permission from Springer Science and Business Media.
}
 \label{Q_OCT}  
  \end{center}
\end{figure}

As discussed in the introduction, entangled photons produced naturally within the specimen under study have been utilised in positron emission tomography (PET) for over half a century. By contrast, the first experiment that engineered a quantum probe field for measurements in biology was reported much more recently in 2009. In that experiment,  quantum correlated photons were used to demonstrate optical coherence tomography (OCT)~\cite{Nasr2009}. 

Classical OCT is a technique that is widely used for medical diagnosis~\cite{Huang1991,Schmitt1999}, where it is employed to generate  high-resolution three dimensional images of such structures as the eye and the retina~\cite{Fercher2003}, and for dermatology~\cite{Welzel2001} and cardiology~\cite{Fujimoto2003}.
 OCT is based on a Michelson interferometer which uses incoherent white light instead of coherent laser light. If the two arms of the interferometer have a length difference that is longer than the coherence length of the light, the output light exhibits no observable interference. In OCT, interference is measured between a reference arm and backscattered light from a specimen. Light scatters from all depths within the specimen; however, the interference is only observed from the light which has propagated a similar distance in both arms. This allows the light to be studied from specific and selectable axial planes within the sample, with axial sectioning given by the coherence length of the light. The coherence length of the light is given by
\begin{equation}
l_c = \frac{4 {\rm ln}2}{\pi} \frac{\bar{\lambda}^2}{\Delta \lambda},
\end{equation}
with $\bar{\lambda}$ the central wavelength and $\Delta \lambda$ the bandwidth~\cite{Serranho2012}. To achieve ultra-short coherence length, and correspondingly narrow resolution, OCT relies on highly broadband light, which is typically centred around 800~nm to allow the deepest penetration (see Section~\ref{BioDamage}) and with a bandwidth that varies from 20 to 300~nm, depending on the application~\cite{Schmitt1999,Serranho2012}. OCT is commonly used to image eyes, for which one must avoid both wavelengths that are heavily absorbed by water, and wavelengths that are visible to the eye. For these applications, the bandwidth is often 20 to 50~nm, resulting in resolution of order 10--30~$\mu$m~\cite{Fercher2003,Serranho2012}. Although wider bandwidth allows finer resolution, wider bandwidths are increasingly sensitive to the degrading effects of dispersion. Interference is only maintained when the sample and reference arms have similar length; but sample dispersion causes the effective path length to differ for different wavelengths, thus causing different spectral components to image planes at different depths. If left uncorrected, this can severely broaden the axial resolution. To address this problem, OCT commonly incorporates dispersion compensation, though this is only possible with precise prior knowledge of the sample dispersion~\cite{Schmitt1999}. Any uncertainty in the sample dispersion, or imperfection in the dispersion compensation, can impose severe limits on the axial resolution.

Quantum OCT adapts the technologies of OCT, but uses entangled photon pairs instead of white light~\cite{Teich2012,Nasr2003,Nasr2004} (see Fig.~\ref{Q_OCT}), which are generated using the method of spontaneous parametric down conversion discussed earlier in Section~\ref{gen_noon_state_subsubsec}. As discussed in Section~\ref{gen_noon_state_subsubsec}, Hong-Ou-Mandel interference occurs when two photons are combined on a balanced beam splitter, producing a two-photon NOON state. However, this interference only occurs when the entangled photons arrive simultaneously, to within the duration of the photon wave packet. Similar to white light interference, this requires the entangled photons to travel a similar distance to the beam splitter. Scanning of the reference arm therefore scans the depth of the measured image, similar to regular OCT (see Fig.~\ref{Q_OCT}). Interference is conventionally considered in field amplitudes, which is a second order process that is described by the operator $\hat{a}^\dagger \hat{a}$ and  modulates intensity (see Section~\ref{quantum_treat_sec1}). By contrast, HOM interference is a fourth order process, described by  $\hat{a}^\dagger\hat{a}^\dagger \hat{a}\hat{a}$, which modulates photon coincidences rather than light intensity (see Sections~\ref{quantum_treat_sec2}~and~\ref{ExampleNOON}). This higher-order process can provide both superior axial resolution and immunity to dispersion~\cite{Nasr2003,Nasr2004}. For similar photon bandwidth, the higher-order HOM measurement improves the axial resolution of low coherence interference by a factor of two~\cite{Teich2012}. Importantly, entangled photons can be produced with bandwidths comparable to low coherence light, with photon wavepacket durations as short as 3~fs demonstrated experimentally. This could allow submicron resolution~\cite{Teich2012,Nasr2008}. Furthermore, the fourth-order statistics are immune to many of the sources of dispersion that can limit conventional OCT.  The technique also makes efficient use of the scattered photons, with each photon contributing the maximum possible information to the image reconstruction.

 This quantum approach to OCT was applied to imaging of onion skin cells in Ref.~\cite{Nasr2009}. That demonstration achieved axial resolution of 7.5~$\mu$m, comparable to many classical OCT systems~\cite{Serranho2012}.  However, shortly afterwards, Ref.~\cite{Lavoie2009} experimentally demonstrated a fourth-order OCT imaging scheme using classically correlated light. This was shown to achieve similar resolution and immunity to dispersion as quantum OCT, and led to a number of quantum-inspired classical interferometry schemes which reproduced the advantages of quantum OCT. On a per-photon basis, the classical techniques fall far short of quantum OCT. As such, the classical techniques demonstrated so far cannot achieve the full performance of quantum OCT. 

To date, quantum OCT experiments have utilised spatially scanned single-photon counting detectors to generate an image over time. In classical OCT, it is common to instead use a CCD or similar high resolution array detector to retrieve an image in real-time. It may be possible, in the near future, to realise similar capabilities with quantum OCT. As discussed in Section~\ref{gen_noon_state_subsubsec}, spontaneous parametric down conversion naturally populates many spatial modes of the electromagnetic field with  entangled photon pairs. Further to this, imaging of the Hong-Ou-Mandel interference used for quantum OCT would only require an array detector that was capable of resolving single photon events with sufficiently high efficiency, spatial resolution and bandwidth.  
  A range of technologies have been developed that offer some of these characteristics to varying degrees (for instance, see Refs.~\cite{edgar2012imaging,fickler2013real,Gatti2008_bad,lemos2014quantum,abouraddy2001demonstration,Armstrong2012}). Quite recently, a sCMOS camera with 23\% quantum efficiency and 7~kHz frame rate has been used to image Hong-Ou-Mandel interference of photon pairs produced via spontaneous parametric down conversion for the first time~\cite{jachura2015shot}.

In addition to the practical applications it may have, quantum OCT also set the foundation for the first significant cross-disciplinary discussions and collaborations between biologists and those working in quantum optics. Scientists working with OCT began to look at the possibilities brought by quantum optics~\cite{Serranho2012}; while the quantum optics community contributed to the design of OCT schemes -- both quantum and classical -- which could have practical benefits in biology~\cite{Lavoie2009,Teich2012}. 
 
\subsubsection{Measurement sensitivity and resolution}
 
Measurement sensitivity $S$ is an important parameter in OCT systems,  limiting the minimum detectable sample reflectivity to $R_{\rm min}=1/S$~\cite{Fercher2003}. The sensitivity of classical OCT is often constrained by detector noise or excess laser noise, though in many cases is limited by quantum shot-noise (see Section~\ref{interferometry}). Within the shot-noise limited regime, it is given by~\cite{Fercher2003} 
\begin{equation}
S=\frac{\eta}{4 \hbar \Omega} \frac{P_{\rm in}}{B}
\end{equation}
with $\eta$ the detection efficiency, $P_{\rm in}$ the input power, $B$ the measurement bandwidth, and $\hbar \Omega$ the energy of a single photon. Since the bandwidth is the inverse of the integration time, this expression simplifies to 
\begin{equation}
S=\frac{\langle n_{\rm sig} \rangle}{4}, 
\end{equation}
where here $\langle n_{\rm sig} \rangle$ is the mean number of input photons within the averaging window. Classical OCT typically uses microwatt to milliwatt optical power, providing sensitivity in the range of $10^7$--$10^{10}$ with a 100~kHz detection bandwidth.

 As with all optical imaging systems, the resolution of OCT can be characterized by the modulation transfer function (MTF), which describes the relative efficiency with which different spatial frequencies of the sample can be detected. Higher spatial frequencies (finer spatial details) are detected with lower efficiency. For instance, a sensitivity of $10^5$ allows features with reflectivity of $10^{-4}$ to be resolved only at spatial frequencies where the MTF is above 0.1 -- and will therefore provide coarser spatial resolution than the nominal resolution limit, which is reached as the MTF approaches 0. Consequently, the photon flux constrains not only the sensitivity and bandwidth of OCT, but can also 
  limit both the sample contrast and spatial resolution~\cite{Fercher2003}.

 \subsubsection{The challenge of flux}
 \label{flux_oct_sec}
 
We saw in the previous section that photon flux can be a critical parameter for OCT. This highlights one of the 
 major challenges that remain outstanding for quantum OCT -- and indeed for most quantum metrology techniques based on entangled photon pairs (see, for example, Section~\ref{NOON_bio}) -- the development of sources and detectors that allow photon flux at similar levels to the flux available with classical techniques. 

The best existing single photon and photon pair sources have flux in the range of megahertz to tens of megahertz. For example, photon pair generation rates exceeding a megahertz have been reported from spontaneous parametric down conversion~\cite{pomarico2012mhz}, and also using the related nonlinear process of four wave mixing~\cite{morris2014photon} within an optical fiber; while single photon fluxes in the range of tens of megahertz have been achieved by exciting quantum dot emitters embedded within microscale optical cavities or optical waveguides (see, e.g., Refs.~\cite{reimer2012bright, gazzano2013entangling,nowak2014deterministic}). At these rates of generation, the illumination power of quantum light sources is in the range of a few picowatts.

For many years, silicon avalanche photodiodes (APDs) represented the state-of-the-art in single photon detectors, providing quantum efficiencies in the range of 65\%, with tens of megahertz single-photon detection rates~\cite{ghioni2007progress}. Recently, however, dramatic progress has been made in alternative single photon detectors based on micro- or nano-scale superconducting devices (see  Ref.~\cite{natarajan2012superconducting} for a good review). In this approach, the device is biased close to the superconducting transition, such that the thermal energy of a single absorbed photon is sufficient to convert it into the non-superconducting state. Superconducting single photon detectors can provide relatively high efficiency, upwards of 90\%, in conjunction with gigahertz single-photon detection rates
(see, for e.g., Ref.~\cite{pernice2012high}). However, since the detection rate depends on the thermalisation time of the devices, larger (and therefore more efficient) devices typically have slower response. For instance, the device in Ref.~\cite{gol2001picosecond} reported a relatively impressive detection rate of 30 GHz,  but with compromised efficiency at the level of 20\%. One major advantage of superconducting photodetectors, is that they allow photon-resolving detection -- that is, they identify not only that a detection event has occurred, but also how many photons arrived at the detector~\cite{natarajan2012superconducting}. While not strictly essential for quantum OCT, this allows separation of incident photon pairs from higher photon numbers. Photon number resolution is an important capability for many other applications of quantum light in biology (see for instance Sections~\ref{ab_im_secasf}, \ref{eye_sec}, and \ref{ent_two_phot_mic_sec}). It is worth noting that, of course, currently there is no known material  that exhibits room temperature superconductivity. Consequently, all superconducting single-photon detectors require cryogenic cooling. This may provide a practical constraint on some biophysical  applications.

Now that we have established the photon flux rates that are available with current entangled photon sources and detectors, let us briefly return to the discussion of the sensitivity and resolution of OCT from the previous section. 
We first observe that a 30~GHz source of entangled photons, combined with a detector capable of resolving this flux, 
would provided quantum OCT with on the order of ten nanowatts of illumination power. This is significantly beneath the micro- to milli-watt powers of classical OCT. Consequently, one should expect that with currently available technology, the sensitivity and/or bandwidth of quantum OCT will be compromised compared to its classical counterpart. The required sensitivity and bandwidth, depend strongly, of course, 
on application. For concreteness we consider, as an illustrative example, an application for which the minimum reflectivity is of order $10^{-4}$, and  spatial frequencies at which the MTF reaches $10^{-2}$ must be resolved. In this case the required sensitivity is $S>10^6$, necessitating the measurement of $\langle n_{\rm sig} \rangle>4\times10^6$ photons. 
Using a 30~GHz photon flux in quantum OCT,  this could be achieved in an averaging time of $\sim$100~$\mu$s, with a corresponding 10~kHz bandwidth. One could also achieve the 100~kHz bandwidth of typical of OCT in cases where reduced sensitivity of $S\sim10^5$ is adequate, such as in a more strongly scattering sample. While it should be clear from this discussion that flux presents a significant issue for quantum OCT based on existing technologies, 
one can see that, combined with the superior per-photon axial resolution and immunity to dispersion discussed above, such an apparatus has the prospect to allow practical applications.

It is worth noting that, in principle, the requirements on the detector bandwidth can be relaxed by using nonlinear detectors such as a photodiode which is excited by two-photon absorption~\cite{Boitier2009,Boitier2013}. Such a detector measures the rate of photon coincidences which, combined with a measurement of the photon intensity, allows a direct measurement of the second-order coherence of the field (see Section~\ref{quantum_treat_sec2}). This measurement need only  estimate the average flux of photon pairs relative to the average photon flux, and therefore does not require  detector bandwidth capable of resolving individual photons. The only limit on the photon flux is that the different photon pairs must arrive separately, such that all detected photon coincidences correspond to twin photons arriving together. The minimum photon separation is thus related to the photon wavepacket duration which, as mentioned above, can be as short as 3~fs~\cite{Nasr2008}. Consequently, this approach could, in principle, allow measurement of HOM interference with photon flux of up to 100~THz,  three orders of magnitude higher than is possible with the fastest linear detectors. A significant challenge is that photon pair detectors demonstrated to date have suffered from 
 low quantum efficiency~\cite{Boitier2009,Boitier2013}. Once this issue is resolved they may see widespread use and, together with  sources of ultra-high flux photon pairs, could drastically expand the capabilities of quantum metrology.


\subsection{Sensing the refractive index of protein solutions with NOON states}\label{NOON_bio}

While the previous demonstration of quantum OCT benefits primarily from the natural immunity to dispersion that is provided by using entanglement photon pairs, the primary advantage of quantum correlated light in most applications is to offer improved measurement precision, as outlined in Section~\ref{semiclass}. This is considered to be an important milestone since biological samples are generally photosensitive, and in some cases increasing the photon flux could damage the specimen (see Section~\ref{BioDamage}).  

As discussed in Section~\ref{NOONState}, an approach that allows enhanced phase precision for a given level of optical power is to utilise a NOON state within a balanced optical interferometer. Entangled photons were applied to biological measurements using exactly this approach in Ref.~\cite{Crespi2012}. In that work, measurements of the refractive index of a solution within a microfluidic device were used to determine the solutions protein concentration. A two-photon NOON state was passed through a Mach-Zehnder interferometer with a microfluidic channel passing through one arm. A standard interferometric phase measurement was then used to infer the refractive index of the fluid within the channel (see Section~\ref{NOONState}). 
%

The experiment of Ref.~\cite{Crespi2012} 
achieved an interferometer visibility beyond the threshold required for supersensitivity, and therefore constitutes an important proof-of-principle demonstration. 
However, the total detection was beneath 50\%. As we will see in following section, this precluded achieving measurement precision superior to the standard quantum limit (Eq.~(\ref{phi_SQL_totalpower})~or~(\ref{phi_SQL_samplepower}) depending on where the power constraint arises within the apparatus).
 As discussed in Section~\ref{gen_noon_state_subsubsec}, the Heisenberg limit of Eq.~(\ref{Heisenberg1}) surpasses the standard quantum limit for constrained total power (Eqs.~(\ref{phi_SQL_totalpower})) by a factor of $\sqrt{2}$; but it coincides exactly with the standard quantum limit for constrained power at the sample (Eq.~\ref{phi_SQL_samplepower}). Larger NOON states would be required to achieve a more substantial precision enhancement. As discussed in Section~\ref{gen_noon_state_subsubsec}, the largest NOON states  produced to date had $N=5$, but co-propagated with a three-orders-of-magnitude larger intensity of non-participating photons. To utilise such sources in practical precision phase measurements, it would be necessary to develop new capabilities in state preparation and in the  filtering of these non-participating photons.

Perhaps more important than the challenges raised in the previous paragraph, the flux of measured photon pairs used in Ref.~\cite{Crespi2012} was 0.1~s$^{-1}$, far below any known damage threshold in biology. Consequently, substantially increased flux would be required for measurements of this kind to compete, in terms of absolute precision, with conventional classical measurement techniques. We discuss these flux considerations in more detail in Section~\ref{flux_cons_noon_sec}.

\subsubsection{Effect of optical inefficiencies}\label{Loss}


As discussed in Section~\ref{QuantumFI}, the Cramer-Rao bound on the precision of phase measurement, or indeed any measurement in general,  is only achievable using a perfect measurement. In practise, such measurements are not possible due to optical loss, limitations in the ability to control the apparatus, and technical noise sources. This is particularly the case for biophysical measurements, as discussed in Section~\ref{prac_cons_bio_sec}. We saw in Section~\ref{inf_bound_limit_sec} that optical inefficiencies lead to a fundamental bound on the precision of phase measurements independent on the quantum state of the probe. Here, we consider the effect of inefficiencies in the specific case where a NOON states is used (see Section~\ref{Sqz_eff_eff_sec} for a similar analysis for squeezed states).


As discussed in Section~\ref{inf_bound_limit_sec}, in a quantum mechanical treatment all linear optical losses, due, for instance, to detector inefficiency, absorption in the specimen, or transmission loss, 
can be modelled as a beam splitter which couples the mode of interest to an unknown environment.\footnote{Nonlinear losses, such as the two-photon absorption discussed in Section~\ref{ClassicalSuperResolution}, cannot be described in this simple manner. However, in almost all optical systems, they are negligible compared to the linear losses.} This discards some of the light, while the unused input to the beam splitter introduces vacuum fluctuations to the detected field (see Fig.~\ref{PhaseSpace}b). Since a coherent state has identical statistics to a vacuum state, except for a coherent displacement (see Fig.~\ref{PhaseSpace}c), the statistical properties of a coherent state are preserved by loss.  This is not true in general: the non-classical statistics of quantum correlated states are degraded by loss. 
The degradation is particularly severe for a NOON state, since loss of a single photon deterministically localizes the remaining light into one arm of the interferometer and completely erases any entanglement~\cite{Dowling2008}.

%

Here, we consider the specific case that is perhaps most relevance to biological measurements, where the potential for damage to the sample determines the maximum photon flux that can be used in the apparatus, and where the optical losses are dominated by the sample -- for instance through absorption or scattering.
A quantitative expression for the sensitivity of phase measurements using NOON states has been derives for this situation in  Refs.~\cite{rubin2007loss, chen2007entanglement}. 
%
For an interferometer with efficiency $\eta$ in the probe arm, and perfect efficiency in the reference arm, detection apparatus, and state preparation, a NOON state of photon number $N$ can achieve an optimum sensitivity of~\cite{rubin2007loss, chen2007entanglement}
\begin{equation}
\Delta \phi_{\rm NOON}^{(1)} = \frac{\sqrt{(\eta^{-N}+1)/2}}{N},
\end{equation}
where here, the superscript $(1)$ denotes that one $N$-photon NOON state has been utilized. 
We see from this expression that the Heisenberg scaling of Eq.~(\ref{Heisenberg1}) is modified by the presence of inefficiency, with the overall degradation in sensitivity depending on $\eta^{-N}$ -- that is, it is exponentially sensitivity to the size of the NOON state.

Of course, it is unlikely that only one NOON state would be used in a practical measurement. Rather, a series of measurements would be made with a sequence of $N$-photon NOON states, and the results averaged to determine the unknown phase. If the total mean number of photons that can be put through the probe arm of the interferometer without sample damage is $\langle n_{\rm sig} \rangle$, then, since the mean number of photons passing through each arm of the interferometers from a single NOON state is $N/2$, it is possible to perform $M=2 \langle n_{\rm sig} \rangle/N$ sequential measurements. Assuming that the measurements are independent, the total sensitivity of the combined measurement is
\begin{eqnarray}
\Delta \phi_{\rm NOON}^{(M)} &=&  \frac{\sqrt{(\eta^{-N}+1)/2}}{\sqrt{M} N}\\
& =& \frac{1}{2 \sqrt{\langle n_{\rm sig} \rangle}} \sqrt{\frac{\eta^{-N} +1 }{N}} \label{sdfsdgsdsf}\\
&=& \frac{\Delta \phi_{\rm SQL}}{\mathcal{E_{\rm NOON}}},
\end{eqnarray}
where we have defined the enhancement factor
\begin{equation}
\mathcal{E}_{\rm NOON} \equiv \sqrt{\frac{N}{\eta^{-N} +1 }},\label{enhance_fact_NOON}
\end{equation}
which quantifies the factor of improvement achieved by the measurement when compared to the probe-power-constrained standard quantum limit 
 $\Delta \phi_{\rm SQL}$ given in Eq.~(\ref{phi_SQL_samplepower}). We see that, while when given the freedom to generate arbitrarily large NOON states, it is possible to achieve phase precision that exhibits Heisenberg scaling with perfect efficiency, for a fixed $N$, NOON states in fact provide the usual $\langle n_{\rm sig} \rangle^{-1/2}$ scaling but with the prospect of a constant enhancement factor in performance, as given by Eq.~(\ref{enhance_fact_NOON}).  
 
 It is interesting to ask what optical efficiency is required for NOON state-based measurements to exceed the standard quantum limit, i.e., when is  $\mathcal{E_{\rm NOON}} >1$?
  It is straightforward to show that the requirement is $\eta > (N-1)^{-1/N}$. From this, we find that a two photon NOON state can only ever equal the standard quantum limit (at $\eta=1$), and cannot exceed it, while a three photon NOON state requires an efficiency of $\eta = 2^{-1/3} \sim 79\%$. Interestingly, the efficiency requirement initially relaxes as the photon number increases, reaching a minimum of $\eta=4^{-1/5} \sim 76\%$ for $N=5$, before increasing monotonically for $N>5$ as the increased fragility of the states overcomes their improved sensitivity. This analysis assumed the power constraint to bin in the sample, for which the two-photon NOON state cannot surpass standard quantum limit even with perfect efficiency (see Section~\ref{gen_noon_state_subsubsec}).  A similar analysis, using Eq.~(\ref{sdfsdgsdsf}) but applying a constraint on total power to derive the standard quantum limit (i.e., using Eq.~(\ref{phi_SQL_totalpower})) shows that, in that case, two-photon NOON states could surpass the standard quantum limit, but only if the total optical efficiency was greater than $3^{-1/2} \sim 58\%$.

\begin{figure}[t!]
 \begin{center}
   \includegraphics[width=12cm]{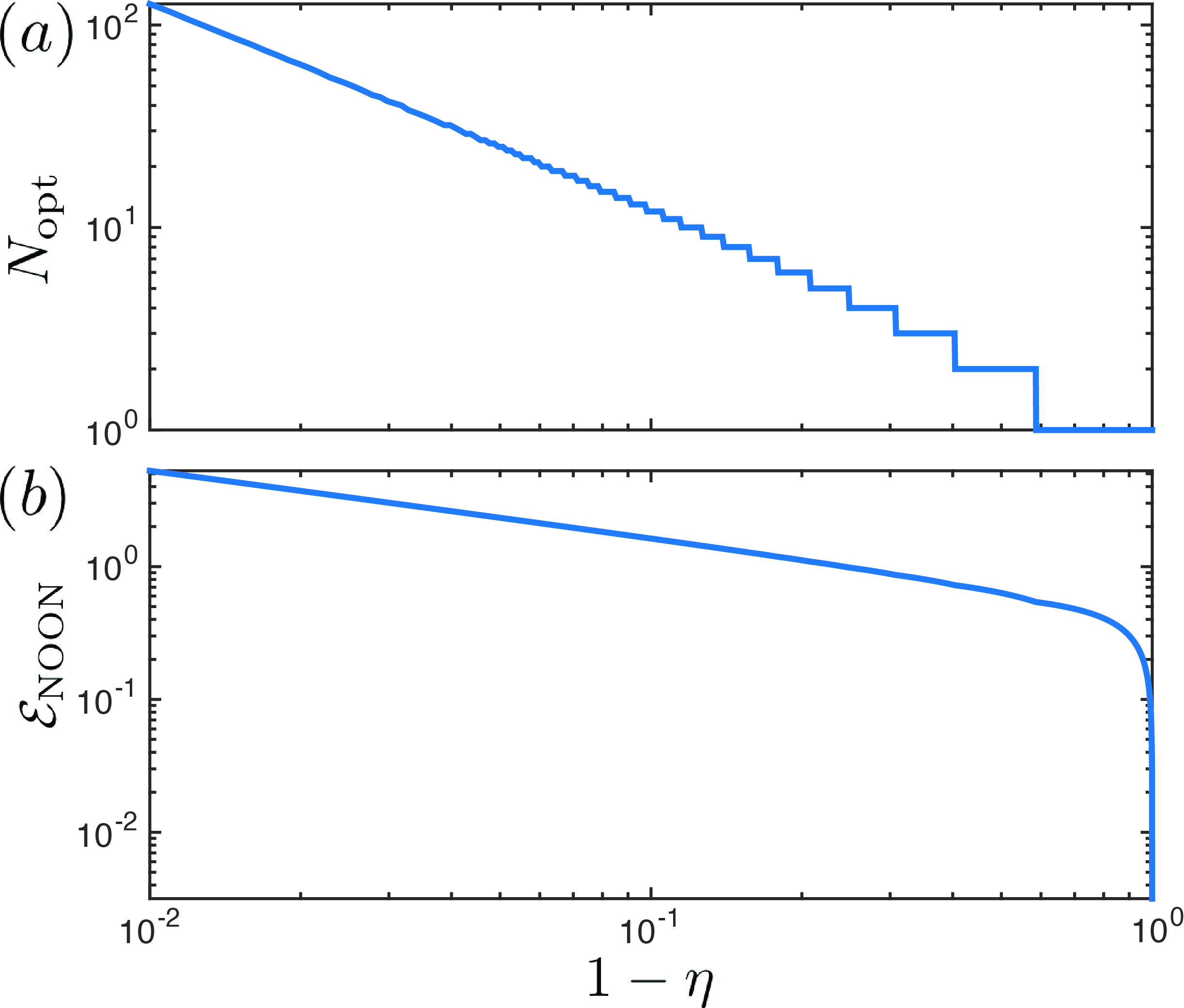}
   \caption{Improvement in phase measurement using NOON states as a function of efficiency in probe arm $1-\eta$. Here, for simplicity, we consider a scenario where the power constraint on the measurement is placed on the power in the probe arm of the interferometer, and apart from the probe arm, all other parts of the interferometer have perfect efficiency.  (a) Optimal number of photons in each NOON state, $N_{\rm opt}$. (b) Enhancement factor, beyond the standard quantum limit given in Eq.~(\ref{phi_SQL_samplepower}). Dashed line: $\mathcal{E}_{\rm NOON}=1$.
}
 \label{NOON_int_N_and_en}  
  \end{center}
\end{figure}
It is interesting to observe that, because of the increasing fragility of NOON states as their size $N$ increases, there is an optimal -- non-infinite -- choice of $N$ to achieve the largest possible enhancement for a given efficiency $\eta$.
It is relatively straightforward to show from Eq.~(\ref{enhance_fact_NOON}) that this optimal $N$ is parametrised by the equation $N_{\rm opt} \ln \eta + \eta^{N_{\rm opt}} + 1 = 0$, modulo the fact that it must be a whole number. 
We plot both this optimal $N_{\rm opt}$, and the optimal enhancement factor achieved at that value of $N$ as a function of inefficiency in Fig.~\ref{NOON_int_N_and_en}.

\begin{figure}[t!]
 \begin{center}
   \includegraphics[width=19cm]{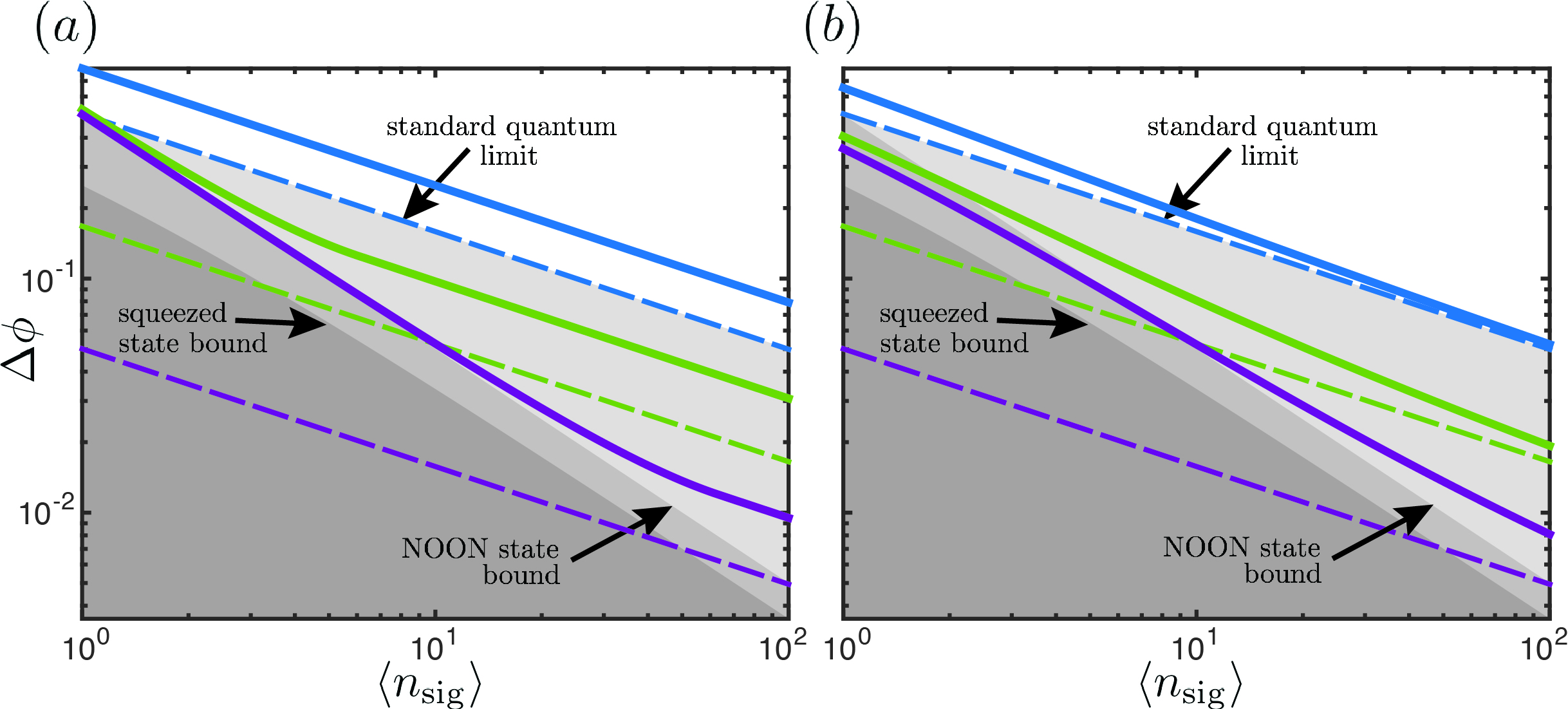}
   \caption{Optimal precision of phase measurements in the presence of losses within the sample, and taking the photon flux in the sample $\langle n_{\rm sig} \rangle$ to be constrained (rather than the total power used for the measurement). (a) Phase sensing via NOON state interferometry. (b) Phase sensing using a bright squeezed state and linear (homodyne) detection. The solid blue, green, and purple curves show the optimal precision that can be achieved with efficiencies of $\eta = \{0.5, 0.9, 0.99 \}$, respectively. The dashed lines show the respective state-independent fundamental bounds introduced by the inefficiency (see Section~\ref{inf_bound_limit_sec}). The light grey, medium grey, and dark grey region indicate phase precision exceeding the standard quantum limit  (Eq.~(\ref{phi_SQL_samplepower})), Heisenberg limit (Eq.~(\ref{Heisenberg1}), with  $\langle n_{\rm sig} \rangle = \langle n_0 \rangle/2$), and squeezed state Cramer-Rao Bound (Eq.~(\ref{CRB_sqz})), respectively.   
}
 \label{Phase_sensing_limitsB}  
  \end{center}
\end{figure}
When the mean number of photons $\langle n_{\rm sig} \rangle$ that may be passed through the sample without damage is less than $N_{\rm opt}/2$, the best measurement strategy is to make a single measurement with a NOON state of size $N = 2 \langle n_{\rm sig} \rangle$. In this limit, the phase precision of NOON state interferometers 
 closely follows the Heisenberg limit of Eq.~(\ref{Heisenberg1}), as shown at sufficiently low $\langle n_{\rm sig} \rangle$ in Fig.~\ref{Phase_sensing_limitsB}a. However, in the more common scenario where the damage threshold  $\langle n_{\rm sig} \rangle$ is larger than $N_{\rm opt}/2$, the optimal strategy is, rather, to make a sequence of measurements with NOON states of size $N_{\rm opt}$. In this regime  (with the transition to it shown by the kink in the curves in Fig.~\ref{Phase_sensing_limitsB}a), the scaling returns to the usual  $\langle n_{\rm sig} \rangle^{-1/2}$ scaling of classical measurements. Fig.~\ref{Phase_sensing_limitsB}a also compares the phase precision that is possible using NOON states at a given efficiency to the fundamental state-independent precision bound due to inefficiencies introduced in Section~\ref{inf_bound_limit_sec} (shown by the dashed lines). It can be seen that NOON state measurements are constrained above this limit at all power levels.

\subsubsection{Photon flux considerations}
\label{flux_cons_noon_sec}

We saw in the previous section that both high optical efficiency and the ability to generate large NOON states are necessary if one wishes to achieve a 
substantial improvement in measurement precision via NOON state interferometry. Furthermore, in practise, the absolute phase precision must compete with classical methods. This generally introduces a stringent requirement on the flux of the source of NOON states, and on the detection apparatus.

For simplicity, here, let us consider the case where the optical efficiency of the apparatus is perfect, with no photons lost in transmission, within the sample, or in detection. If a sequence of $M$ independent measurements are then made within one second, each with an $N$-photon NOON state, the Heisenberg limit of Eq.~(\ref{Heisenberg1}) becomes  
%
\begin{equation}
\Delta \phi_{\rm Heisenberg}^{(M)} = \frac{1}{N \sqrt{M}}. \label{precision_noon_repeat}
\end{equation}
We wish to compare this ideal NOON-state precision with the precision that could be achieved with a realistic classical measurement. The maximum precision of a classical measurement is tied to the threshold for optical damage of, or photochemical intrusion to, the specimen, which
%
%
varies by many orders of magnitude between biological specimens, and is also highly wavelength-dependent (see Section~\ref{BioDamage}). Damage has been reported at microwatt powers, though many experiments safely use power that is above a milliwatt.  Let us consider a classical experiment that passes
$\langle n_{\rm sig} \rangle \gtrsim 1~{\rm \mu W}/\hbar \Omega \sim 10^{12}$ photons per second through the specimen, where $\Omega$ is the laser frequency.  
From Eq.~(\ref{phi_SQL_samplepower}), the sample power constrained standard quantum limit of phase precision is then $\Delta \phi_{SQL} \lesssim 10^{-6}$~rad~Hz$^{-1/2}$.
%
To compete with this precision, we find from Eq.~(\ref{precision_noon_repeat}) that NOON interferometry would require a sample rate 
\begin{equation}
M \gtrsim \frac{10^{12}}{N^2}. \label{raate_eq_as}
\end{equation}
As discussed in Section~\ref{flux_oct_sec}, the best entangled photon sources currently have photon production rates in the range of megahertz, while 
the best existing single photon counters and photon resolving detectors have bandwidths in the range of gigahertz. Even using the largest yet reported NOON state of $N=5$~\cite{Afek2010} in Eq.~(\ref{raate_eq_as}) yields a required trial rate of order $\sim$100~GHz. This sets a challenging task for technology development before NOON state interferometers become broadly relevant for precision phase measurement in biological applications. It is important to note, however, that existing capabilities may already find relevance if addressing exceptionally fragile or light-sensitive samples. Such fragility is seen, for instance, in UV imaging~\cite{zeskind2007nucleic}, where the short wavelength allows an extremely small optical focus, though at cost of highly toxic photochemical interactions (see Section~\ref{BioDamage}).  Furthermore, as we discuss for some specific examples in Sections~\ref{Q_OCT_section},~\ref{ab_im_secasf},~and~\ref{ent_two_phot_mic_sec}, 
 entangled photons can provide other significant advantages unrelated to enhanced measurement precision.

\subsection{Squeezed light enhanced particle tracking}
\label{SQZ_light_Qmet_sec}

\begin{figure}
 \begin{center}
   \includegraphics[width=8.25cm]{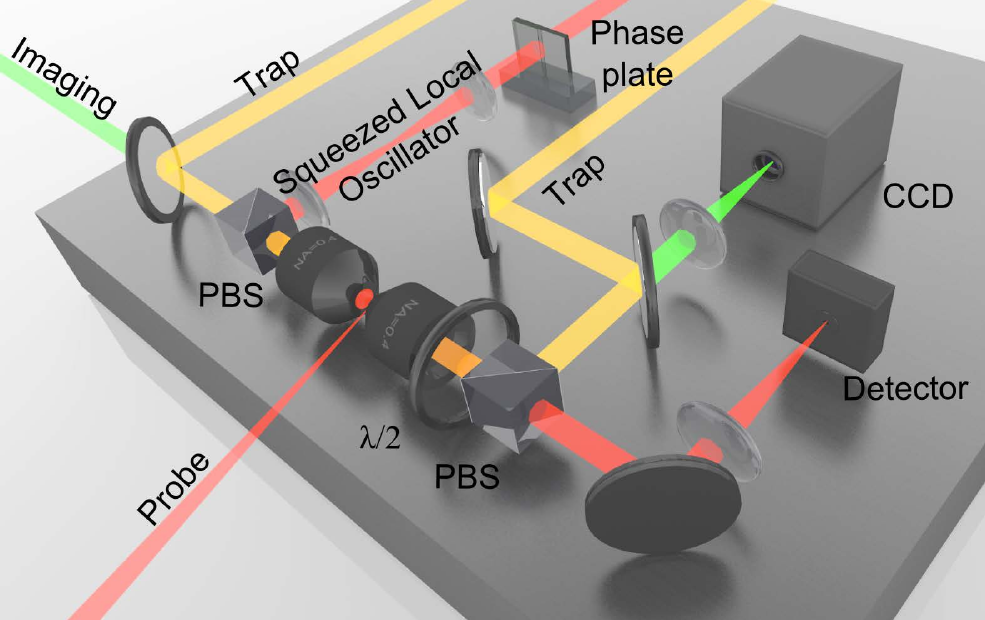}
   \caption{The apparatus which applied squeezed light to enhance particle tracking microscopy. The probe field acts as a dark-field illumination to generate scattered light. The scattered field is then combined with the squeezed local oscillator and detected to allow 1D tracking of the weakly scattering particle. Reproduced with permission from Ref.~\cite{Taylor2014_image}.
}
 \label{Sqz_Layout}  
  \end{center}
\end{figure}

In Section~\ref{SqueezedState} we introduced an alternative to NOON state interferometry, where squeezed states of light are used to enhance the measurement precision beyond what is possible without quantum correlations. As observed in that section, one advantage of this approach is that squeezed states can be combined straightforwardly with a coherent laser field, with the laser providing high photon flux while the squeezed state enhances precision. Unlike measurements with NOON states, the measurement can compete in absolute terms with practical classical measurements without requiring a high flux of quantum correlated photons associated with the squeezing. A further advantage is that measurements with squeezed states generally do not require single-photon or photon resolving detectors, but rather use high bandwidth photodiodes similar to those used in most classical measurements.  References~\cite{Taylor2013_sqz,Taylor2014_image} demonstrated 
 quantum enhanced precision in optical tweezers based biophysics experiments, taking advantage of these attributes of experiments with squeezed light.
 
 Particle tracking in optical tweezers has revolutionized the field of biophysics by allowing characterization of single-molecule dynamics. This has enabled a vast array of discoveries, including both the dynamics and magnitude of the forces applied by biological motors~\cite{Finer1994,Svoboda1993},  the stretching and folding properties of DNA and RNA~\cite{Bustamante2005,Greenleaf2006}, the dynamics of virus-host coupling~\cite{Kukura2009}, and the mechanical properties of cellular cytoplasm~\cite{Yamada2000,Senning2010,Norrelykke2004}.

 The particle position in optical tweezers is estimated from a laser beam deflection measurement. Such deflection measurements can be enhanced via the use of squeezed light~\cite{Treps2002,Treps2003} (see Section~\ref{SqueezedState}), and Ref.~\cite{Tay2009} proposed use of a similar technique to enhance the precision of  optical particle tracking. However, the relevant frequency range of optical tweezers based biophysics experiments is typically in the hertz to kilohertz range (see Section~\ref{BioFreq}), while most squeezed light sources provide an enhancement in the megahertz frequency regime (see Section~\ref{SqueezedState}). To address this, an optical lock-in detection scheme was developed to shift the low frequency particle tracking information into the megahertz squeezing band without loss of precision~\cite{Taylor2013_lockin}.
 
 Following these developments, Ref.~\cite{Taylor2013_sqz} integrated squeezed light into optical tweezers to track nanoscale particles below the quantum shot noise limit, using the apparatus shown in Fig.~\ref{Sqz_Layout}. This experiment relied on an interferometric measurement of a weak scattered field~\cite{Taylor2013darkfield}, and can thus be treated with a similar framework to interferometric phase measurement. We introduce this treatment in more detail in Section~\ref{single_param_sec}. Quantum enhanced precision was demonstrated by producing the local oscillator field in a bright squeezed state (see Section~\ref{large_alp_lim_sec}), which allowed the uncertainty in the measurement to be reduced by a factor of 0.54, or -2.7~dB when compared to the quantum noise limit (see Section~\ref{QNL_sec}). It is important to note that this approach allowed the quantum noise limit to be surpassed, and therefore achieved a precision that could not be achieved using uncorrelated photons in the same apparatus; but it did not surpass the standard quantum limit. As a result, a different apparatus could -- in principle -- achieve improved precision without recourse to quantum correlations. The primary reason this experiment could not surpass the standard quantum limit was the inefficient collection of scattered photons from the trapped particle. For nanoscale particles, these photons are scattered into a $4\pi$-steradian solid angle. No optical system has yet been demonstrated for any optical tweezer apparatus that is capable of collecting this full angular scattering range in such a way that precision measurements can be made. As such, while not as fundamental as the standard quantum limit, the quantum noise limit provides an appropriate benchmark for comparison to existing optical tweezers.

 Since the approach demonstrated in Ref.~\cite{Taylor2013_sqz} enhances the precision of optical tweezers based particle tracking, it could in principle be applied in any of the applications of optical tweezers. It is important to note, however, that thermal noise in the motion of the trapped particle limits the sensitivity of many optical tweezers applications, including biomolecular force sensing. For applications limited by thermal noise, quantum enhanced precision does not typically provide a practical benefit; although it is often desirable to redesign such experiments using smaller markers, for which shot noise is generally much more relevant (see Section~\ref{sen_rec_sec}). As such, this technology is primarily useful for shot-noise limited measurements such as studies of high-frequency processes, or in such experiments as microrheology where the thermal motion itself provides the signal~\cite{Taylor2013tutorial}.


\subsubsection{Quantum enhanced microrheology}

\begin{figure}
 \begin{center}
   \includegraphics[width=10cm]{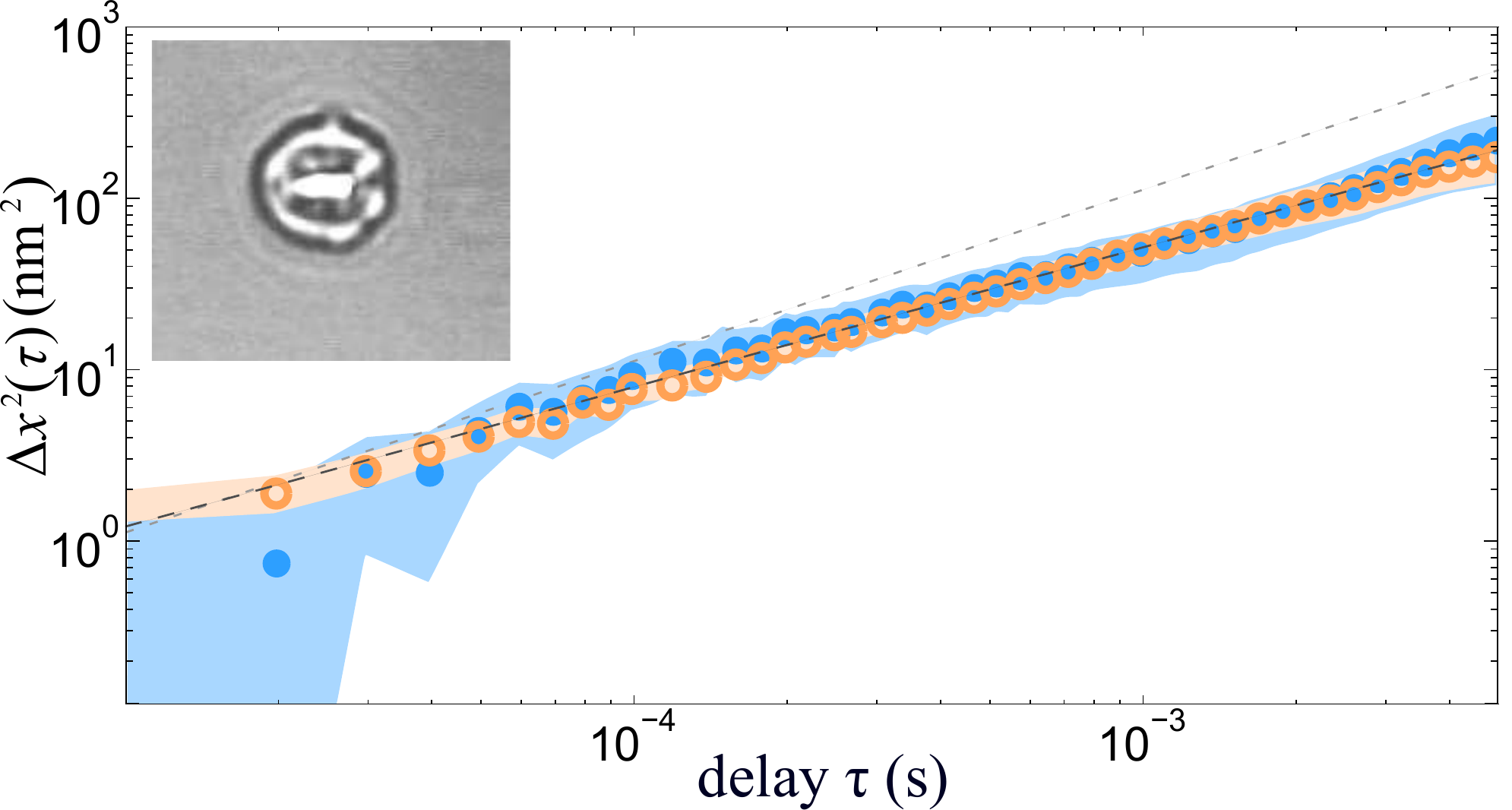}
   \caption{ The mean squared displacement ($\Delta x^2 (\tau)$) of thermally driven particles within yeast cells was measured using both squeezed (orange) and coherent (blue) light. The data points represent the mean of the measured displacements, while the shaded regions represent the standard error. In this case, the use of squeezed light improved the precision by 2.4~dB, thus reducing the measurement uncertainty in the  characterization of the thermal motion.  Reproduced with permission from Ref.~\cite{Taylor2013_sqz}, Macmillan Publishers Ltd: Nature Photonics, Copyright (2013).
}
 \label{MSDyeast}  
  \end{center}
\end{figure}

 The quantum enhanced nanoparticle apparatus discussed above
  was used to track the thermal motion of naturally occurring lipid granules within {\it Saccharomyces cerevisiae} yeast cells, with squeezed light yielding a factor of 0.54 (or 2.4~dB) improvement in precision~\cite{Taylor2013_sqz}.
 The cellular cytoplasm is crowded with large molecules, polymer networks, and various other organelles which confine the particle and influence the thermal motion~\cite{Weiss2004crowding}. The mechanical properties of the cellular cytoplasm could be determined by characterizing this thermal motion, with  quantum enhanced sensitivity allowing a 22\% improvement in precision over that achievable with coherent light (Fig.~\ref{MSDyeast}). This level of enhancement allowed dynamic changes in the intracellular mechanical properties to be measured with 64\% improved temporal resolution. Alternatively, it would allow the optical power to be reduced by 42\% without compromising sensitivity. Importantly, since this apparatus used a bright squeezed state, it was able to achieve absolute precision which was comparable to many similar classical microrheology experiments~\cite{Norrelykke2004,Selhuber2009}, though state-of-the-art microrheology experiments achieve over an order of magnitude greater displacement sensitivity~\cite{Grebenkov2013}.

\subsubsection{Quantum enhanced scanning probe microscope}
\label{scan_probe_q_sec}

The quantum optical tweezers of Ref.~\cite{Taylor2013_sqz} was later applied to spatially resolve cytoplasmic structure in Ref.~\cite{Taylor2014_image}. Spatial imaging was achieved using a technique called photonic force microscopy~\cite{Florin1997,Friese1999}, in which spatial variations in the local environment are sampled by a small particle as it slowly moves through the cell. A profile of the structure is then constructed from the influence of the structure on the thermal motion of the particle. This approach to imaging is closely analogous to that taken in atomic force microscopy, though here a trapped bead acts as the probe tip. In both cases, the use of a scanning probe means that the spatial resolution is limited by the signal-to-noise ratio rather than the diffraction limit~\cite{Friese1999,Rohrbach2004}. Thus, the use of squeezed light to improve sensitivity can also enhance resolution.

\begin{figure}
 \begin{center}
   \includegraphics[width=9cm]{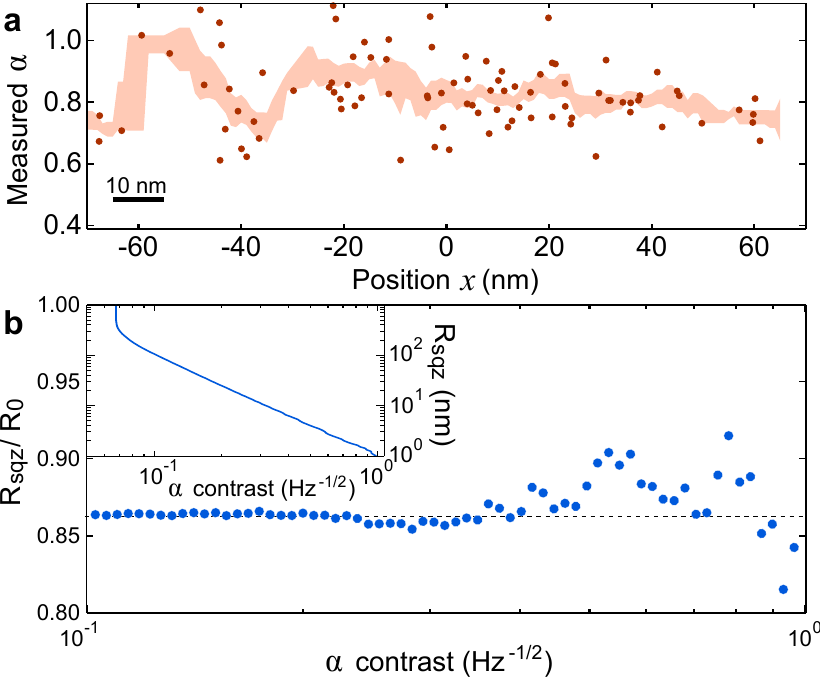}
   \caption{Quantum imaging results from Ref.~\cite{Taylor2014_image}. {\bf a} An example one-dimensional profile of the viscoelastic structure within a yeast cell, with a peak around -55~nm indicating reduced molecular crowding, and a dip around -40~nm indicating a confining boundary. Each dot represents a separate measurement, while the shaded region indicates the running mean and standard error over a 10~nm window. {\bf b} The enhancement in spatial resolution achievable was characterized for a fixed contrast in the measured parameter $\alpha$. As the spatial averaging increases, the contrast in the measured parameter $\alpha$ improves, as shown for squeezed light in the inset. Here the squeezed light was shown to improve the imaging resolution by 14\%. Reproduced with permission from Ref.~\cite{Taylor2014_image}.
}
 \label{QImagingPic}  
  \end{center}
\end{figure}

Photonic force microscopy was used in Ref.~\cite{Taylor2014_image} to construct profiles of spatial structures within a living yeast cell (Fig.~\ref{QImagingPic}). In this experiment, particles were only tracked along the $x$ axis, while the particles diffuse along all three. As such, the constructed profiles follow the projection of an unknown trajectory onto the $x$ axis. For any practical application of this technique, three-dimensional imaging is required. However, this experiment demonstrated both sub-diffraction-limited quantum metrology and quantum enhanced spatial resolution for the first time in a biological context. Spatial structure within a living yeast cell was observed at length scales down to 10~nm, far below the diffraction limit (see Section~\ref{ResolutionBio}) and comparable to that achieved in similar classical photonic force microscopy experiments~\cite{Scholz2005,Bertseva2009}. Squeezed light was found to enhance the spatial resolution by 14\% over that achievable with coherent light. This was the first demonstration that squeezed light could be used to improve resolution, though such a resolution enhancement has been proposed for far-field optical imaging~\cite{Sokolov2004,Beskrovnyy2005}. The absolute enhancement in resolution was relatively small, due to the relatively modest degree of available squeezing. If combined with state-of-the-art squeezed light with over 10~dB of measured squeezing~\cite{Mehmet2010,Stefszky2012,Eberle2010}, this technique is predicted to allow up to an order of magnitude improvement in resolution over similar classical imaging techniques~\cite{Taylor2014_image}.


The experiments reported in Refs.~\cite{Taylor2013_sqz,Taylor2014_image} were the first experiments to apply squeezed light in biological applications. They demonstrated an enhancement in particle tracking precision over that possible with coherent light using a device which was comparable to those used in biophysics experiments, though constrained by the use of low numerical aperture objectives~\cite{Taylor2013QNL}. With some improvements to the design, it is feasible that a quantum optical tweezers apparatus based on this approach could outperform classical technology. This could benefit a range of important biophysical applications of optical tweezers.

\subsubsection{Single-parameter quantum imaging}
\label{single_param_sec}

It is not obvious, at first glance, how one would determine the quantum limits to measurements such as the nanoparticle tracking measurements discussed above. Even in the simple case of a linear interaction, the optical field interacts in a complex way with the particle. The action of the particle is to deform the optical mode, that is, to transfer optical power from the input mode into other modes via its scattering. So, in what mode is the information contained?, and how would one optimally access it?

\begin{figure}
 \begin{center}
   \includegraphics[width=12cm]{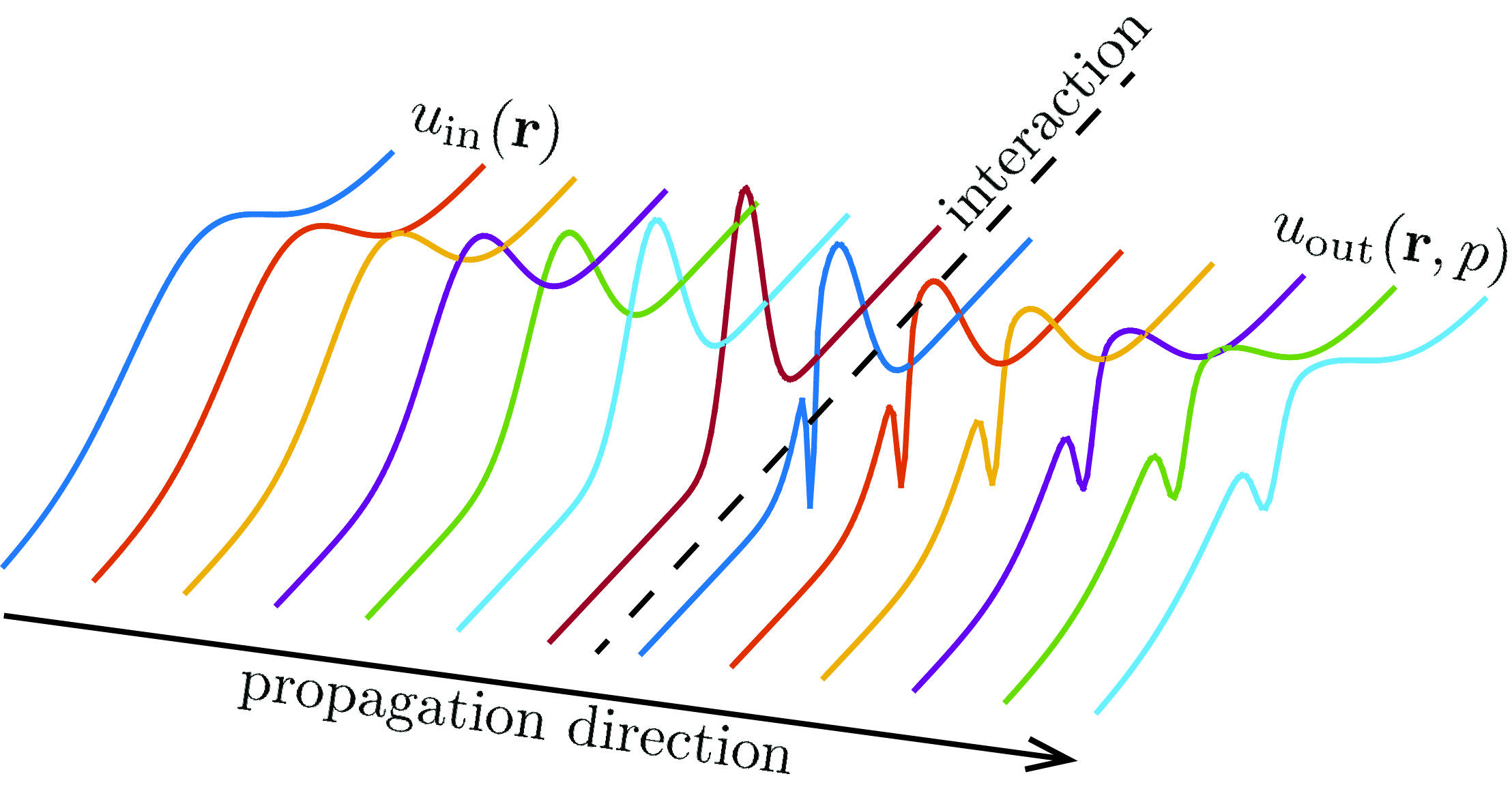}
   \caption{General quantum imaging problem. An initial occupied optical mode with shape $u_{\rm in}({\bf r})$ is transmitted towards a sample. The interaction with the sample modifies the modeshape to $u_{\rm out}({\bf r},p)$, with the modification described by some general parameter $p$ (this could, for example, be the position of a particle in the mode). the aim is then to determine the magnitude of the parameter $p$ with the maximum signal-to-noise.
}
 \label{Gauss_perturb}  
  \end{center}
\end{figure}

It turns out that nanoparticle tracking is a nice example of a single-parameter quantum imaging problem.
The essential idea of such schemes -- at least in the limit where this field has a bright coherent occupation -- is shown in Fig.~\ref{Gauss_perturb}. We saw in Section~\ref{FieldModes} that in a quantum description, the optical electric field is decomposed into a complete set of orthogonal spatial modes, with the contribution from a given spatial mode $m$ given by (see Eq.~(\ref{FieldModesEqn})):
\begin{equation}
\hat E_m(t) = i \sqrt{\frac{\hbar \Omega_m}{2 \epsilon_0 V_m}} \left [ \hat a_m u_m({\bf r}) e^{- \Omega_m t} - \hat a^\dagger_m (t) u_m^*({\bf r}) e^{i \Omega_m t} \right ].
\end{equation}
The exact choice of mode basis is arbitrary, so long as it is orthonormal and complete. The basis where $\langle \hat E_m(t) \rangle = 0$ for all but one mode is a natural choice. In an experiment involving a bright Gaussian laser field, for example, this corresponds to choosing a basis where the Gaussian mode of the laser is one of the modes.

In Fig.~\ref{Gauss_perturb}, we show a scenario where an initially Gaussian occupied field mode $u_{\rm in}({\bf r})$ is perturbed by an interaction with a sample, modifying the modeshape of the occupied mode to $u_{\rm out}({\bf r},p)$, where $p$ is a general parameter that describes the modification -- for instance, in the case of optical particle tracking considered here, $p$ corresponds to the position of the particle. One way to understand this interaction is that the sample acts to scatter light from one optical mode into others. The natural question to ask is then: how well can the perturbation $p$ be determined, both in principle and in practise? This question can be answered in the relevant limit of a small perturbation by Taylor expanding the output mode shape as a function of the perturbation $p$:
\begin{equation}
u_{\rm out} ({\bf r},p) = u_{\rm out} ({\bf r},0) + \frac{\partial u_{\rm out} ({\bf r},p)}{\partial p} \bigg |_{p=0} p + \frac{1}{2} \,\frac{\partial^2 u_{\rm out} ({\bf r},p)}{\partial p^2} \bigg |_{p=0} p^2 + ...
\end{equation}
In the limit of a small perturbation, higher-order terms in $p$ may be neglected, and this can be well approximated as 
\begin{equation}
u_{\rm out} ({\bf r},p) = u_{\rm out} ({\bf r},0) + \frac{p}{\mathcal{N}} u_{\rm sig}({\bf r}), \label{sdfgsdf}
\end{equation}
where 
\begin{equation}
u_{\rm sig}({\bf r}) \equiv \mathcal{N} \, \frac{\partial u_{\rm out} ({\bf r},p)}{\partial p} \bigg |_{p=0},
\end{equation}
and $\mathcal{N}$ is a  coefficient which is chosen so that $u_{\rm sig}({\bf r})$ is properly normalised (i.e., $\langle |u_{\rm sig}({\bf r}) |^2 \rangle=1$). It can be shown using the normalisation property of $u_{\rm out} ({\bf r},p)$ ($\langle |u_{\rm out} ({\bf r},p)|^2 \rangle = 1$) that, in the small perturbation limit we consider here, $\langle  u^*_{\rm out} ({\bf r},0) u_{\rm sig}({\bf r}) \rangle=0$. We see from this that to first order, the action of the perturbation $p$ is to scatter light from the unperturbed mode $u_{\rm out} ({\bf r},0)$ into an orthogonal signal mode $u_{\rm sig}({\bf r})$, with the amplitude of the scattered field linearly proportion to the level of perturbation.

In principle, one must worry about the effect of the perturbation not only on the coherently occupied mode, but also on all other modes of the electromagnetic field. However, in the  generally relevant  limit that the perturbation is very small and the optical field is very bright (i.e., its coherent amplitude $|\alpha|^2 \gg 1$), the scattering process discussed in the previous paragraph is the dominant effect, with all other effects of the perturbation being negligible.  The effect of the perturbation $p$ is then well approximated as a simple transfer population from the coherently populated mode into the signal mode, such that
\begin{equation}
\hat a_{\rm sig} \rightarrow  \alpha_{\rm sig} + \hat a_{\rm sig} \label{coh_pop_sig_q_im}
\end{equation}
where the coherent amplitude $\alpha_{\rm sig}$ is determined from the coherent amplitude of the incident field $\alpha$ and the overlap of the signal and output modes as:
\begin{equation}
\alpha_{\rm sig} = \alpha \left \langle u_{\rm out}^*({\bf r}, p) \, u_{\rm sig} (\bf r) \right \rangle \approx \frac{\alpha p}{\mathcal{N}}. \label{alpha_sig_asdgsd}
\end{equation}
Here, to arrive at the final approximate expression we have used Eq.~(\ref{sdfgsdf}) and the orthonormality of the signal and unperturbed output modes.

To justify the above approach, where all perturbations other than the scattering of coherent population are negligible, let us consider the particular example of nanoparticle tracking that is the main focus of this section. In this case, the ratio of scattered photons to incident photons is $n_{\rm scat}/n_{\rm in}=\sigma_{\rm scat}/4 \pi w^2$, where $\sigma_{\rm scat} $ is the scattering cross-section of the particle and $w$ is the waist of the optical field. For dipole scattering from a dielectric particle $\sigma_{\rm scat} = (8  \pi/3) k^4 a^6 (m^2-1)^2/(m^2 +2)^2$, where $a$ is the radius of the particle, $k = 2 \pi/\lambda$ is the wavenumber of the field with $\lambda$ being the optical wavelength in the medium, and $m$ is the ratio of the refractive index of the particle to that of the medium surrounding it.
 If the particle to be tracked is  silica and has radius $a=150$~nm, the medium is water, and the optical wavelength is $\lambda=750$~nm, then
 $\sigma_{\rm scat}  \approx 3\times10^{-15}$~m$^2$. For a laser beam focussed to a waist of $w=1~\mu$m, we therefore find that $n_{\rm scat}/n_{\rm in} \approx 3 \times~10^{-4}$. So, less than one photon per thousand is scattered from the coherently populated mode into all other spatial modes, even when the particle is centred in the trap ($p=0$). Since the coherent fields used in nanoparticle tracking experiments typically have an intensity in the range of tens to hundreds of milliwatts (around $10^{17}$ photons per second), even these low levels of scattering can introduce significant coherent occupation of spatial modes other than the input mode $u_{\rm in} ({\bf r})$. However, other processes, where for instance vacuum fluctuations from an unoccupied field are scattered into a different unoccupied field, or into the input mode, are essentially negligible.

\begin{figure}
 \begin{center}
   \includegraphics[width=8cm]{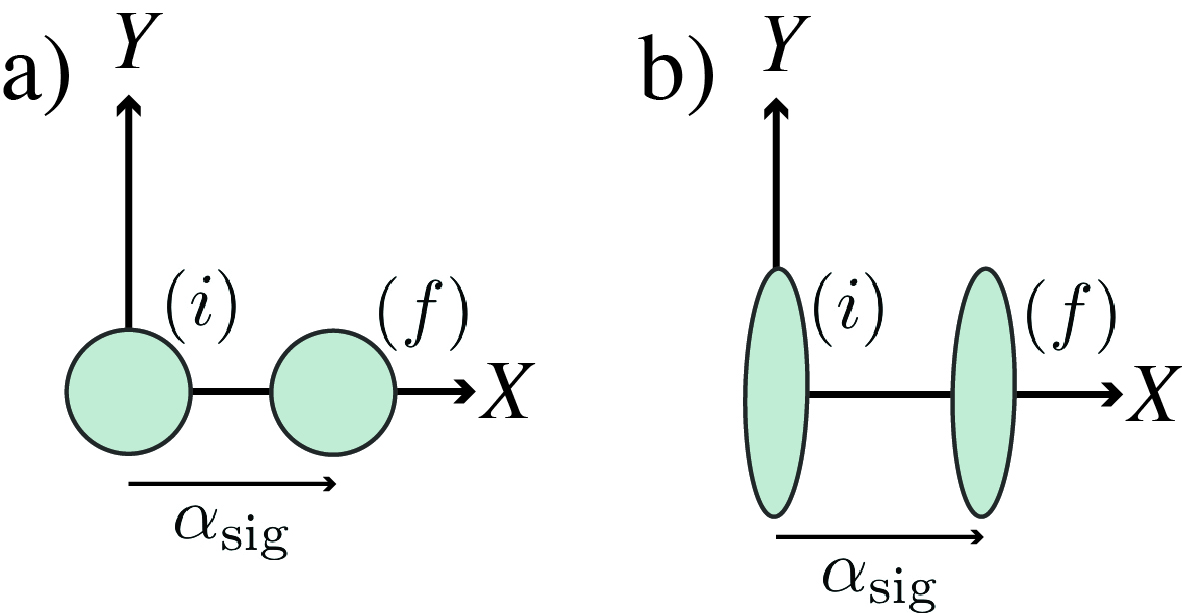}
   \caption{Phase space representations of the displacement introduced on  (a)  bright coherent (b) and phase squeezed fields by motion of a nanoparticle. ($i$) and ($f$) label the states of the fields prior to, and after, the phase shift, respectively.
}
 \label{Bright_SQZ_bio}  
  \end{center}
\end{figure}

Returning to the problem at hand, from Eq.~(\ref{alpha_sig_asdgsd}) we can see that the first-order effect of the perturbation is to displace the signal mode in phase space by an, in general, complex parameter that depends linearly on both the magnitude of the perturbation $p$ and the amplitude of the incident field, and also depends, through $\mathcal{N}$, on the effect of the perturbation on the spatial mode of the output field. The perturbation can then be determined by via an estimate of $\alpha_{\rm sig}$. This effect is shown for the case of a real displacement in the phase-space diagrams in Fig.~\ref{Bright_SQZ_bio}. It is clear from these diagrams that an improved estimate is possible by populating the signal mode with a squeezed vacuum with appropriately orientated squeezed quadrature, rather than a vacuum state. Indeed, comparison of Eqs.~(\ref{phase_sqz_a})~and~(\ref{coh_pop_sig_q_im}) (compare also Fig.~\ref{PhaseSpaceBrightSQZ} to Fig.~\ref{Bright_SQZ_bio}) shows that, to first-order, there is a direct analogy between quantum imaging of a single parameter and phase measurement, with the substitution $\phi \rightarrow - i p/\mathcal{N}$. The standard quantum limit and Heisenberg limit of phase sensing of Eqs.~(\ref{phi_SQL_totalpower}),~(\ref{phi_SQL_samplepower}),~and~(\ref{Heisenberg1}) can then be directly applied to single-parameter quantum imaging,  with $\Delta \phi \rightarrow \Delta p/|\mathcal{N}|$. 

While here we have limited our analysis to the case of single-parameter estimation, it should be clear that an analogous approach can be taken in the more general and common situation where the output modeshape depends on multiple orthogonal parameters $p_j$ as $u_{\rm out} ({\bf r}, {\bf p})$, with ${\bf p} \equiv \{ p_1, p_2, ... \}$. A first-order Taylor expansion of $u_{\rm out} ({\bf r}, {\bf p})$ can be used to obtain signal modes associated with each of the parameters, thereby identifying the optimal approach to measuring them and the quantum limits to such measurements.

\subsubsection{Effect of optical inefficiencies}
\label{Sqz_eff_eff_sec}

In Section~\ref{Loss} we saw that optical losses can severely degrade the precision of quantum measurements, in the particular case of NOON state interferometry. One should expect, therefore, that losses will also degrade measurements with squeezed states. This is indeed the case, and is the topic of this section.

Let us consider phase measurement using a bright squeezed state, as described without the introduction of losses in Section~\ref{SqueezedState}. In Section~\ref{SqueezedState}, we found that with access to a perfect measurement apparatus, the precision of such measurements has Heisenberg scaling. However, as discussed in Sections~\ref{flux_oct_sec}~and~\ref{flux_cons_noon_sec}, the photon number resolving detectors required to achieve such a measurement significantly constrain the flux of photons that can be used in the measurement, and therefore the absolute sensitivity. Here, therefore, let us restrict ourselves to  the commonplace practise of using linear detectors such as semiconductor photodiodes. By bright, we mean that the field in the probe arm, prior to the sample, may be described in the Heisenberg picture of quantum mechanics via the annihilation operator $\hat a = \alpha + \delta \hat a$ where $\alpha = \langle \hat a \rangle$ and $\delta \hat a$ includes all of the fluctuations in the field. Upon experiencing the signal phase shift $\phi$ and inefficiencies $1-\eta$, this becomes
\begin{eqnarray}
\hat a_{\rm out} &=& \sqrt{\eta} (\alpha + \delta \hat a) e^{i \phi} + \sqrt{1-\eta} \, \hat a_v\\
&\approx& \sqrt{\eta} (\alpha + \delta \hat a) (1 + i \phi) + \sqrt{1-\eta} \, \hat a_v,
\end{eqnarray}
where, as discussed in Section~\ref{Loss}, we have modelled the loss via a beam splitter that both attenuates the field and introduces vacuum noise $\hat a_v$, and we have assumed that $\phi \ll 1$ to reach the second equation. The phase quadrature of the output  field, which encodes the action of the signal phase shift,  can then be calculated from Eq.~(\ref{Yquad}) as
\begin{equation}
\hat Y_{\rm out} \equiv i (\hat a^\dagger - \hat a) = \sqrt{\eta} \left (2 \alpha \phi + \hat Y \right ) + \sqrt{1-\eta} \, \hat Y_v.
\end{equation}
Rearranging this expression, the phase $\phi$ is given by
\begin{equation} \label{asdgnjss}
\phi = \frac{1}{2 \alpha \sqrt{\eta}} \bigg [ \underbrace{\hat Y_{\rm out}}_{\text{signal}} - \underbrace{\sqrt{\eta}\, \hat Y - \sqrt{1-\eta} \, \hat Y_v}_{\text{noise}} \bigg ] .
\end{equation}
Using an ideal linear homodyne detector, we are able in principle, and to good approximation in practice, to directly measure the output field $\hat Y_{\rm out}$. The variance of the phase measurement is given by the sum of the variances of the second and third terms in Eq.~(\ref{asdgnjss}) (the ``noise" terms), so that the standard deviation
\begin{equation}
\Delta \phi = \frac{1}{2 \alpha} \sqrt{V_{\rm sqz} + \frac{1 - \eta}{\eta}}, \label{phase_sens_sqzdaef}
\end{equation}
where $V_{\rm sqz}$ is the variance of the phase quadrature of the input field. We see, therefore, that consistent with our previous analysis in Section~\ref{SqueezedState},
reducing the variance of the input phase quadrature improves the precision of the measurement, as does increasing the amplitude ($\alpha$) of the field. We also see here that the effect of loss is to degrade the precision; and as expected, the measurement precision cannot exceed the absolute bound on precision introduced by inefficiencies that was derived for any quantum state of light in Section~\ref{inf_bound_limit_sec}.\footnote{It should be noted that, here, by treating an ideal homodyne measurement of the squeezed field, 
 we are implicitly considering the case where the power in the sample is constrained, since an ideal noise-free local oscillator field is infinitely bright (this is not as impractical as it sounds -- in practise the local oscillator must just be significantly brighter than the squeezer field). Therefore, Eq.~(\ref{phase_sens_sqzdaef}) should not be compared to the total power constrained fundamental bound in Eq.~(\ref{fund_loss_limit}), but rather the sample-power constrained bound discussed later in Section~\ref{inf_bound_limit_sec}.}
%
The quantum noise limit (see Section~\ref{QNL_sec}) of a given measurement with an inefficient apparatus can be calculated by taking $V_{\rm sqz} = 1$ in Eq.~(\ref{phase_sens_sqzdaef}), which, using Eq.~(\ref{mean_n}), can indeed be seen to reproduce the sample-power-constrained version of our earlier result in Eq.~(\ref{QNL}). In quantum metrological experiments with squeezed light, including the nanoparticle tracking experiments we have been discussing in this section, the quantum noise limit can therefore be established simply by replacing the squeezed field with a coherent or vacuum state while, of course, ensuring that the measurement is quantum noise limited. It should be clear from this discussion that, unlike the standard quantum limit, any level of squeezing ($V_{\rm sqz}<1$) is sufficient to surpass the quantum noise limit even in the presence of inefficiencies.

We saw in the previous paragraph that the precision of phase measurements with squeezed light can be improved both by increasing the level of squeezing, and by increasing the coherent amplitude of the field.
However, both of these actions also increase the number of photons in the probe field (see Eq.~(\ref{mean_n})). To compare squeezed light based measurements to other approaches to precision phase measurement, we must determine the optimal sensitivity under a constraint on input power. The total number of photons in the input field can be quantified by Eq.~(\ref{mean_n}). Rearranging this expression for $\alpha$, and assuming that the amplitude quadrature variance $V(\hat X) = 1/V_{\rm sqz}$ (that is, that the initial squeezed state is a minimum-uncertainty state -- see Eq.~(\ref{HeisenbergX})), 
we find that
\begin{equation}
\alpha =  \sqrt{\langle n_{\rm sig} \rangle - \frac{1}{4} \left (V_{\rm sqz} + \frac{1}{V_{\rm sqz}} - 2 \right )} .
\end{equation}
Substituting this into Eq.~(\ref{phase_sens_sqzdaef}), we can express the uncertainty in the phase measurement in terms of the total photon number passing through the sample $\langle n_{\rm sig} \rangle$ and the squeezing as
\begin{equation}
\Delta \phi = \sqrt{\frac{V_{\rm sqz} + ({1 - \eta})/{\eta}}{4 \langle n_{\rm sig} \rangle - V_{\rm sqz} - {V_{\rm sqz}^{-1}} + 2}}.\label{phase_sens_sqzdaef2}
\end{equation}
In the usual bright squeezing limit where $4 \langle n_{\rm sig} \rangle \gg V_{\rm sqz} + {V_{\rm sqz}^{-1}} - 2$ which we considered in Section~\ref{large_alp_lim_sec}, and with high squeezing and non-negligible loss  $ V_{\rm sqz} \ll (1-\eta)/\eta$,
 the phase precision achievable with squeezed states asymptotes to 
\begin{equation}
\Delta \phi_{\rm SQZ} = \frac{1}{2 \sqrt{\langle n_{\rm sig} \rangle}} \sqrt{\frac{1-\eta}{\eta}} = \Delta \phi_{\rm SQL} \sqrt{\frac{1-\eta}{\eta}}. \label{sqz_ohase_asasd}
\end{equation}
This exactly corresponds  to the absolute state-independent precision limit imposed by inefficiencies of Section~\ref{inf_bound_limit_sec}~\cite{Demkowicz2012}. This is illustrated in Fig.~\ref{Phase_sensing_limitsB}b, where, for each efficiency shown, the phase sensing precision using squeezed states (solid curves) asymptotes to the corresponding state-independent precision limit (dashed curves) for sufficiently high $\langle n_{\rm sig} \rangle$.
We see, therefore, that phase measurements using bright squeezed states and linear optical detection, while being one of the simplest approaches to quantum-enhanced measurement, are optimal in this regime. Below this asymptotic limit, however, more complex states are optimal; see Refs.~\cite{PhysRevLett.102.040403,PhysRevA.83.021804,Escher2011}.

\begin{figure}[t!]
 \begin{center}
   \includegraphics[width=12cm]{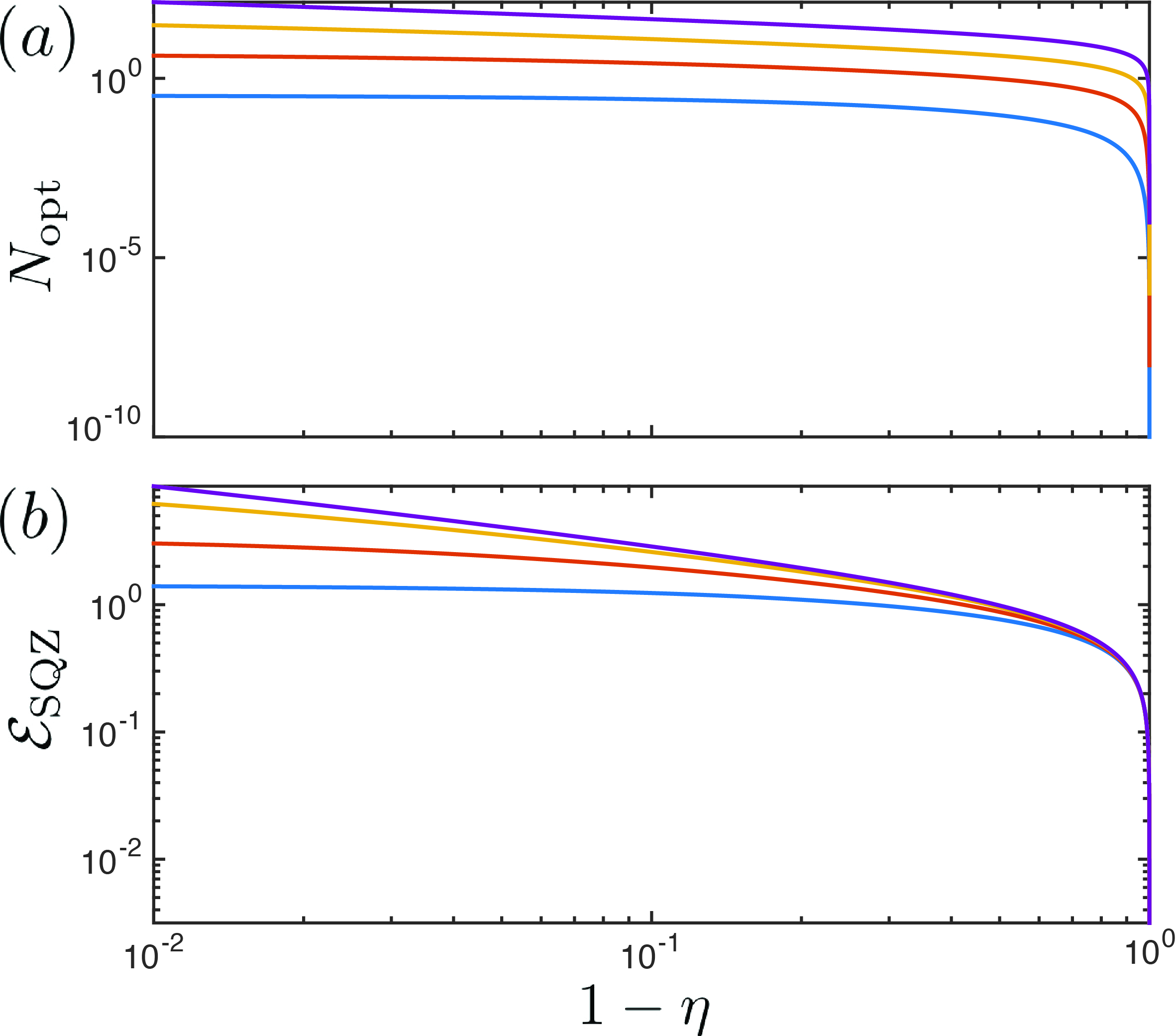}
   \caption{Improvement in phase measurement using squeezed light as a function of efficiency in probe arm $1-\eta$. Here, for simplicity, we consider a scenario where the power constraint on the measurement is placed on the power in the probe arm of the interferometer, and apart from the probe arm, all other parts of the interferometer have perfect efficiency.  (a) Optimal non-classical photon number $N_{\rm opt}$ for squeezed state interferometry. (b) Enhancement factor, beyond the standard quantum limit given in Eq.~(\ref{phi_SQL_samplepower}). The total number of photons available for the measurement is varied between the four curves shown in both (a) and (b). From the top to bottom curve: $\{\langle n_{\rm sig} \rangle =  1000, 100, 10, 1\}$. Dashed line: $\mathcal{E}_{\rm SQZ}=1$.
}
 \label{SQZ_int_N_and_en}  
  \end{center}
\end{figure}
It is straightforward to determine the squeezed variance that optimises the phase sensitivity of Eq.~(\ref{phase_sens_sqzdaef2}) for fixed power in the probe field, with the result:
\begin{equation}
V_{\rm sqz}^{\rm opt} = \frac{\eta + \sqrt{4 \eta (1 - \eta) \langle n_{\rm sig} \rangle + 1 }}{4 \eta \langle n_{\rm sig} \rangle + \eta + 1}. \label{opt_V_asdsdf}
\end{equation}
Figure~\ref{Phase_sensing_limitsB}b shows this optimal phase sensitivity as a function of mean probe photon number $\langle n_{\rm sig} \rangle$  for several different efficiencies. These curves show that, in general, the optimal squeezed state phase precision has scaling with $\langle n_{\rm sig} \rangle$ that lies in-between standard quantum limit scaling and Heisenberg scaling.

A squeezed field enhancement factor $\mathcal{E}_{\rm SQZ}$ can  be defined in analogy to the NOON state enhancement factor $\mathcal{E}_{\rm NOON}$ introduced in  Section~\ref{Loss}, quantifying the improvement in precision that can be obtained compared with the standard quantum limit. It is given by
\begin{equation}
\Delta \phi_{\rm SQZ}(V^{\rm opt}_{\rm sqz}) = \frac{\Delta \phi_{\rm SQL}}{\mathcal{E}_{\rm SQZ}},
\end{equation}
where $\Delta \phi_{\rm SQZ}(V^{\rm opt}_{\rm sqz})$ is determined by Eqs.~(\ref{phase_sens_sqzdaef2}) with the squeezed state variance given in Eq.~(\ref{opt_V_asdsdf}).
Also analogously to NOON state interferometry, we can define an optimal photon number $N_{\rm opt}$ in the fluctuations of the squeezed field -- or in other words -- an optimal balance between quantum correlated photons and photons contributing to the coherent amplitude $\alpha$:
\begin{equation}
N_{\rm opt} = \langle \delta \hat a^\dagger \delta \hat a \rangle =  \frac{1}{4} \left (V_{\rm sqz}^{\rm opt} + \frac{1}{V_{\rm sqz}^{\rm opt}} - 2 \right ).
\end{equation}
Unlike the case of NOON states, this optimal quantum photon number depends, through the optical variance $V_{\rm sqz}^{\rm opt}$, on the total number of photons $\langle n_{\rm sig} \rangle$ passing through the sample. It is plotted as a function of efficiency and for various $\langle n_{\rm sig} \rangle$ in Fig.~\ref{SQZ_int_N_and_en}a. Fig.~\ref{SQZ_int_N_and_en}b shows the squeezed light enhancement factor $\mathcal{E}_{\rm SQZ}$ for the same parameters.

\subsubsection{Precision comparison between squeezed and NOON state  single-parameter measurements}

We have now determined the phase precision that can be achieved in the presence of loss using both an ideal NOON state interferometer (Section~\ref{Loss}), and a squeezed state-based measurement constrained to standard linear detection. As such, it is possible to compare the sensitivities achievable via each technique. All else being equal, we hope that such a comparison will inform decision making in the design of future apparatus for quantum enhanced  biophysical measurements, providing guidance as to the parameter regimes where each of the two approaches are most effective.

\begin{figure}[t!]
 \begin{center}
   \includegraphics[width=12cm]{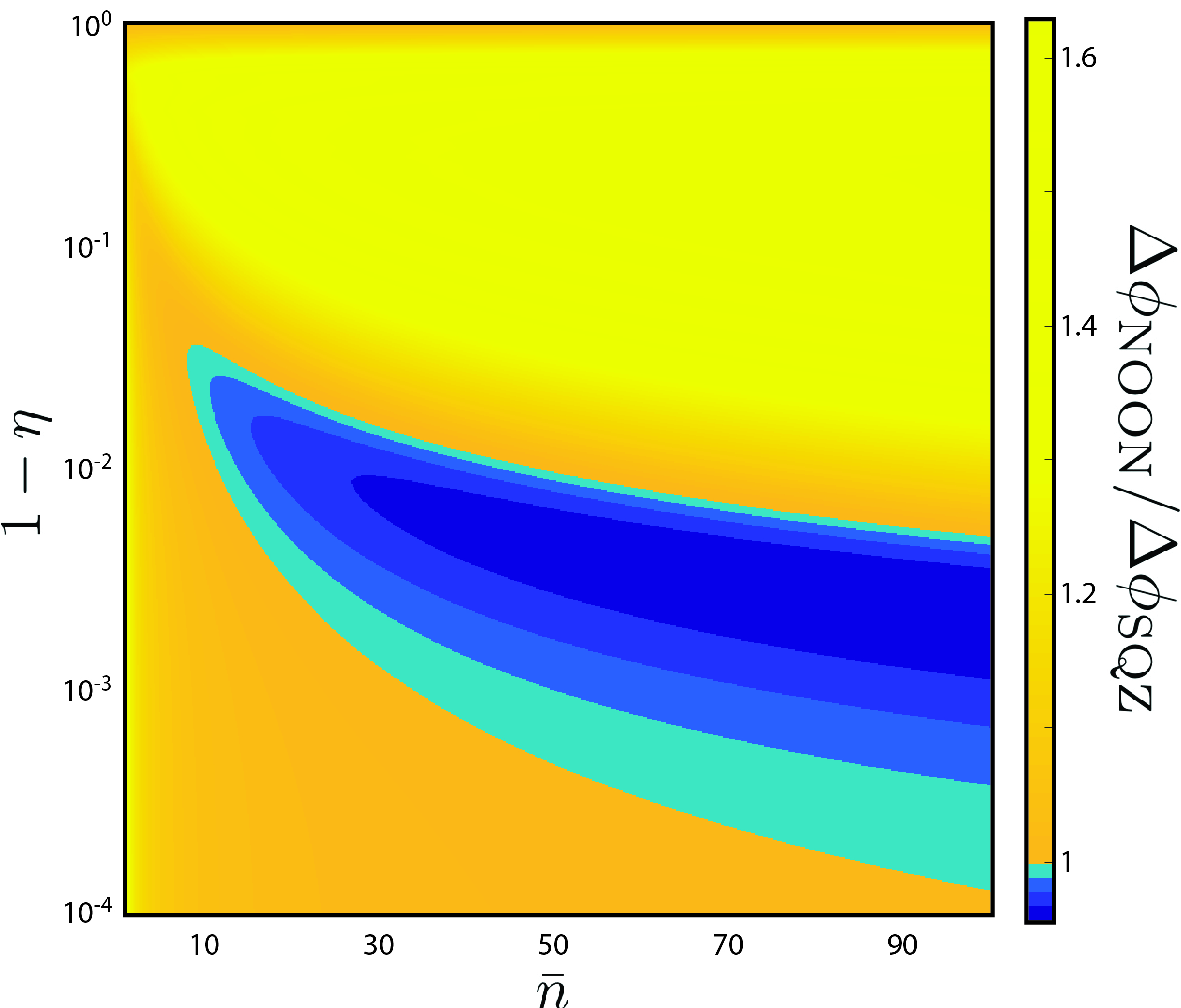}
   \caption{Comparison between the sensitivity of an optimal NOON state based phase measurement and a squeezed state based measurement using conventional linear detectors as a function of detection efficiency and power in the probe arm of the interferometer. Here, we consider the case where the power transmitted through the sample is constrained and inefficiency  is only introduced in the probe arm of the interferometer.}
 \label{SQZ_NOON_comp}  
  \end{center}
\end{figure}

The ratio of the achievable precision using  NOON and squeezed states is shown in Fig.~\ref{SQZ_NOON_comp} as a function of probe photon number $\langle n_{\rm sig} \rangle$ and efficiency $\eta$. From this figure, we see that NOON and squeezed state interferometry, in principle, achieve quite comparable phase sensitivities over the parameter space plotted, with the ratio $\Delta \phi_{\rm NOON}/\Delta \phi_{\rm SQZ}$ varying over a range of 0.96 to 1.63.
It is notable that, for the NOON state approach to be superior requires both
 efficiency greater than 97\% and a photon number in the probe arm $\langle n_{\rm sig} \rangle > 8$ (or $N>16$); while even with $\langle n_{\rm sig} \rangle = 100$ (or $N =200$), NOON state interferometry only improves the precision achievable with squeezed states by at best 4\% at an efficiency of around $99.8$\%. Even setting aside considerations of flux (see Sections~\ref{flux_oct_sec}~and~\ref{flux_cons_noon_sec}),
 since, as we have discussed previously,  the largest optical 
NOON state reported to date had $N=5$~\cite{Afek2010}, and the squeezed state protocol requires  only  standard linear detection on a high efficiency, low noise, semiconductor photodiode, it appears reasonable to predict that squeezed states will provide a better approach to most practical quantum-enhanced biophysical measurements over the near future.

\subsection{Absorption imaging}
\label{ab_im_secasf}

While, as we saw in the previous two sections, squeezed states of light appear to offer the most versatile  near-term approach to single- or few-parameter quantum measurement problems; to date,  entangled photon pairs have allowed significantly more progress to be made on full-scale quantum imaging problems. The key reasons for this are, firstly, that techniques to produce entangled photons, such as spontaneous parametric down conversion, often naturally produce them simultaneously in a large number of electromagnetic modes as required for full-scale imaging problems (see Sections~\ref{gen_noon_state_subsubsec}~and~\ref{QImaging}); and secondly, that in imaging problems the number of photons arriving at each pixel is reduced commensurately with the number of pixels so that, on a per-pixel basis, classical techniques generally utilise lower photon flux than they do for single- or few-parameter estimation problems. This second factor reduces the per-pixel photon flux required for practical quantum enhanced measurements 
(see Sections~\ref{flux_oct_sec}~and~\ref{flux_cons_noon_sec}).


In this Section we review applications of entangled photon pairs in absorption imaging. Absorption imaging constitutes a staple application of optical microscopy, and is routinely used in a vast array of biological applications. Recently, heralded single photons have been used to image weakly absorbing objects with quantum enhanced precision~\cite{Brida2010,Brida2011}, and to suppress background noise in the measurement process~\cite{Morris2015}. 

\subsubsection{Quantum enhanced absorption imaging}
\label{Q_ab_im_sec}

Absorption measurements are distinct from the phase or amplitude measurements considered earlier, since the presence of absorption necessarily couples  vacuum fluctuations into the detected optical field (see Section~\ref{Loss}). However, the detection statistics of such measurement can also be understood within the simple semiclassical framework of Section~\ref{semiclass}. When light is incident on an absorbing sample, the loss can be estimated from the resulting reduction in photon flux. Any classical light source is subject to statistical fluctuations which enforce a minimum uncertainty in the incident photon number of $\langle n_{\rm sig} \rangle^{1/2}$, given the average flux $\langle n_{\rm sig} \rangle$ (see Sections~\ref{interferometry}~and~\ref{CoherentState}). Consequently, the sample must absorb at least $\langle n_{\rm sig} \rangle^{1/2}$ photons to be statistically observable, with photon shot noise  precluding measurement of an absorption coefficient lower than $\langle n_{\rm sig} \rangle^{-1/2}$. This limit was overcome in Refs.~\cite{Brida2010,Brida2011} by using spontaneous parametric down-conversion to produce entangled photon pairs in two spatially separate beams (see Section~\ref{gen_noon_state_subsubsec}). One of the beams of entangled photons was passed through the sample, while the other was directed upon a single photon detector. Each photon impinging on the sample was then heralded via measurement of its twin photon, thereby eliminating the statistical uncertainty in the incident photon number. In the limit that all incident photons are perfectly heralded, absorption can be measured from the loss of a single photon. This then allows measurement of absorption coefficients as small as $\langle n_{\rm sig} \rangle^{-1}$. 

In real experiments, optical inefficiencies limit the efficiency with which the incident photons can be heralded (see Section~\ref{eye_sec} for further discussion of this). If the heralding photon is lost, a photon can be incident without being heralded; while loss in the experiment prior to the object will be falsely attributed to absorption in the sample. Nevertheless, the experiment in Ref.~\cite{Brida2010} succeeded in improving the measurement signal-to-noise by $30\%$ within a  multi-mode parallel optical imaging experiment. While this is itself impressive, it is far from the limits of the technique. Improvements in technology have the potential to allow far greater enhancement, with order-of-magnitude enhancement possible for sufficiently high optical efficiency. 
Furthermore, this method also has the advantage that it does not require that each pair of twin photons is distinguishable from other pairs. Many techniques based on entangled photon pairs require sufficiently low light flux that the probability of two pairs of twin photons being temporally co-located is negligible; with examples including  quantum OCT and NOON state sensing (discussed in Sections~\ref{Q_OCT_section} and \ref{NOON_bio}). We saw this in  Section~\ref{gen_noon_state_subsubsec} for the specific example of two-photon NOON state via
spontaneous parametric down conversion, which required sufficiently weak pumping that the nonlinear interaction strength, $\epsilon$, was very small.
In heralded photon imaging, it is only important that the total number of photons arriving at each pixel within the camera integration time can be precisely determined. This can be achieved, in principle, using a photon number resolving heralding camera. 
As such, this method is scalable to high photon flux, and could therefore realistically outperform state-of-the-art classical absorption imaging techniques in biologically relevant regimes.

\subsubsection{Ghost imaging of weakly absorbing samples}

In a recent experiment, an approach similar to that described above has been used to image a wasp wing, among other weakly absorbing samples~\cite{Morris2015}.
Unlike the experiments above, and most other experiments discussed this review, this experiment did not aim to demonstrate quantum enhanced performance, but rather addressed another fundamental question: how many photons are required to form an image? 
This is a question of relevance to both classical and quantum imaging. 
When imaging with only a few photons, any background light and electronic noise can quickly wash out the signal. One might think that this noise could be suppressed by gating the camera such that it only records signals when a photon is actually incident on the sample. However, as discussed in Sections~\ref{interferometry}~and~\ref{CorrelationSection}, classical light sources exhibit probabilistic photon arrival times which makes it fundamentally impossible to perfectly predict the exact time when the photons will arrive. By contrast, as discussed above,  entangled photons can be heralded to perfectly predict the arrival time. This procedure involves the generation of entangled photon pairs, followed by detection of one photon on a reference detector. Whenever this photon is detected, it indicates with certainty that another photon is propagating towards the main experiment. By gating the camera with this heralding signal, it is possible to greatly suppress the background counts, and thereby improve the background noise level when imaging with very low light intensity. Although this may be seen as a less fundamental enhancement than violation of a quantum limit as was reported for the exeriments in Section~\ref{Q_ab_im_sec}, it is no less relevant. Rather, it achieves a practical advantage over state-of-the-art technology, and therefore provides capabilities that are otherwise unachievable.
 
  \begin{figure}
 \begin{center}
   \includegraphics[width=12cm]{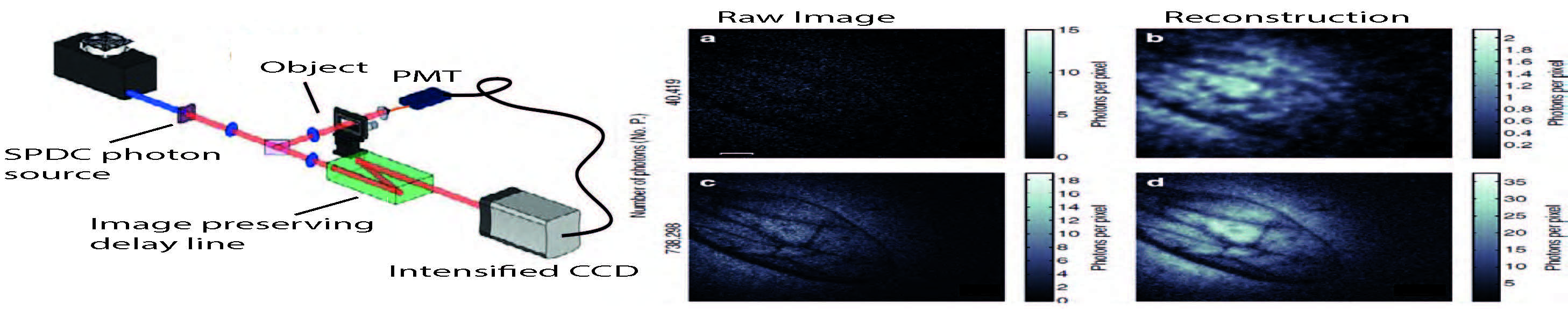}
   \caption{Quantum ghost imaging with few photons. Left: Entangled photon pairs are generated, and with a photon propagating in two separate spatial modes of the electromagnetic field, each towards a single photon detector. If one photon is transmitted through the specimen and detected, a trigger pulse is sent to the intensified CCD which then measures the transverse location of its incident photon. Due to spatial correlations between the entangled photons which are introduced by energy and momentum conservation when the photon pair is generated, 
   this location corresponds to the position at which the photon passed through the specimen. {\bf a}, {\bf c} Experimental images of a wasp wing formed with this method with different numbers of photons. {\bf b}, {\bf d} Reconstructed images based on the measured data and the photon noise statistics. The shape of the specimen can be estimated even when no details are clearly visible in the direct image ({\bf b}). Scale bar 400~$\mu$m. Figure from Ref.~\cite{Morris2015}.
}
 \label{GhostBio}  
  \end{center}
\end{figure}
The experiment was conducted using a prominent quantum imaging techniques termed {\it ghost imaging}~\cite{Pittman1995}, and is shown in Fig.~\ref{GhostBio}. In ghost imaging, entangled photon pairs are generated using spontaneous parametric down-conversion. One photon propagates in a reference arm toward an imaging detector  that is capable of spatially resolved single photon detection (see the discussion of such detectors in Section~\ref{Q_OCT_section}),
 while the other photon propagates through an absorbing specimen and onto a non-spatially resolved single photon detector, in what is essentially the reverse of the arrangement discussed in Section~\ref{Q_ab_im_sec} above.
 When a photon is detected after the specimen, the position of its twin reference photon on the imaging detector is recorded.
 Even though the reference photon does not pass through the sample itself, the strong position correlations between photons produced via spontaneous parametric down-conversion  (see Section~\ref{gen_noon_state_subsubsec}) allow the position of the photon that did pass through the specimen to be inferred.
Each detection event on the imaging detector therefore records the position at which a photon was transmitted through the absorbing specimen. Consequently, the imaging detector can form an image exactly as it would if it were behind the specimen. 

In Ref.~\cite{Morris2015} the imaging detector was an intensified CCD camera, and was rapidly gated to detect the reference photon only in a short interval of time around the detection time of photons transmitted through the specimen, and thereby reduce the measurement noise. The technique was used to image both a US Air Force resolution target, and a wasp wing (shown in Fig.~\ref{GhostBio}a--d). Following image acquisition, the shape of the underlying object was estimated using the assumption that the image is sparse in spatial frequencies, and taking into account the probabilistic nature of photon shot noise. This procedure allowed the objects to be reconstructed even when the direct image is completely obscured by the photon shot noise (Fig.~\ref{GhostBio}a,b.). This result was the main focus of the paper. 

Quantum metrology played a more technical role, allowing the background noise to be suppressed. We note that, while when initially proposed in the 1990s, ghost imaging was considered to be a purely quantum imaging protocol~\cite{Pittman1995}, later experiments demonstrated similar capabilities with classical photon correlations in the form of pseudorandom laser speckle~\cite{Ferri2005}. This led to a lively debate in the quantum optics community regarding which aspects were truly quantum~\cite{Erkmen2010}, with the eventual conclusion that thermal light could reproduce nearly all features of ghost imaging with entangled photons, but with degraded contrast~\cite{Erkmen2010,Saleh2005}. From this perspective, the experiments of Ref.~\cite{Morris2015} can be viewed as one, but by no means the only, technique to suppress background noise. 
The experimental method they have developed could be beneficial to a range of absorption imaging applications where 
 ultra-low photon flux is required.

\subsubsection{Super-resolution in ghost imaging}

Ghost imaging, as introduced above, allows an apparent violation of the diffraction limit.
%
 This  can be understood most easily within context of classical ghost imaging, where classical twin beams are generated by passing a laser through a dynamically and spatially varying turbid medium, and separating the output speckle field into two beams using a beam splitter~\cite{Ferri2005}. Since the positions of speckle foci are randomly modulated, this imaging technique can be thought of as effectively performing a random raster scan through three-dimensional space. In the case of ghost imaging,  the achievable resolution has been shown to correspond to the minimum size of the laser speckle~\cite{Ferri2005}, independent of the numerical aperture of the lens used to image the field onto a detector (in contrast to the diffraction limit described in Eq.~(\ref{DiffractionLimitEq})).
The sequential nature of this imaging procedure also allows it to violate the Fourier limit, which imposes a trade-off between the resolution of simultaneous images at near-field and far-field~\cite{Ferri2005}. It does not, however, allow the sub-wavelength resolution which is highly sought after in biological microscopy. In fact, the resolution limit due to speckle is given by approximately $\lambda z /D$, where $D$ is the diameter of the classically fluctuating light source and $z$ is the distance from it to the image~\cite{Ferri2005,Erkmen2010}. If we consider the light source itself to be an element which focuses light into speckles, its $NA$ is given approximately as $2 D /z$, and the resolution is constrained by diffraction as described in Eq.~(\ref{DiffractionLimitEq}). Consequently, the diffraction limit is not violated in a treatment of classical ghost imaging which considers the focusing power of both the imaging lens and the light source, while a similar result also holds for ghost imaging with entangled photons~\cite{Erkmen2010}.

\subsection{Optical stimulation with number states of light}
\label{eye_sec}

Each of the experiments described above relied on non-classically correlated light to very gently probe a biological system, with the aim of observing it while introducing minimal disruption. In Ref.~\cite{Teich1990}, Teich and Saleh discussed a completely different approach to quantum metrology in which non-classical light is used to stimulate retinal rod cells.  Rod cells are biological detectors which produce electrical signals in response to optical stimulation, and are sensitive at the single-photon level~\cite{Baylor1979,Rieke1998}.

   \begin{figure}
 \begin{center}
   \includegraphics[width=8cm]{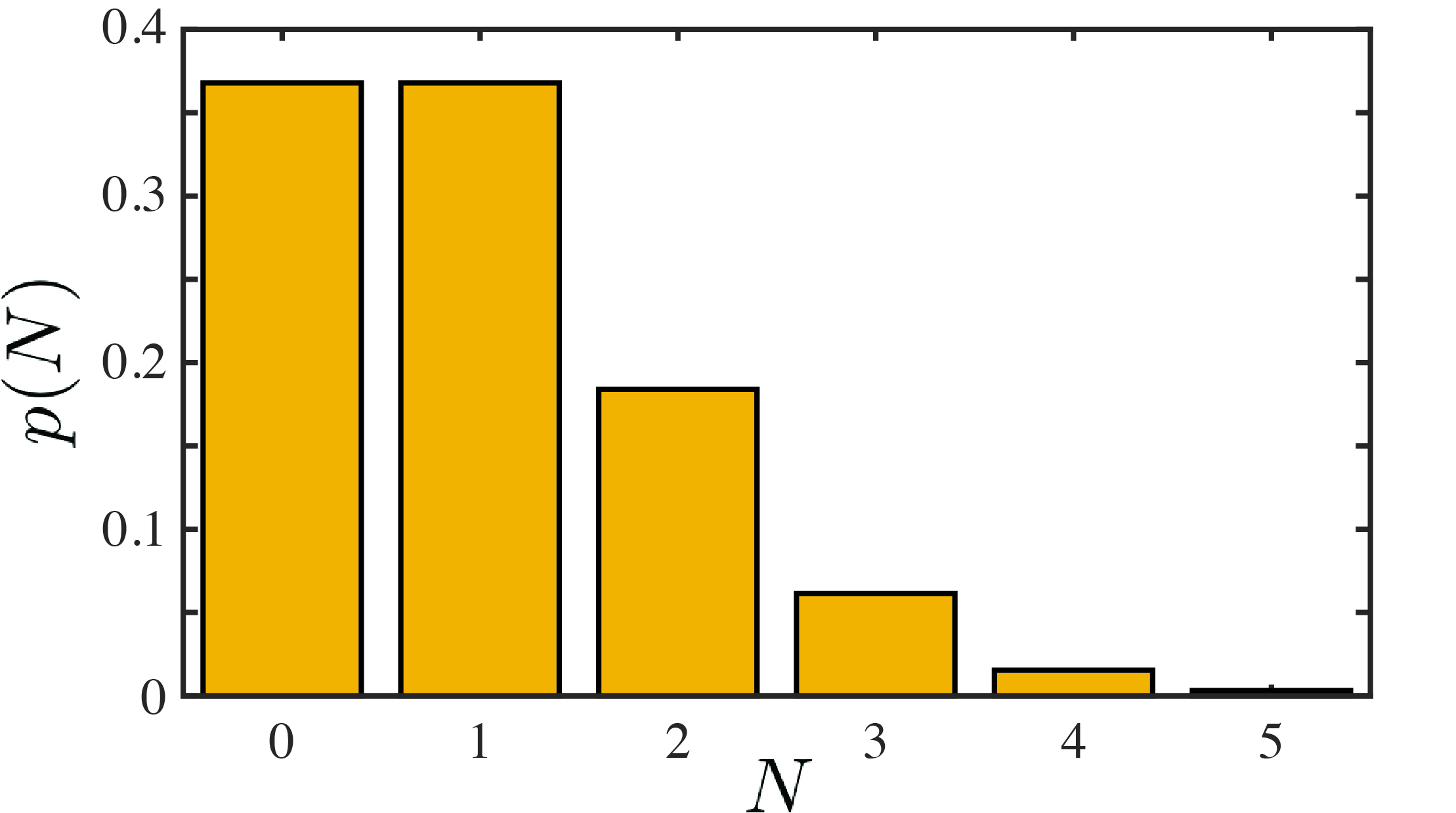}
   \caption{Photon number probability distribution for a coherent state with $\langle n \rangle$ = 1.
}
 \label{Coherent_p_dist}  
  \end{center}
\end{figure}
Generally, the electrical response of rod cells to illumination is been characterized using classical light sources. As has been discussed at various points throughout this review, the photon statistics of such light sources inevitably introduces uncertainty to the illuminating photon number. As an illustration, Fig.~\ref{Coherent_p_dist} shows the photon number distribution of a coherent state with mean photon number $\langle n \rangle =|\alpha|^2=1$, derived from Eq.~(\ref{photon_num_dist_coh_sdgs}). As can be seen, when illuminated with a pulse of this intensity, a rod cell with perfect efficiency would be equally likely to register zero or one photon, with non-negligible probabilities of even three of four photon events. 
Consequently, a single-photon response can only be unambiguously observed if the response of the rod cell allows it to be 
be resolved from both the zero-photon response and higher-order responses. In some species, the electric response of rod cells to optical stimulation allows discrimination of the single-photon response~\cite{Baylor1979,Rieke1998}. In other species, however, the electric signal from the rod cells does not allow the single-photon response to be separated reliably, and the single-photon response can only be inferred statistically.  As proposed in Ref.~\cite{Teich1990}, and utilized in Section~\ref{Q_ab_im_sec} in the alternative application of absorption imaging,  non-classical states of light can be used to suppress the statistical uncertainty in the optical stimulation, thus improving characterization of the retinal response.

 Such a scheme was demonstrated very recently, with single photons used to stimulate retinal rod cells from {\it Xenopus laevis} toads~\cite{Phan2014}. In that work, similarly to many of the other experiments discussed in this review, parametric down-conversion was used to generate entangled photon pairs, here at the visible wavelength of 532~nm. One of these photons was collected and directed onto the rod cell. This photon was heralded via detection of the second photon, thereby reducing the statistical uncertainty in the photon arrival statistics into the cell.
   This allowed the single-photon sensitivity of the cells to be directly confirmed in a species for which the electrical response itself does not allow photon resolution, and enabled a precise determination of the quantum efficiency of the cell. It further allowed the transient response of the cell to a single photon to be characterized without relying on statistical inference to estimate which detection events corresponded to single photons.

This preliminary work established the possibility of using non-classical optical stimulation for biological experiments. Such techniques could potentially allow important advances in the study of eyes, particularly when performed with a more intricate neural network. While the single-photon sensitivity of rod cells is already well established in many species~\cite{Baylor1979}, it remains unclear whether this translates to single-photon sensitivity in an eye. Retinal rod cells are not perfectly quiet detectors, but rather have non-negligible dark noise of approximately 0.01~s$^{-1}$~\cite{Rieke1998}. Given that a human has around 100 million rod cells, it is necessary for the the retina to filter the signals to remove the dark counts. It is expected that the signal from a single incident photon would be indistinguishable from dark noise, and would most likely be filtered from the perceived signal~\cite{Rieke1998}. To test this prediction, pulses of light containing definite numbers of photons could be sent into eyes to test whether the light can be observed. Such experiments have previously been performed using weak classical pulses of light, and have found that people can reliably discern pulses with a mean of 70--100 photons~\cite{Rieke1998}. However, the lower limit to human vision remains unknown.  
Experiments along these lines could reveal fundamentally new information about visual processing at low light levels.
	
 An alternative approach would be to electrically record activity in the ganglion nerve cells in response to retinal rod cell stimulation.	Ganglion cells are the first layer of visual filtering, combining and processing signals from multiple rod cells~\cite{Gollisch2010}.  One complicating factor in studying this system is that random photon statistics contribute significant variations to the resulting neural signal transmitted to the brain~\cite{Barlow1971,Saleh1985}. Consequently, Ref.~\cite{Teich1990} proposed the use of non-classical stimulation of rod cells to study the responsive properties of ganglion nerve cells in a retinal network, which could unambiguously discern the temporal and spatial filters used, and the resulting photon sensitivity of the retinal network.

 We further note that non-classical stimulation experiments utilize a biological sample as a photodetector, which maps the photonic state to an electrical state in the detector (see the discussion of photodetection in Section~\ref{CorrelationSection}). This provides a method to map non-classical correlations into a biological system. If these correlations could be observed after introduction to the cells, the decoherence of the quantum state could be quantified. This would open new possibilities for biological studies; in particular, it could be relevant to studies of photosynthesis where experimental evidence suggests that quantum coherence could be present in the light harvesting process~\cite{Panitchayangkoon2011,Hildner2013}. Although these experiments have presented evidence for coherence, so far there has been no test which could provide an unambiguous witness of this quantum coherence. Additionally, there have been further suggestions that entanglement may also be present and provide enhanced collection efficiency~\cite{Sarovar2010}, which has inspired lively debate~\cite{Tiersch2012}. Observation of the evolution of a quantum state could provide a witness to the coherence properties of photosynthesis, and might therefore provide important new insights into such systems~\cite{Li2012}. 
 
 \subsubsection{Effect of inefficiency}
 
So far in this section we have seen that, using conditioning detection on one of the twin photons emitted from a parametric down conversion source, it is possible in principle to entirely remove the statistical shot noise associated with weak optical fields, and thereby better characterise the optical response of photosensitivity biological systems such as the retina. However, in practise inefficiencies in both the transmission of the light to the sample and in the conditioning detection introduce some statistical fluctuations. We examine these sources of fluctuation in what follows. For these considerations, it is important to know the photon number distribution in both the reference field, used for conditional detection, and the probe field, used to illuminate the sample. Assuming these fields are generated via spontaneous parametric down conversion, the two probability distributions are identical and can be seen from
 Eq.~(\ref{state-para_down}) to be
\begin{equation}
p(N) = \left (1-\epsilon \right ) \epsilon^{N}, \label{p_dist_PDC}
\end{equation}
with the mean number of photons $\langle n \rangle$ in each field given by
\begin{equation}
\langle n \rangle \equiv \sum_{N}  N  p(N) = \frac{\epsilon}{1-\epsilon}.
\end{equation}

  \begin{figure}
 \begin{center}
   \includegraphics[width=17cm]{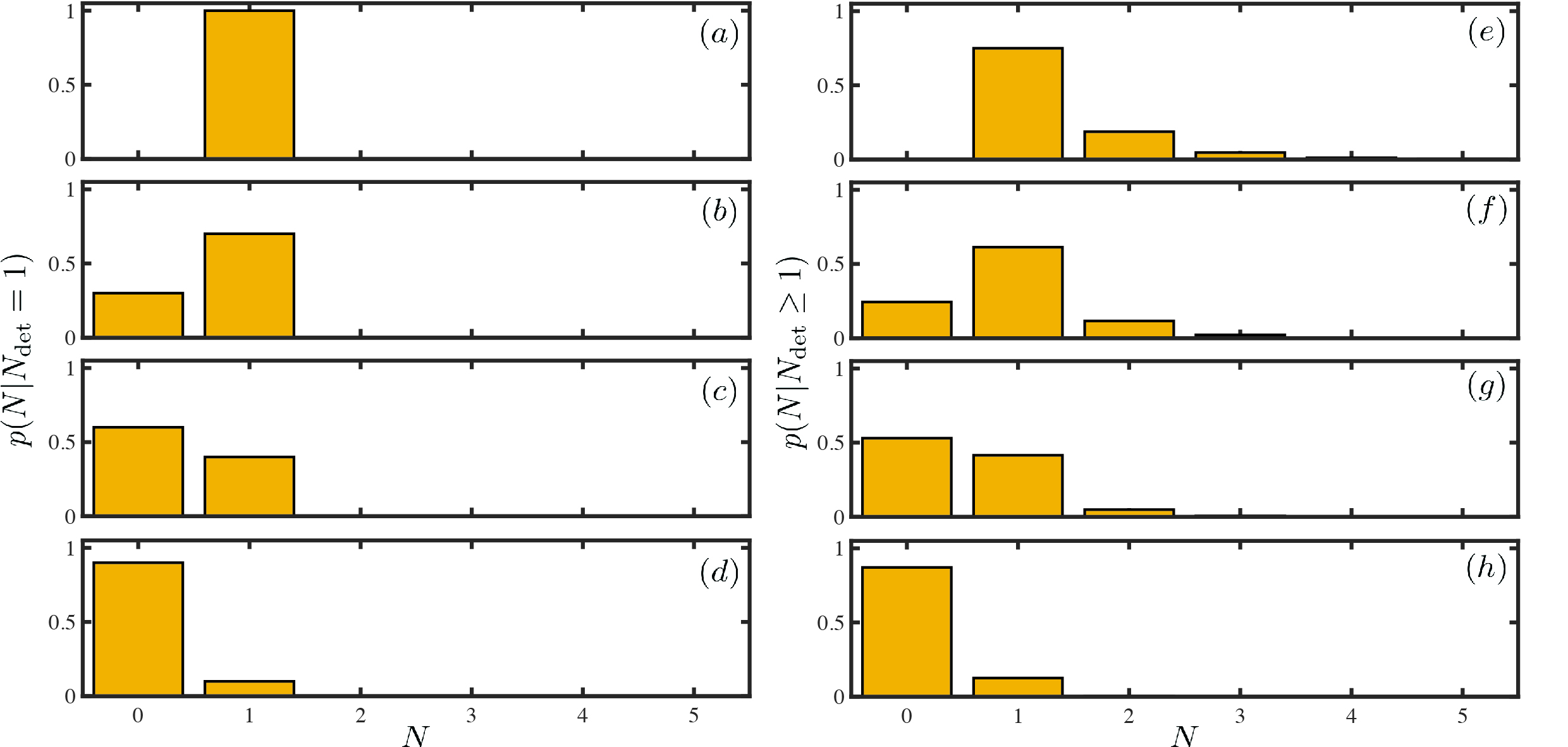}
   \caption{Conditional photon number distributions at the sample in the presence of loss prior to the sample using perfectly efficient photon number resolving detection ({\it left column}), and perfectly efficient bucket detection ({\it right column}). Here, we have chosen $\epsilon=0.5$, corresponding to $\langle n \rangle =1$ in each of the two entangled fields  prior to the introduction of loss. For (a)--(d) (and (e)--(h)) the efficiency of transmission to the sample is $\eta = \{1, 0.7, 0.4, 0.1 \}$, respectively.
}
 \label{p_N_given_det_loss_sample}  
  \end{center}
\end{figure}
Let us first consider the case where the conditioning detector has perfect efficiency, so that 
photons in the reference field  are detected without loss. A detection event at the detector then modifies the probe probability distribution prior to any losses to some conditional probability distribution $p(N| N_{\rm det})$ to be defined shortly (see Eqs.~(\ref{num_rs_dec_cond})~and(\ref{bucket_dec_cond})). If the probe field then experiences 
 a fractional loss of $1-\eta$, photons are probabilistically lost from the field, and the conditional probability distribution is modified, in general, to~\cite{waks2004direct}
%
\begin{equation}
p(N| N_{\rm det})
\rightarrow \eta^N \sum_{N'=N}^\infty \left( \!\! \begin{array}{c}
N' \\
N \\
\end{array}  \!\! \right )  (1-\eta)^{N'-N} \, p(N' | N_{\rm det}), \label{loss_photon_dist}
\end{equation}
%
where the vector represents the usual binomial coefficient.

As discussed in Section~\ref{flux_oct_sec}, there are two common forms of single photon detectors: photon resolving detectors which both register that a detection event occurred, and identify how many photons were involved; and so-termed ``bucket detectors" that only register the event. This is an important distinction, since as can be seen from Eq.~(\ref{p_dist_PDC}), parametric down conversion has a finite probability to generate more than one photon into each of the reference and probe fields. Let us first consider conditioning that is achieved via a perfect photon number resolving detector was used for conditioning. Prior to any loss, since the photon number in the probe and reference fields is perfectly correlated (see Eq.~(\ref{state-para_down})), detection of  $N_{\rm det}$ photons would condition that state of the probe field to  
a photon number state with probability distribution  
%
\begin{equation}
p(N| N_{\rm det}) =  \Bigg \{ \!\! \begin{array}{c}
\,\, 0, \hspace{5mm} N \neq N_{\rm det} \\
\,\, 1,   \hspace{5mm} N = N_{\rm det}  \\
\end{array} . \hspace{8mm}   \text{(photon number resolving detector)} \label{num_rs_dec_cond}
\end{equation}
%
Alternatively, if  a perfect bucket detector was used in the reference field, the sole effect of a detection event would be to rule out the possibility of having zero photons in the probe arm, prior to any inefficiencies. From Eq.~(\ref{p_dist_PDC}), the renormalised probability distribution in the probe arm would then be 
\begin{equation}
p(N| N_{\rm det} \ge 1) =  \Bigg \{ \!\! \begin{array}{c}
0, \hspace{5mm} N=0 \\
\left (1-\epsilon \right ) \epsilon^{N-1},   \hspace{5mm} N>0  \\
\end{array} . \hspace{20mm}   \text{(bucket detector)} \label{bucket_dec_cond}
\end{equation}

The probe photon number distributions with both photon number resolving and bucket detection can then be determined in the presence of probe inefficiencies via Eq.~(\ref{loss_photon_dist}). The results of these calculations are shown, for various probe efficiencies in the left and right columns of Fig.~\ref{p_N_given_det_loss_sample}, respectively. It can be seen from Fig.~\ref{p_N_given_det_loss_sample}a that, indeed, perfect photon number resolving detection with $N_{\rm det}=1$ combined with perfect probe efficiency, conditions a single-photon state in the probe.  As one would expect, losses introduce some probability that the probe arrives at the sample in a vacuum state; though notably, we see that with perfect photon number resolving detection, there is no probability of illuminating the sample with a number of photons greater than $N_{\rm det}$. By contrast, even with no probe loss bucket has some finite probability of illuminating the sample with more than $N_{\rm det}$. 

Let us now consider the case where the probe is transmitted into the sample with perfect efficiency, but the conditioning detection is imperfect. Similar to our prior analysis (Eq.~(\ref{loss_photon_dist})) of the effect of inefficiencies on the photon probability distribution in the probe field, 
 loss of magnitude $1-\eta$ in the reference field will result in the photon number distribution $p(N_{\rm det})$ arriving at the reference detector~\cite{waks2004direct}:
\begin{equation}
p(N_{\rm det}) = \eta^N \sum_{N=N_{\rm det}}^\infty \left( \!\! \begin{array}{c}
N \\
N_{\rm det} \\
\end{array}  \!\! \right )  (1-\eta)^{N-N_{\rm det}} p(N) \label{PNdet}
\end{equation}
where $p(N)$ is given in Eq.~(\ref{p_dist_PDC}).
To determine the probe probability distribution given a detection event on the reference field, sampled from this modified probability distribution, we must resort to Bayesian statistics. Bayes' rule states that 
the conditional probability distribution of the probe $p(N|N_{\rm det})$ given photon number resolving detection of $N_{\rm det}$ photons in the conditioning arm   is given by
\begin{equation}
p(N|N_{\rm det}) = \frac{P(N_{\rm det}|N) p(N)}{p(N_{\rm det})}, \label{bayesrule}
\end{equation}
where $p(N_{\rm det}|N)$ is the conditional photon number probability distribution in the reference field given that the probe field contains $N$ photons. This 
 can be found straightforwardly from Eq.~(\ref{PNdet}) as
\begin{equation}
p(N_{\rm det}|N) = \eta^N  \left( \!\! \begin{array}{c}
N \\
N_{\rm det}  \\
\end{array}  \!\! \right )  (1-\eta)^{N-N_{\rm det}}.
\end{equation}
%
%
The conditional photon number probability distribution arriving at the sample given bucket detection of $N_{\rm det} \ge 1$ photons can be determined from Eq.~(\ref{bayesrule}) as
\begin{equation}
p(N|N_{\rm det}\ge1) = \frac{1}{1- p (N_{\rm det}=0)}\sum_{N_{\rm det} =1}^\infty p (N_{\rm det}) p(N|N_{\rm det}). \label{cond_inf_ref_field}
\end{equation}

  \begin{figure}[b!]
 \begin{center}
   \includegraphics[width=17cm]{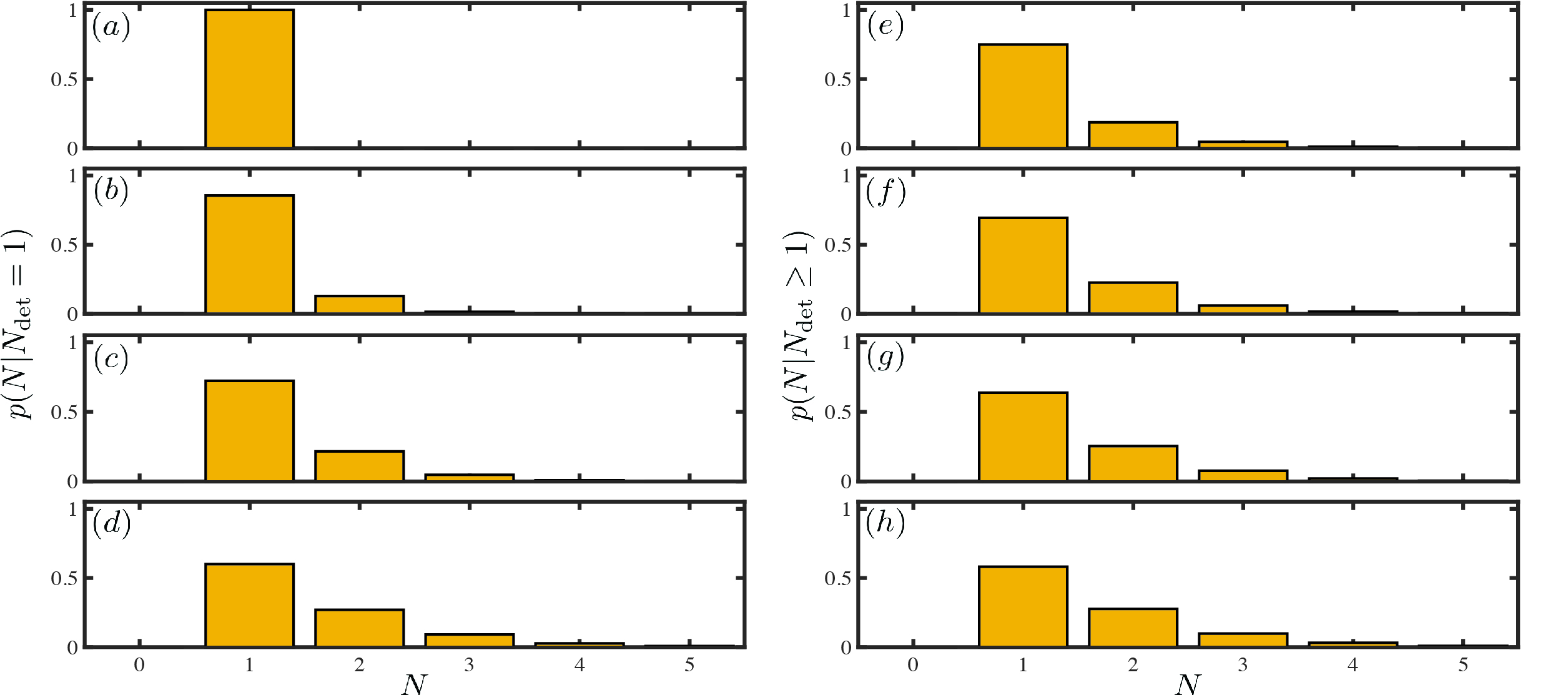}
   \caption{Conditional photon number distributions at the sample in the presence of loss prior to the conditioning detection using photon number resolving detection ({\it left column}), and bucket detection ({\it right column}). Here, we have chosen $\epsilon=0.5$, corresponding to $\langle n \rangle =1$ in each of the two entangled fields  prior to the introduction of loss.  For  (a)--(d) (and (e)--(h)) the detection efficiency  is $\eta = \{1, 0.7, 0.4, 0.1 \}$, respectively.
}
 \label{p_N_given_det_loss_cond_detect}  
  \end{center}
\end{figure}
The conditional photon number probability distributions of the illumination given inefficient conditioning via photon number resolving and bucket detection (as calculated from  Eqs.~(\ref{bayesrule})~and~(\ref{cond_inf_ref_field}), respectively) are shown in Fig.~\ref{p_N_given_det_loss_cond_detect} as a function of inefficiency. We see that, unlike the case where loss is present before the sample considered earlier, here the inefficiency {\it increases} the conditional probabilities of illumination with photon numbers larger than $N_{\rm det}$. This makes sense, since in the presence of detection inefficiency, detection of one photon in the reference field implies that {\it at least} one photon is present in the probe field.

\section{Promising technologies for future biological applications}\label{Future}

It should be clear from the previous section that 
 quantum measurements of biological systems are now a reality. In this section we briefly review some prominent examples of other quantum measurements that may soon 
 be used to realize  important applications in biology. 

\subsection{Entangled two-photon microscopy}
\label{ent_two_phot_mic_sec}

\begin{figure}[b!]
 \begin{center}
   \includegraphics[width=12cm]{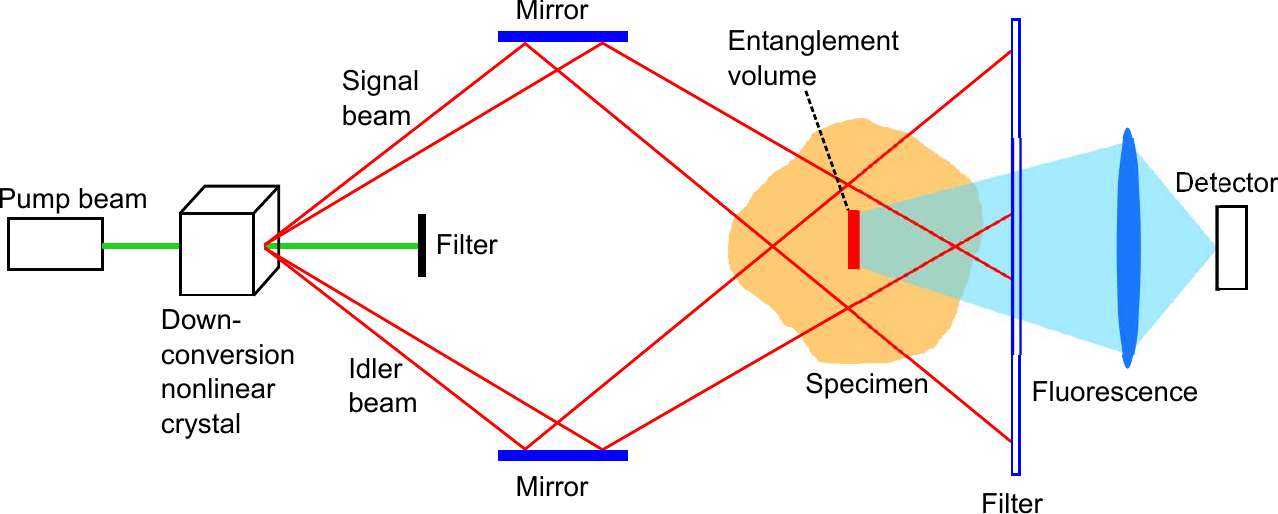}
   \caption{An entangled two-photon microscope. A pair of entangled optical fields are generated with a nonlinear crystal, and recombined at the specimen. The entanglement enhances two-photon absorption where the fields recombine, as can be observed by measuring the resulting fluorescence. Adapted from Ref.~\cite{Teich1997}.
}
 \label{Entangled2photon}  
  \end{center}
\end{figure}

Two-photon microscopy is an area in which entangled photons could provide a substantial practical advantage~\cite{Teich1997,Fei1997} (see Fig.~\ref{Entangled2photon}). In two-photon microscopy, two-photon absorption within a sample is studied via its specific fluorescent signature. Since two-photon absorption is a third-order nonlinear optical process, it will typically occur only near the central focus of the illumination (this was the basis of one of the super-resolution imaging techniques discussed in Section~\ref{ClassicalSuperResolution}). When compared to one-photon absorption, this not only sharpens the spatial resolution, but also suppresses fluorescence away from the focal plane. However, without quantum correlations, two-photon absorption is an extremely inefficient process  requiring a very high input flux of photons.  As a result, the peak power is typically maximized by use of high-peak-intensity pulsed lasers which can damage the specimen~\cite{Helmchen2005}.

If quantum correlated photon pairs are used instead of classical light the two-photon absorption rate can be vastly enhanced, with the absorption process depending linearly rather than quadratically on the photon-flux density~\cite{Pevrina1998}. Qualitatively, this enhancement occurs because the rate of two-photon absorption is proportional to the flux of coincident photon pairs; for uncorrelated light, this flux scales as the square of the power, while entangled photons always arrive in coincident pairs such that the flux scales linearly with the power. This can also be expressed with the photodetection framework introduced in Section~\ref{CorrelationSection}. If the entangled photon pairs are produced in two separate spatial modes via spontaneous parametric down conversion, as discussed in Section~\ref{gen_noon_state_subsubsec}, and the strength of the nonlinear interaction is chosen so that $\epsilon \ll 1$,
the overall state that is generated can be well approximated by:
\begin{equation}
| \Psi \rangle = \sqrt{1-\epsilon} |0\rangle_1 |0\rangle_2 + \sqrt{\epsilon} |1\rangle_1 |1\rangle_2,\label{ETPA_state}
\end{equation}
where higher photon number terms have been neglected (c.f., Eq.~(\ref{state-para_down})). For this state, two-photon absorption from the two modes occurs with a rate of
\begin{equation}
R^{(2)}_{1,2}\propto\langle \hat a^\dagger_1 \hat a^\dagger_2 \hat a_2 \hat a_1 \rangle = \epsilon  .
\end{equation}
This scales linearly with the incident power, since $\langle \hat  n_1 \rangle =  \langle \hat n_2 \rangle = \epsilon$. Using this expression in Eq.~(\ref{g2_general}), one can see that  the second order coherence function is given by $g^{(2)}_{12}=1/\epsilon$.  In the limit that $\epsilon$ is small $g^{(2)}_{12}$ can be very large, which indicates a large enhancement in two-photon absorption over the use of coherent light. At high flux, however, the two-photon absorption rate becomes comparable for coherent and entangled illumination due to saturation of the two-photon transition.
 Therefore, to achieve substantial enhancement requires that the flux is low relative to the saturation threshold; a condition which is almost always met.

 The enhanced absorption of entangled photon pairs allows multi-photon fluorescence microscopy to proceed with intensities more suited to biological samples. For instance, recent demonstrations in non-biological organic chemistry have found that the two-photon absorption and two-photon fluorescence measured with an entangled photon flux of 10$^7$~s$^{-1}$ is comparable to similar measurements with 10$^{17}$~s$^{-1}$ coherent photons, allowing a 10 order of magnitude reduction in power from tens of mW to a few pW~\cite{Lee2006,Guzman2010}. In principle, an entangled light microscope could also be modified to illuminate with a NOON state rather than the state in Eq.~(\ref{ETPA_state}), which would then also allow superior resolution to that achievable classically~\cite{Saleh2005} (see Section~\ref{QLithography}). However, this enhancement has not yet been demonstrated in any imaging context.

 Furthermore, entangled photons reveal information about the nonlinear absorption mechanism which is inaccessible to classical light sources. Two-photon absorption can occur either via a permanent dipole transition or via a virtual state transition~\cite{Jagatap1996}. While these transitions are classically indistinguishable, the different mechanisms have a markedly different response to entangled photons; dipole transitions are not enhanced by the entanglement~\cite{Upton2013}, while entangled photons that are phase-matched to virtual states are absorbed at a vastly enhanced rate~\cite{Pevrina1998,Saleh1998}. Even among transitions mediated by virtual states, the entangled photon absorption cross section is not proportional to the classical absorption cross section, as the enhancement depends on the detuning and linewidth of the virtual state~\cite{Roslyak2009}. The two-photon transition amplitudes contributed by the entangled photons can also interfere, producing ``entanglement-induced transparency'', analogous to electromagnetically induced transparency~\cite{Fei1997}. An entangled two-photon microscope may prove the only tool capable of probing the properties of virtual states.

 Although these techniques have not yet been applied in biological measurements, these preliminary results suggest that entanglement could be extremely promising in future two-photon microscopy applications. It could both enhance the visibility in two-photon fluorescent microscopy, and reveal classically inaccessible information.

\subsection{Quantum super-resolution in fluorescence microscopy}
\label{quantum_super_flour_mic_sec}

While the non-classical states used in most quantum metrology experiments require sophisticated state preparation, systems such as fluorescent particles naturally emit non-classically correlated light. These quantum correlations are ignored in classical experiments, though recent results have shown that measurement of the correlations provides additional information that can be used to enhance spatial resolution~\cite{Schwartz2012}. In classical fluorescent microscopy, the optical diffraction limit restricts the resolution with which fluorescent particles can be distinguished to approximately half the wavelength of light (see Section~\ref{ResolutionBio}).  While a number of techniques have been developed to overcome this limit, almost all rely on optical nonlinearities in the fluorescent particle (see, also, Section~\ref{ResolutionBio}). The only exception is structured illumination, which only requires the illuminating field to be controllably modulated to shift high spatial frequency patterns into the measurable range. This approach allows the resolution to be improved by a factor of two, and has become an important tool in biological microscopy~\cite{Langhorst2009}, enabling for instance direct imaging of molecular motor dynamics in living cells~\cite{Kner2009}. Broadly speaking, while other techniques use nonlinearities to provide more resolution information, structured illumination can be though of as a technique that makes full use of the aperture of the microscope objective to capture information.

In a similar manner, photon statistics in the fluorescent emission carries information. However, this information is usually neglected. In fluorescence microscopy, a fluorescent particle absorbs energy as it is excited to a higher state, and is then imaged as it decays back to the ground state by re-radiating at another wavelength. This mechanism generally only allows emission of one photon at a time, which results in photon anti-bunching and associated non-classical photon correlations (see Section~\ref{CorrelationSection}). When multiple photons are simultaneously measured, they must have originated from separate fluorescent centers. This provides an additional method to discriminate the emitted fields of closely spaced emitters which can be used to enhance resolution.

 This concept has been applied in two separate approaches to resolve fluorescent particles below the diffraction limit. In Ref.~\cite{Cui2013} the positions of fluorescent particles was estimated from both the intensity profile and the coincident detection events. The intensity profile and the coincident detection events both contributed information which could be used to statistically estimate the fluorescent center locations. Such an estimation procedure is a class of image deconvolution~\cite{Park2003} (see Section~\ref{ResolutionBio}), with the measurement of photon coincidences providing information that is not available in classical deconvolution methods. In principle this allows improved resolution, though to date the achievable improvement in the statistical estimation has not been quantified.

\begin{figure}
 \begin{center}
   \includegraphics[width=8.25cm]{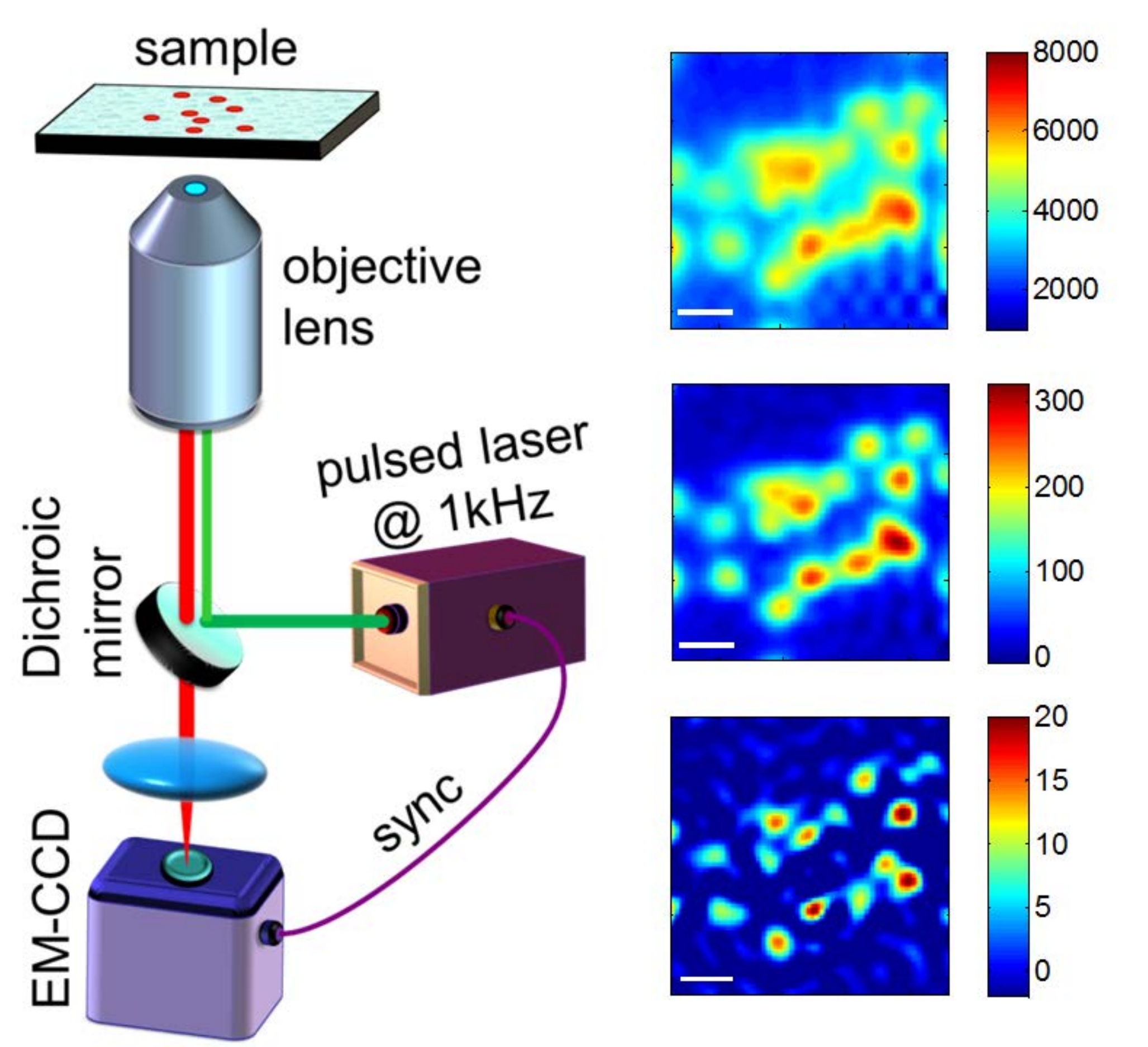}
   \caption{Super-resolution via measurement of photon coincidence statistics. Fluorescent particles are illuminated with short pulses of laser light, and the resulting fluorescence is measured on a high efficiency camera. Each pulse is sufficiently fast that a particle can only be excited once per pulse, and each coincident photon must originate from a different fluorescent particle. An analysis of the $N$ photon coincidences then allows a $N^{1/2}$ enhancement in resolution, with example data shown at the right. This shows the $N=1$, which is simply the intensity profile, the $N=2$ and the $N=3$ data. As $N$ increases the resolution is visibly improved, though the contrast is degraded. Reprinted with permission from Ref.~\cite{Schwartz2013}. Copyright (2013) American Chemical Society.
}
 \label{SuperFluorescent}  
  \end{center}
\end{figure}
 
 A more sophistocated approach was also demonstrated in Ref.~\cite{Schwartz2013}. In that work a large collection of fluorescent particles were excited with pulsed light and simultaneously measured in wide-field imaging. With this configuration,  the fluorescent particles could only emit one photon each per excitation pulse. Consequently, multi-photon detection events could occur only when multiple emitters were separated by less than the diffraction limit of the microscope. For locations with only a single emitter, the measured light is highly anti-bunched, with a second order coherence function which ideally reaches $g^{(2)}(0)=0$ (see Section~\ref{CorrelationSection}). By contrast, the light from multiple co-located emitters is only weakly anti-bunched. The anti-bunching can be quantified by the parameter $(1-g^{(2)}(0))$, which drops to zero for uncorrelated light.  By measuring both the one- and two-photon coincidences at each pixel, the spatial profile of both anti-bunching and  intensity can be determined. Together, these allow the {\it square} of the  single-photon emission probabilities to be mapped out. When compared to the intensity image, which represents the profile of single-photon emission probabilities, this  allows a resolution enhancement of $2^{1/2}$~\cite{Schwartz2012}.

 This approach can  be extended to higher order correlations, with measurements of 1 to $N$ coincident detection events at each pixel allowing the $N^{\rm th}$ power of the  single-photon emission probabilities to be mapped, with a resulting $N^{1/2}$ enhancement in resolution~\cite{Schwartz2012}. This was demonstrated experimentally with photon coincidences measured up to $N=3$, which allowed the spatial resolution to be enhanced from 272~nm to 181~nm~\cite{Schwartz2013}. A very recent extension of this technique  has shown that theoretically unlimited resolution is achievable with improved analysis~\cite{Monticone2014}. This is because each fluorescent center can only emit a single photon, so provided one can estimate the number of fluorescent centers within a region of the image, one can reasonably assume that there are no higher number photon coincidences. Under this assumption, measurement of the 1 to $N$ photon coincidences is sufficient to reconstruct the photon emission probabilities to arbitrarily high order. In this scenario, the resolution becomes limited only by the statistical uncertainty in the determination of  the 1 to $N$ photon coincidences~\cite{Monticone2014}.

 Since this approach involves parallel imaging, it can in principle proceed at high speed and with extremely narrow resolution. This enhancement could hold practical significance, just as structured illumination is already used in important biological applications. Furthermore, applying this method in conjunction with nonlinearities or structured illumination could potentially combine the resolution enhancements, which would allow this technique to resolve smaller features than any directly comparable classical technique.

\subsection{Super-resolution quantum lithography}\label{QLithography}

The field amplitudes of non-classical states of light follow the usual rules of diffraction, and cannot be focused into a smaller spot size than specified in Eq.~(\ref{DiffractionLimitEq}). It is therefore not possible to use entanglement to enhance resolution in far-field optical imaging with linear measurement. However, quantum correlations in non-classical states can allow distinctly non-classical characteristics in multi-photon or nonlinear measurement schemes, as discussed in Section~\ref{CorrelationSection}. For instance, although an $N$ photon NOON state can only be focused into a diffraction-limited region described by $x_{\rm min}$, coincident detection of all $N$ photons can be localized to  $x_{\rm min} N^{-1}$~\cite{DAngelo2001} (see Section~\ref{NOONState}). By comparison, coincident absorption of $N$ photons is classically localized to $x_{\rm min} N^{-1/2}$. Consequently, it is in principle possible to achieve a resolution enhancement of $N^{-1/2}$ in multi-photon absorption microscopy, or in lithography of a multi-photon absorbing substrate~\cite{PhysRevLett.85.2733,DAngelo2001}.

 If, however, single photon losses are present, the entanglement of the NOON state is quickly lost (see Section~\ref{Loss}). In principle, the enhanced resolution could still be observed in the instances that no photons were lost, i.e., 
 by post-selection of events for which $N$ photons are detected. However, high losses and difficulties in finding materials with suitably high multi-photon absorption have posed severe constraints. Although the narrowed region of multi-photon coincidences was observed with photon-counting detectors in the year 2001~\cite{DAngelo2001}, to date no experiment has succeeded in achieving super-resolution on a multi-photon absorber, as required for its application in imaging or lithography~\cite{Boyd2012}. 

 In the above proposal, only the photons which are co-localized are included in the measurement. It has been shown that the centroid of the entangled field can be estimated from measurement of all the photon arrivals, which therefore improves efficiency by a factor of approximately $N$ while maintaining the resolution of  $x_{\rm min} N^{-1}$~\cite{Tsang2009}. This has recently been demonstrated experimentally~\cite{Rozema2014}. This constitutes a demonstration of quantum imaging below the diffraction limit, with potential applications in biological imaging. However, it is important to note that optical centroid measurement is not a fundamentally diffraction-limited problem. For comparison, the centroid of a coherent field can be estimated to $x_{\rm min} N^{-1/2}$~\cite{Treps2002,Treps2003}, and is routinely estimated with precision many orders of magnitude smaller than $x_{\rm min}$ in applications such as optical tweezers and atomic force microscopy.

\subsection{Differential interference microscopy}
\label{diif_inf_mic_sec}

Quantum metrology can also provide enhanced sensitivity in differential interference microscopy, which is a technique that is widely used for biological imaging. In such measurements, a light beam is split and focused through two slightly different sections of a sample. The phase difference between the paths is then measured to determine differences in the refractive index along differing paths. This can then be used to reconstruct the spatial profile of the refractive index within the sample~\cite{Kam1998}.

It is well known that quantum correlated light can enhance the precision of phase measurements, as we have seen in Section~\ref{PhaseTheory}.
This suggests that phase contrast microscopy can also be enhanced  with quantum correlated light.  The feasibility of this was demonstrated in Ref.~\cite{Ono2013}, where two-photon NOON states were used to allow sub-shot noise imaging. In this case, the two photons each occupied one of the two linear polarizations, and were separated with polarizing optics. Both polarization modes were then focused through a sample, and slight differences in the two paths were measured via their effect on the relative phase. A full image could then be reconstructed by raster scanning over the sample. In this case, however, the light was not tightly focused and the resolution achieved was approximately 20~$\mu$m; far inferior to comparable classical techniques.

A closely related approach was later taken in Ref.~\cite{Israel2014}. In this case, the relative phase between two orthogonal polarization states was measured to image sample birefringence. This allowed polarization imaging of a small transparent sample; in this case, a quartz crystal. Additionally, this experiment also used lenses with a numerical aperture of 0.6, and achieved micron-scale resolution. While not state-of-the-art, this resolution is comparable with standard microscopes, and shows that such imaging techniques are compatible with high resolution imaging.
 
 In both of these experiments, the sensitivity was limited by the presence of relatively high levels of loss which degraded the input states (see Section~\ref{Loss}). Consequently,  post-selection was required to observe quantum enhanced precision. However, this limitation is purely technical, and can be resolved  with improved optics and state generation. An extension of this technique which applies high flux quantum correlated light with classically optimized spatial resolution could be expected to be of great benefit for biological research.

\section{Spin based probes of relevance to biology}\label{spinsection}

 While the discussion so far has focused entirely on optical measurements, quantum metrology is also performed with non-optical systems. In particular, non-classical spin states are a promising technology for future applications. Although such states have not been discussed in this review, a quantum treatment of spin-based measurements follows a similar treatment to that in Section~\ref{PhaseTheory}, with precision limits given by the quantum Fisher information (Section~\ref{QuantumFI}). A comprehensive review of spin based quantum metrology is beyond the scope of this work, and would warrant an entire review in its own right. Here we briefly overview some technologies which are likely to have future quantum metrological applications with biological systems.

\subsection{Atomic magnetometers}\label{Magnetometer}

 The development of high sensitivity magnetometers capable of measuring biomagnetic fields has both advanced biological research and enabled new methods in medical diagnosis. In particular, such studies have provided rich information about the function of the heart~\cite{Fenici2005} and brain~\cite{Hamalainen1993}, which both produce a relatively large biomagnetic field. These measurements have typically been carried out with a superconducting quantum interferometric device (SQUID), which uses interference in a superconducting circuit to sense magnetic fields with extreme precision~\cite{Clarke1996}; mostly because for a long time the SQUID was the only magnetometer sensitive enough for  useful biomedical applications. To maintain its superconducting circuit a SQUID requires cryogenic cooling, making it bulky and expensive. This limits the clinical applications described above to large and well-funded institutions~\cite{Johnson2013}.
 
  With recent advances, atom based magnetometers can now provide the sensitivity of a SQUID while operating at room temperature~\cite{Kominis2003,Dang2010,Budker2007}. As such, atomic magnetometers are used in the above mentioned biological applications, measuring both the dynamics and spatial profile of the biomagnetic field generated by a beating heart~\cite{Bison2003,Bison2003_dynamic,Bison2009} and neural activity in the brain in response to stimuli~\cite{Xia2006,Johnson2013}. Magnetic resonance imaging (MRI) is another important application which has previously required cryogenic cooling of superconductors~\cite{Hendee1984}; either in the form of superconducting magnets to produce the fields, or a SQUID to measure the low-field MRI signal. Now, atomic magnetometers have been applied to allow cryogen-free low-field MRI~\cite{Savukov2009}.

 \subsubsection{Quantum limits to sensitivity}

 Atomic magnetometers are based on optical manipulation and readout of the spin state of an atomic ensemble in the presence of a magnetic field.  Light that is near-resonant with an optical transition spin polarizes the hyperfine levels of the ground state, which subsequently undergo Larmor spin precession in the magnetic field. The phase and amplitude of the transmitted light then encodes information about the spin precession, from which the magnetic field can be estimated.

  \begin{figure}
 \begin{center}
   \includegraphics[width=8.25cm]{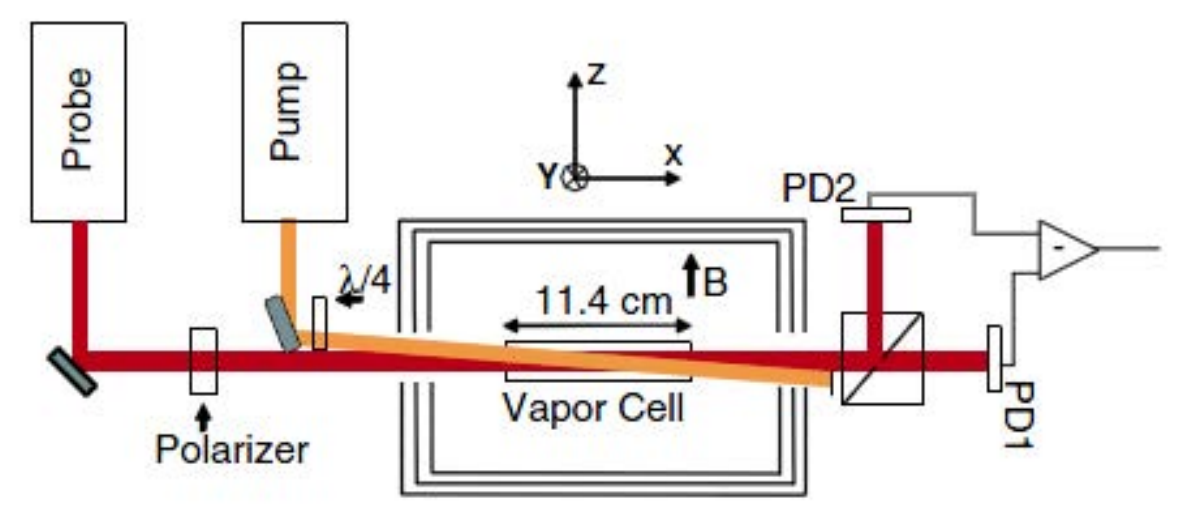}
   \caption{Layout for QND atomic magnetometry. A pump field is used to spin polarize an atomic vapor. After this a probe field is passed through the cell, and the magnetic field estimated from the Faraday polarization rotation.  Reproduced with permission from Ref.~\cite{Shah2010}, Copyright (2010) by The American Physical Society.
}
 \label{MagnetometerLayout}  
  \end{center}
\end{figure}
 
 The sensitivity of this measurement is fundamentally limited by quantum noise in both  the optical readout and the spin of the atomic ensemble. Quantum noise in the spin states, referred to as projection noise, results from the projection of the atomic spin onto the measurement axis. The mean spin is oriented orthogonal to the measurement axis to provide optimal estimation of the spin precession. Projection noise then follows from the statistical fluctuations in the measurement. The optical readout can also be limited by quantum shot noise. While the dominant source of quantum noise depends on the details of the measurement, the contributions of projection noise and optical shot noise are comparable when the measurement is optimized~\cite{Auzinsh2004}.

 \subsubsection{Enhancement with entangled atoms}

 There has been much interest in the possibility to improve sensitivity beyond the projection noise and shot noise limits. Optical readout of the Faraday polarization rotation is a quantum non-demolition (QND) measurement of the spin polarization state along one axis~\cite{Shah2010}. In a QND based magnetometer, each measurement both provides information about the magnetic field and  projects the system onto a spin-squeezed state, increasing the sensitivity of subsequent measurements.

   \begin{figure}
 \begin{center}
   \includegraphics[width=8.25cm]{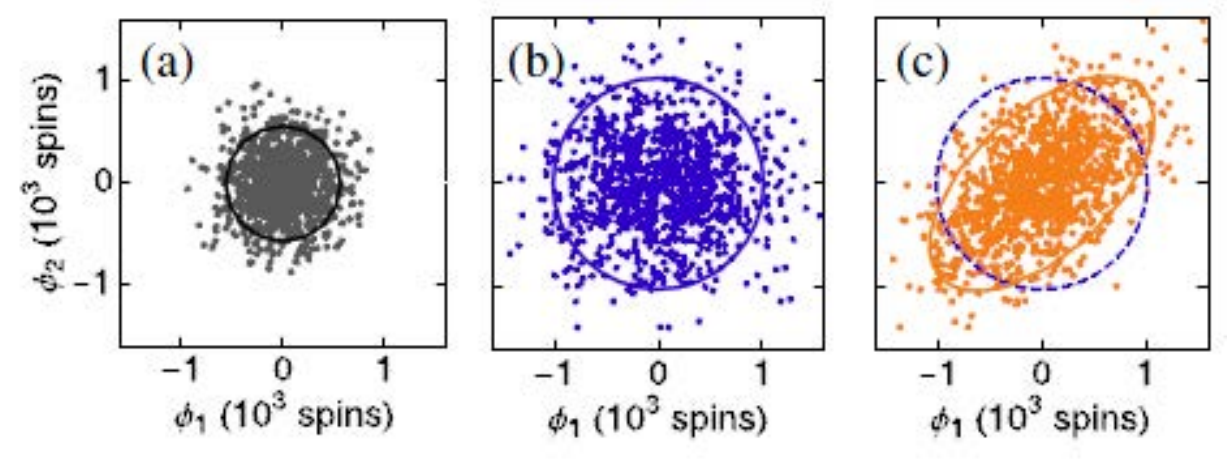}
   \caption{Measurements with spin squeezing. The measurement statistics of  QND measurements were studied, first with no atoms in the trap to characterize the readout noise (a) and then independently prepared coherent spin states (b), with the measured distribution corresponding to the projection noise. The solid curves indicate 2$\sigma$ radii for Gaussian fits.  (c) When a single coherent spin state is prepared and sampled sequentially, the successive measurements squeeze the spin variance below the projection noise limit (dashed circle) along one axis. 
Reproduced with permission from Ref.~\cite{Sewell2012}, Copyright (2012) by The American Physical Society.   
}
 \label{SpinSqueezing}  
  \end{center}
\end{figure}

 An early theoretical analysis suggested that quantum correlations within the spin ensemble would be destroyed by spin relaxation, rendering spin squeezing effective only for measurements much shorter than the spin-relaxation time~\cite{Auzinsh2004}. Later analysis has shown that the spin relaxation preserves quantum correlations, and that spin squeezing could be used to  vastly improve the sensitivity on any timescale~\cite{Kominis2008}. This has then been applied experimentally, with QND measurements used to induce spin squeezing in an atomic ensemble, resulting in improved sensitivity to magnetic fields~\cite{Koschorreck2010,Sewell2012}. One variation of this approach is to use QND measurements to induce anticorrelated noise in two separate vapor cells, such that the resulting two-mode squeezing (or equivalently, entanglement) suppresses the measured projection noise~\cite{Wasilewski2010}. This approach enabled an absolute sensitivity approaching that achieved in state-of-the-art atomic magnetometers. Another related approach applied QND measurements to a scalar atomic magnetometer and achieved the most sensitive scalar magnetic field sensitivity to date, even without demonstrating spin squeezing~\cite{Sheng2013}.

\subsubsection{Use of quantum correlated light}

In addition to the enhancement by spin squeezing, non-classical states of light have also been used to overcome the quantum noise on the optical readout~\cite{Wolfgramm2010,Horrom2012,Wolfgramm2013}. The optical  magnetometer in Ref.~\cite{Wolfgramm2010} achieved sensitivity better than the shot-noise limit using a polarization-squeezed probe (see Section~\ref{SqueezedState}) tuned near the atomic resonance. In that work, however, the atomic vapor was not spin polarized, but instead kept in a thermal state, which resulted in poor absolute sensitivity. A similar experiment was reported later in  Ref.~\cite{Horrom2012}, with the absolute sensitivity vastly improved and the frequency range of squeezing extended into the biologically relevant sub-kHz regime.  
 Such an enhancement could be particularly important once it is applied in a QND based magnetometer, because it acts both to improve the measurement sensitivity and also to improve the spin squeezing. Overall, this improves the  fundamental limit on sensitivity for a QND based magnetometer~\cite{Auzinsh2004,Wolfgramm2010}. To achieve this, it is important that there is no increase in spin decoherence induced by the probe when using quantum correlated light. To test for additional decoherence,  Ref.~\cite{Wolfgramm2013} applied two-photon NOON states to perform optical readout in atomic magnetometry. By characterizing the photon scattering and atomic excitations, they could verify that entangled light induced less decoherence in the spin state than coherent light which achieved the same sensitivity. 

\begin{figure}
 \begin{center}
   \includegraphics[width=8.25cm]{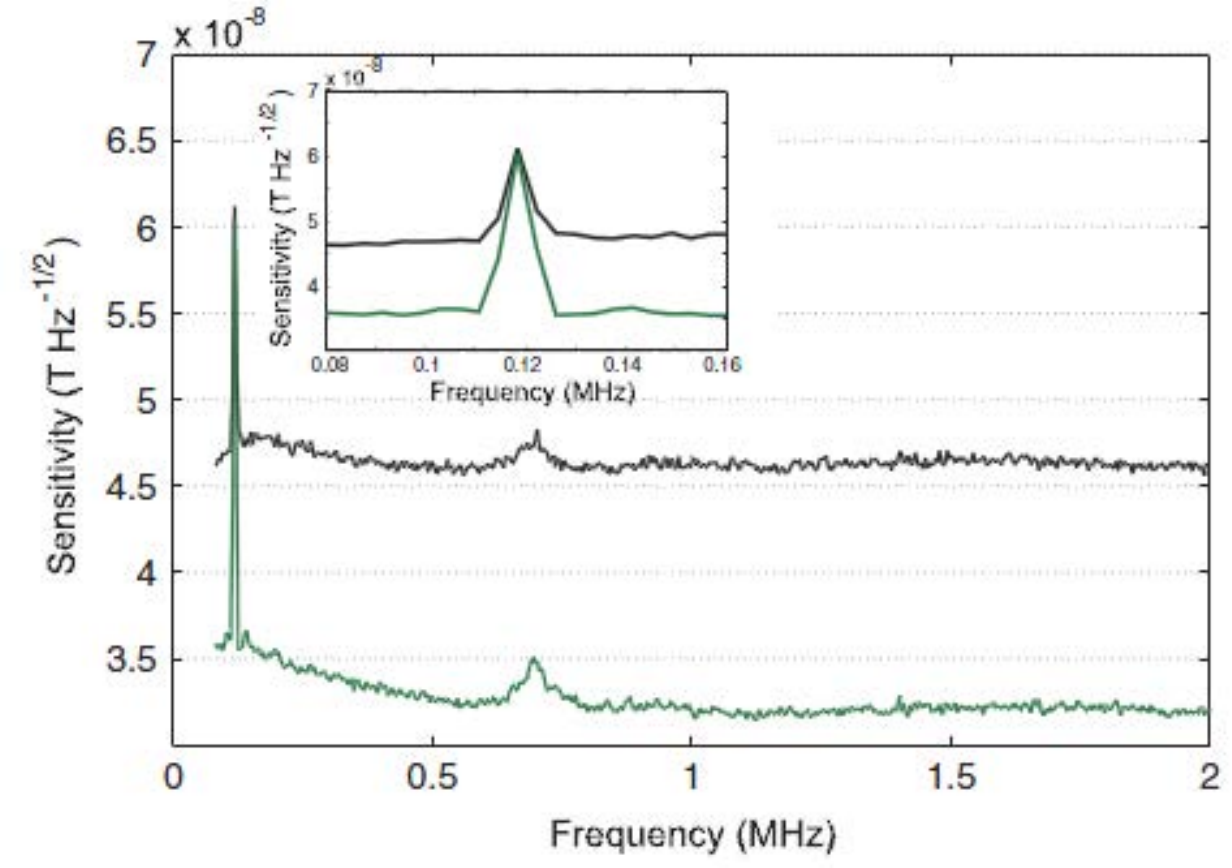}
   \caption{Magnetic field measurements with squeezed and coherent light. Reproduced with permission from Ref.~\cite{Wolfgramm2010}, Copyright (2010) by The American Physical Society.
}
 \label{SqzMagnetometer}  
  \end{center}
\end{figure}

By using squeezed light in a QND measurement to condition the spin-states of the atomic ensemble, it should be possible to achieve extremely high sensitivity. Once this is achieved in state-of-the-art magnetometers, the magnetic field sensitivity can be expected to outcompete SQUIDs, and have important biological and clinical applications.


\subsection{Nitrogen vacancy centres in diamond}

One of the key factors that enables spin polarisation, spin coherence, and quantum correlations to be established in the atomic magnetometers discussed in the previous section is that dilute gases of atoms are naturally well isolated from their environments and therefore from possible quantum decoherence channels. However, this limits applications in biology to scenarios where the atomic ensemble is also well isolated from the biological specimen, typically confined within a millimetre to centimetre sized gas cell~\cite{Mhaskar2012}. This prohibits the nano- or micro-scale resolution crucial for a great many biological applications (see Section~\ref{ResolutionBio}). An alternative approach is to use solid-state artificial atoms such as quantum dots or fluorophores as quantum probes, which are typically embedded within a crystalline host material. These can allow the close proximity required for nanoscale resolution, either by placing the biological specimen in direct contact with the substrate~\cite{LeSage2013}, or embedding a probe nanocrystal within the specimen~\cite{McGuinness2011,Kucsko2013}. 

While this close proximity allows nanoscale resolution, it also introduces significant environmental coupling to the state which results in strong quantum decoherence. Substantial advances have been made in reducing this decoherence by controlling the material properties of the host, most particularly its structural and chemical purity and surface quality; by employing control techniques such as dynamical decoupling to suppress the effect of environmental spin noise~\cite{BarGill2013}; and by identifying artificial atoms that display intrinsic shielding from their environments. Even so, cryogenic conditions are still required to achieve appreciable spin coherence with the majority of solid-state quantum probes~\cite{Chernobrod2005,Michaelis2000}. Nitrogen vacancy centres in diamond are a notable exception, exhibiting long spin coherence times even at room temperature~\cite{BarGill2013}. Combined with optical addressability, biocompatibility~\cite{Chao2007}, and photostability~\cite{Jelezko2006}, this presents unique opportunities for biological applications in quantum metrology which we briefly introduce in this Section.


\begin{figure}
 \begin{center}
   \includegraphics[width=8.25cm]{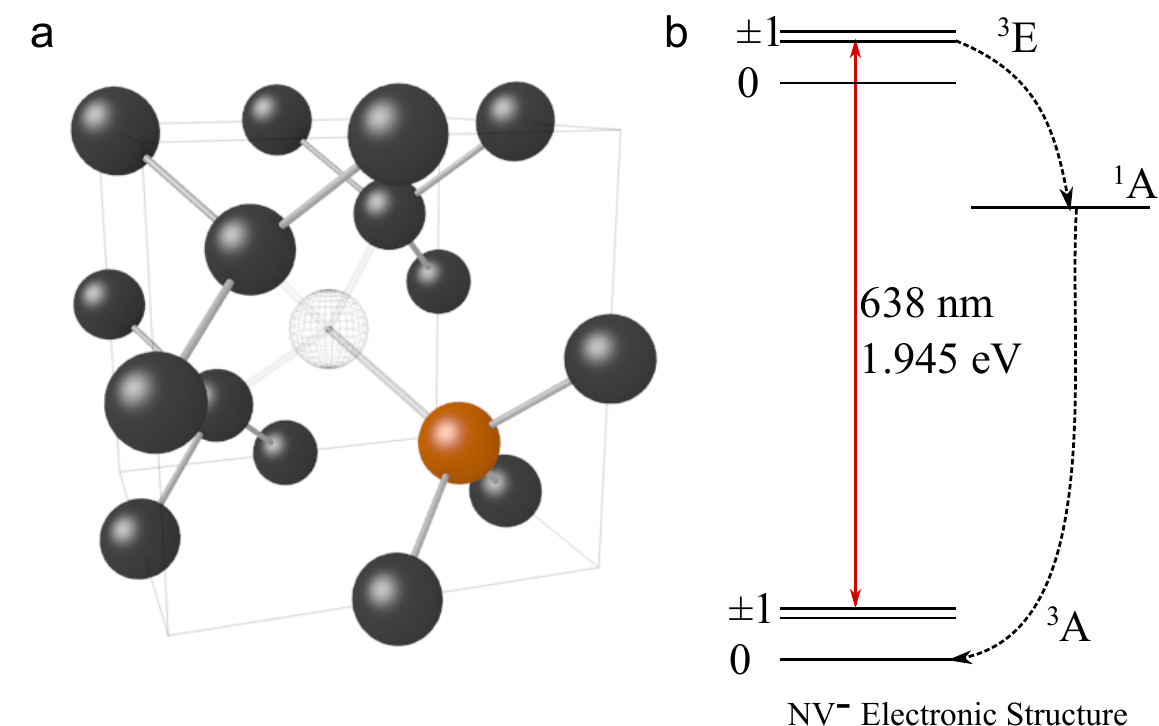}
   \caption{Schematic of the lattice structure of an NV centre. {\bf a} Carbon atoms are shown in black, with one Carbon replaced with a Nitrogen (orange) and one resulting vacancy. {\bf b} The resulting energy structure for the NV$^{-}$ state. The $m_s = \pm 1$ excited states can decay non-radiatively to the $m_s=0$ ground state, which therefore allows state preparation into the $m_s=0$ ground state. 
}
 \label{NVstructure}  
  \end{center}
\end{figure}

\subsubsection{Basic properties of nitrogen vacancy centres}

The nitrogen vacancy (NV) centre naturally forms as a defect in diamond when two nearest-neighbour carbon atoms are displaced, with one site filled with a nitrogen atom and the other left vacant (see Fig.~\ref{NVstructure}). The defect has two charge states, a neutral NV$^0$ state and a negative NV$^-$ state. The NV$^-$ state is particularly favourable for quantum metrological applications, featuring spin-triplet ground and excited states that can be conveniently accessed via optical transitions at wavelengths near 637~nm as shown in Fig.~\ref{NVstructure}(b). Very long spin coherence times of 0.6~s and 1.8~ms have been achieved at 77 K~\cite{BarGill2013} and room temperature~\cite{Balasubramanian2009}, respectively. The optical transitions allow both control and read-out of the spin state of the NV. Perhaps most significantly, due to non-radiative transitions that connect the $m_s = \pm 1$ excited states to the $m_s=0$ ground state, broadband optical illumination is sufficient to spin polarise the NV into its $m_s=0$ ground state. Microwave fields can then be used to coherently drive transitions between the $m_s=0$ and $m_s = \pm 1$ ground states; while significant differences between the fluorescence from the $m_s=0$ and $m_s = \pm 1$ levels allows accurate read-out of the NV spin state~\cite{Aharonovich2011}.

\subsubsection{{\it In vivo} thermometry and field sensing}

\begin{figure}
 \begin{center}
   \includegraphics[width=8.25cm]{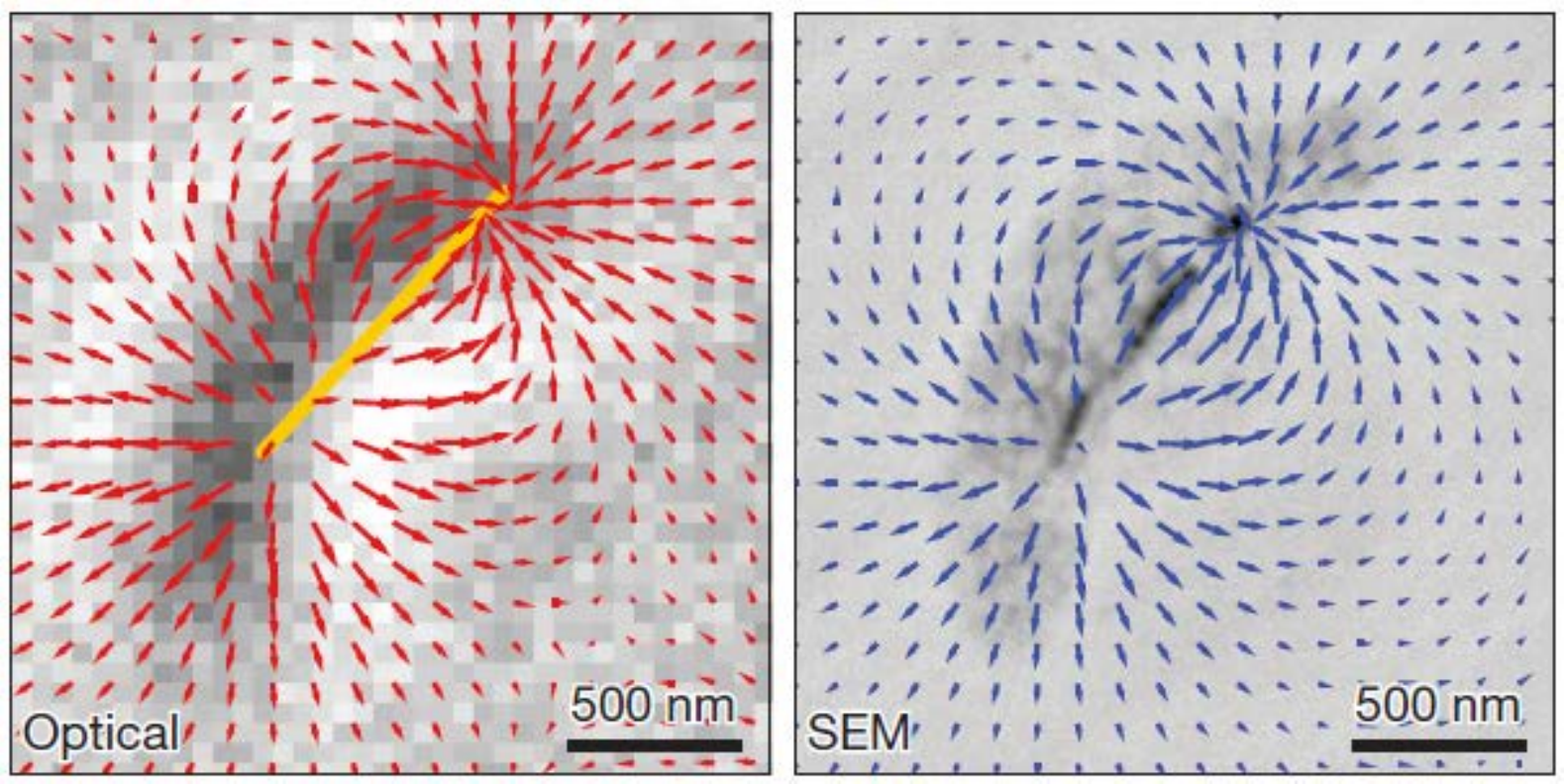}
   \caption{Magnetic field imaging of a live magnetotactic bacterium cell, with the left anr right panels respectively showing the measured and simulated magnetic field projections in the x-y plane, superimposed on the optical and backscattered electron images, respectively.  Reprinted by permission from Macmillan Publishers Ltd: Nature, Ref.~\cite{LeSage2013} Copyright (2013)
}
 \label{MagneticImagingCell}  
  \end{center}
\end{figure}

The ability to initialise, control, and readout spin states, are the key requirements to realise both magnetometry and electrometry with nitrogen vacancy centres~\cite{Doherty2013}. The exquisite sensitivity possible is clearly illustrated by the demonstration of nanoscale magnetic imaging of a single electron spin at room temperature~\cite{Grinolds2013}. In that experiment, Zeeman-shifts in the NV electronic spin resonance frequencies due to the electron spin were read-out using optically-detected magnetic resonance (ODMR). Similar approaches have recently been applied in three experiments to perform NV-based magnetometry of biological systems. In the first, individual fluorescent nanodiamonds were introduced within living human HeLa cells and their orientation was tracked via continuous ODMR~\cite{McGuinness2011}. In the second, ODMR was used to characterise the magnetic field of living magnetotactic bacteria~\cite{LeSage2013}.  The bacteria was placed upon a diamond chip in close proximity to a thin NV layer, allowing the magnetic field of the bacteria to be characterised with 400~nm resolution (see Fig.~\ref{MagneticImagingCell}). In the third and final experiment, cancer biomarkers expressed by tumour cells were quantified using correlated magnetic and fluorescence imaging on a millimetre field-of-view diamond chip~\cite{glenn2015single}

While the ODMR measurements in Ref.~\cite{McGuinness2011} were used to track the nanodiamond rather than extract meaningful information about the magnetic properties of the cell, the same reference experimentally demonstrated the technique of quantum decoherence sensing~\cite{Hall2009,Cole2009}. Such sensing seeks to take advantage of the extreme sensitivity of the coherence time of qubits to learn something of the nature of their environment. Ref.~\cite{McGuinness2011} probed the coherence time of the spin of an NV-nanodiamond within an HeLa cell using Rabi and spin-echo sequences. Subsequent work has used this approach to image microbiological structures within a diamond microfluidic channel~\cite{Steinert2013}. Here, the approach is to introduce labels with large magnetic moments within the channel. In the absence of an external magnetic field, these labels freely diffuse. When in close proximity to an NV-spin, the stochastic magnetic field fluctuations they introduce act to decrease the NV coherence time. In Ref.~\cite{Steinert2013} Gadolinium ions (Gd$^{3+}$) were chosen as the label due to their very large magnetic moment. This allowed microfluidic detection of as few as 1000 statistically polarised spins; as  well as imaging of 150~nm thick microtomed sections of HeLa cells with 472~nm resolution, though this latter demonstration was performed with pre-frozen cell sections embedded in PDMS (polydimethylsiloxane) and was therefore not compatible with experiments in living biology.

Nitrogen vacancy centres have also been utilised for nanoscale thermometry within living cells with a sensitivity of 9~mK~Hz$^{-1/2}$~\cite{Kucsko2013}. In this experiment, the separation of the $m_s = 0$ and $m_s = \pm 1$ ground states of the NV-spin was accurately determined using a spin-echo sequence. This separation is linked to the temperature of the nanodiamond due to thermally induced lattice strain. The measurement used a nanodiamond containing an ensemble of approximately 500 uncorrelated NV spins, with the sensitivity correspondingly enhanced by a factor of $\sqrt{500}$. This enabled superior sensitivity to  all previous nanoscale thermometers, offering the prospect to study the nanoscale energetics of biophysical processes such as gene expression and tumour metabolism.  

\subsubsection{Quantum correlations}

To date all biological experiments with NV centres have relied either on a single NV or an ensemble of uncorrelated NVs. As discussed in Section~\ref{PhaseTheory}, in quantum metrology quantum correlations and entanglement can be used to suppress measurement noise, and ultimately improve how the measurement sensitivity scales with particle number. Such techniques could be particularly relevant to nanodiamond NV sensors, where the number of NV centres in the nanodiamond are constrained both by its physical dimensions and by NV-NV interactions that cause decoherence~\cite{Jarmola2012}. Quantum correlations between the 500 NV centres contained within the nanodiamond thermometer discussed above, for instance, could in principle yield over an order-of-magnitude improvement in sensitivity. Techniques capable of generating such correlations in an NV ensemble are currently under development~\cite{BarGill2013}. For example, one method which is being considered is to use strain-induced resonance frequency shifts to couple multiple NV spins to the motion of a micromechanical oscillator~\cite{Bennett2013}. This discussion is far from exhaustive; for a more detailed review see Ref.~\cite{Doherty2013}.

\section{Conclusion}\label{conclusion}

The field of quantum metrology broadly encompasses the use of quantum correlated states  to enhance or enable measurements.  This is motivated by two distinct goals. One aim is to establish the fundamental consequences of quantum mechanics on measurements. Toward this end, the quantum limits on measurements and the strategies required to overcome them are both studied. The other aim of quantum metrology is to harness quantum effects to provide a practical benefit in measurement applications.  The primary motivation for these experiments is to out-compete their classical counterparts in some manner. Currently, the best example of this in optics is in gravitational wave observatories, where squeezed light is used to achieve sensitivity below the quantum noise limit~\cite{LIGO2011,LIGO2013}. This is now routinely employed because it has proven to improve performance in absolute terms~\cite{Grote2013}. In the past few years, however, a range of proof-of-principle applications have emerged in biology and biophysics.

This review has introduced quantum metrology from the context of biological measurements, with a specific focus on the practical advantages which can be gained, and it has provided an overview of both the advances already achieved and the future potential for biological quantum metrology. We have sought to introduce the quantum measurement theory in the most accessible possible manner to the benefit of biophysicists and biologists who may wish to adapt quantum measurement techniques to their applications, and conversely, to introduce the challenges of biological imaging and measurement to the uninitiated quantum physicist. Thereby, we hope that this review will act, at least to some degree, as a bridge between the communities, guiding and accelerating future implementations of quantum measurement techniques to biological systems.


Biological measurements with improved performance  have  been an important goal of the field of quantum metrology since the 1980s
~\cite{Slusher1990}, encompasing a wide range of potential applications. Recent advances in quantum technologies outlined in this review now allow such applications, with proof-of-principle demonstrations already achieved using a wide variety of non-classical states. Entangled photon pairs, photonic NOON states, bright squeezed states, and single photon states have enabled enhancements in tissue imaging, protein sensing, measurement of the mechanical properties of living cells, and the study of the photodetection response of retinal rod cells (see Section~\ref{Progress}). Though these experiments have not yet outperformed state-of-the-art classical measurements, they have established the potential for quantum metrology to provide practical benefits to future experiments. Furthermore, they have  demonstrated that quantum metrology is relevant to a wide variety of different measurement applications.

A range of other technologies have been developed that are yet to be applied in a biological context, but have the potential to provide significant capabilities in the near future.
As we have seen in this review, these include microscopes which use entangled photons to probe two-photon transitions, super-resolution techniques  which enhance resolution by studying the photon antibunching of fluorescent emission, both phase-contrast and absorption microscopy, magnetometry with quantum correlated atomic spins, and nanoscale field and temperature imaging with NV centres (see Sections~\ref{Future} and \ref{spinsection}). This list is not exhaustive, with other quantum measurement technologies also holding promise for future applications in biology. 

This work was supported by the Australian Research Council Centre of Excellence for Engineered Quantum Systems CE110001013, the Air Force Office of Scientific Research, and the Asian Office of Aerospace Research and Development. We appreciate illuminating discussions with Simon Haine, Carl Caves, and Josh Combes. W.P.B. acknowledges the Australian Research Council Future Fellowship FT140100650.

\bibliographystyle{elsarticle-num}
\bibliography{biblPR}

\end{document}